\documentclass[journal]{IEEEtran}

\ifCLASSINFOpdf

\else

\fi

\usepackage{cite}
\usepackage{amsmath,amssymb,amsfonts}
\usepackage{algorithmic}
\usepackage{graphicx}
\usepackage{textcomp}
\usepackage{comment}
\usepackage{color,soul}
\pdfoutput=1
\allowdisplaybreaks[1]
\usepackage[subrefformat=parens,labelformat=parens]{subfig}
\usepackage{float}

\begin{document}

\title{Hardware Trojan Detection Using Controlled Circuit Aging}

\author{V. R. Surabhi$^{1}$, P. Krishnamurthy$^{1}$, H. Amrouch$^{2}$, K. Basu$^{3}$, J. Henkel$^{2}$, R. Karri$^{1}$, and F. Khorrami$^{1}$
\thanks{This work was supported in part by the Office of Naval Research under Grant N00014-18-1-2672.}
\thanks{$^{1}$Authors are with the Dept. of Electrical \& Computer Engineering, NYU Tandon School of Engineering, Brooklyn, NY 11201.
        {\tt\small \{virinchi.roy, prashanth.krishnamurthy,  rkarri, khorrami\}@nyu.edu}}%
\thanks{$^{2}$Authors are with the Department of Computer Science, Karlsruhe Institute of Technology, Karlsruhe, Germany.
        {\tt\small \{amrouch, henkel\}@kit.edu}}
\thanks{$^{3}$Author is with the Department of Electrical and Computer Engineering, University of Texas at Dallas, TX 75080 USA.
        {\tt\small kanad.basu@utdallas.edu}}
}

\maketitle

\begin{abstract}
This paper reports a novel approach that uses transistor aging in an integrated circuit (IC) to detect hardware Trojans. 
When a transistor is aged, it results in delays along several paths of the IC.
This increase in delay results in timing violations that reveal as timing errors at the output of the IC during its operation. We present experiments using aging-aware standard cell libraries to illustrate the usefulness of the technique in detecting hardware Trojans. Combining IC aging with over-clocking produces a pattern of bit errors at the IC output by the induced timing violations. We use machine learning to learn the bit error distribution at the output of a clean IC. We differentiate the divergence in the pattern of bit errors because of a Trojan in the IC from this baseline distribution. We simulate the golden IC and show robustness to IC-to-IC manufacturing variations. The approach is effective and can detect a Trojan even if we place it far off the critical paths. Results on benchmarks from the Trust-hub show a  detection accuracy of $\geq$99\%.
\end{abstract}


\IEEEpeerreviewmaketitle

\section{Introduction}
\label{sec:introduction}
Manufacturing of ICs is expensive and requires special fabrication equipment that becomes outdated in a short time. To reduce the cost of IC manufacturing, this task is typically outsourced to  offshore IC foundries. In a related trend, embedded systems source specialized intellectual property (IP) cores from different vendors. Design and assembly of the IP cores is done using third party computer aided design, integration, and test tools. As an IC design travels through the complicated supply chain, the IC could be corrupted at one of the stages. Example threats due to the corruption include passing off low quality ICs as good, infiltrating the supply chain with imitation ICs, and insertion of Hardware Trojans in ICs. Hardware Trojans, which are typically triggered by rare events, may alter function, deny service, or leak information.  Such infected ICs affect the critical information systems in finance, military, and health care. Functional and structural tests to weed out manufacture-time defects are ineffective against Trojans for the following reasons:
\begin{enumerate}
\setlength\itemsep{0in}
\item Structural tests produced by automatic test pattern generation (ATPG) may not detect a Trojan, since the behaviour of the Trojan may not be in the fault list~\cite{Wolff2008TowardsTT}.
\item Functional tests do not uncover Trojans since they trigger on rare events.
\item A brute force application of all inputs does not scale. For example, a 64-bit circuit will need $2^{64}$ inputs.
\end{enumerate}
Reverse engineering can  authenticate the IC but it does not guarantee that the unchecked ICs do not have Trojans  \cite{Agrawal2007TrojanDU}.

There are different types of Trojans, depending on their functionality. Some leak information, some change function, and some discharge power, etc. Therefore, a robust approach is needed to detect Trojans. Prior work
 \cite{chakraborty2009mero, huang2016mers, aarestad2010detecting, forte2013temperature, stellari2014verification}
has considered side channel and ATPG test pattern analysis. 
Power side channel fingerprinting is not a reliable method for detecting Trojans within a large circuit. In classic VLSI test, even if the test patterns cover all corner cases, they may not trigger a Trojan. For example, the trigger could be inputs applied in a particular sequence, which is extremely challenging to replicate. 
We explore  aging for Trojan detection. In transistor aging, we do not need the Trojan to be triggered because the underlying aging mechanism will naturally occur during the circuit's operation. From results (see Section~\ref{sec:experimental}), aging can detect small Trojans (occupying 0.22\% of the circuit) and even if it is about 4000 paths far from critical path. 

In this work, transistor aging along with over-clocking is used to expose the Trojan effects on various circuit properties. When aging and over-clocking are jointly applied to an IC, they produce a pattern of bit errors at the output. The output bit error patterns for clean IC are used train a Support Vector Machine to learn the clean IC output pattern distribution to determine presence of Trojan at the test time. The efficacy of our approach is shown on  gate-level simulations on Advanced Encryption Standard (AES)  and Rivest-Shamir-Adleman (RSA) crypto circuits with different Trojans that show different challenges using aging-induced standard cell libraries. This method applies to any circuit but the crypto circuits are used for secure data transmission and hence are subject to attacks. The Trojans appear at different locations corresponding to the rank of the critical path. The experimental results show an accuracy of over 99\% on all the circuits when considered with varying number of inputs.

The paper is organized as follows. 
 Section~\ref{sec:rel} discusses prior work on Trojans. Section~\ref{sec:aging} describes transistor aging and its effects. Section~\ref{sec:cell_lib} overviews the creation of aging-aware standard cell libraries. 
Section~\ref{sec:methodology} introduces our approach and Sections~\ref{sec:setup} and \ref{sec:experimental} outline the experimental setup and results. Section~\ref{sec:conclusions} draws the conclusions.

\section{RELATED WORK}
\label{sec:rel}

\subsection{Hardware  Trojans}

Globalization has led to a distributed IC manufacturing environment.
The globally-distributed IC design cycle has led to a lot of vulnerabilities, including Hardware Trojans. A Hardware Trojan is a malicious modification to the circuit, which is unknown to the designer and can have consequences like incorrect functionality, loss of secret information, etc. Hardware Trojans are critical threats to military, finance, transportation and corporate or consumer electronics \cite{tehranipoor2010survey}. 
A Hardware Trojan has two parts -- trigger and payload. Trigger signal activates the Trojan. Payload is the effect of the Trojan.  A trigger is a  signal in the circuit which is rarely activated.  As a result,  the payload is dormant during normal function of the circuit. Hence, the Trojan  is difficult to  detect. Trojans can be classified based on  five attributes: insertion phase, abstraction level, location, trigger and payload \cite{karri2010trustworthy, tehranipoor2011introduction}. 

Trojans can be inserted in various stages of a design flow. Semiconductor companies use third party EDA tools, third party IPs (3PIPs), and untrusted foundries. Insertion of Trojans at various  stages of EDA design flow  has been demonstrated by \cite{basu2019cad}. 
Insertion of Trojans during High-level Synthesis was proposed  by \cite{pilato2018black}.  
\cite{king2008designing} designed  a  malicious processor by modifying the open source Leon processor. The Trojans allow a user to  violate Operating System exceptions and execute  a malicious firmware. Don't care states in a  design  were  utilized to trigger Hardware Trojans by  \cite{dunbar2014designing}.  \cite{zhang2011case} triggers Trojans by exploiting silicon wear-out.

\subsection{Trojan Detection}
\label{sec:detection}
Trojan detection methods are usually applied either at the design stage or post-manufacturing stage to verify them~\cite{xiao2016hardware}. Pre-silicon detection approaches are used to validate 3PIP cores before integrating them to a design. Pre-silicon verification is performed using functional  validation, structural analysis or formal verification. Functional validation methods use functional tests to activate a Trojan and validate  the response against a ``golden'' Trojan-free circuit response. Since Trojan triggers are rarely turned on, researchers have developed test generation techniques that can activate those rare triggers \cite{chakraborty2009mero, huang2016mers}. However,  functional tests fail to detect a non-functional Trojan which  does  not  alter the function of the circuit and transmits secret data. Structural analysis involves identifying redundant statements and circuits in the HDL code \cite{zhang2011case, hicks2010overcoming}. 

Post-silicon Trojan detection involves either destructive or non-destructive testing. Destructive testing implies reverse-engineering and de-layering to detect the presence of malicious circuitry \cite{xiao2016hardware}. Although this approach is costly, time consuming and renders the IC useless, it guarantees Trojan detection in the single IC.  Non-destructive  methods use  functional testing (similar to pre-silicon Trojan detection) and side-channel analysis. The value of a side-channel parameter  will differ between a Trojan activated circuit and a ``golden'' circuit. 
In \cite{jin2008hardware}, path delay information of the IC at each output is considered to generate a path delay fingerprint. The path delay fingerprints help distinguish between clean IC and Trojaned IC. There are millions of paths in ICs nowadays, it is not practical to measure the delay for all the paths. Also, the method does not work well for the Trojans that leak information through side channels. Temperature tracking using on-chip thermal sensors during run-time is an option \cite{forte2013temperature,PKA20,AKPHKK17}. However, if the Trojan activity lasts for a short duration, the slow rate of thermal variation cannot detect the Trojan. Furthermore, adversary can place the Trojan in an active area of the chip. This makes the Trojan detection difficult. In \cite{aarestad2010detecting}, leakage current measurement detects Trojans since additional gates consume extra leakage power. In \cite{stellari2014verification}, Picosecond Imaging Circuit Analysis (PICA) is used to measure optical emissions of the ICs and compare them with a trusted emission image of a ``golden'' IC. In both the methods (i.e., power and radiation), access to ``golden IC'' is required and as the feature size of IC shrinks, the deviation from ``golden'' IC due to process variations become  pronounced, compensating for the deviations introduced by the Trojans. Our method does not need a ``golden IC''. We simulate the circuit for the golden IC and show robustness of the technique to IC-to-IC manufacturing variations. The approach is aging-based, non-destructive and can detect a Trojan even if it is dormant, removing the above limitations.

\section{TRANSISTOR AGING}
\label{sec:aging}
Semiconductor technology has advanced to the nanometer regime wherein electric fields are stronger with every new generation to allow the transistor to switch faster. When an IC is turned ON, the Bias Temperature Instability (BTI) effect comes into play and increases the threshold voltage ($\Delta V_{th}$) of transistors. The magnitude of ($\Delta V_{th}$) depends on the supply voltage ($V_{dd}$). High $V_{dd}$ causes a large increase in $V_{th}$, thus increasing the delay of paths of circuit leading to a noticeable performance degradation. Modern ICs use fast voltage regulator (switching between voltage levels in less than a micro second) to implement effective power management schemes in which the overhead of voltage scaling is minimized. The high frequency of voltage switching does not allow for degradation accumulated at high $V_{dd}$ to settle down (i.e. to recover) at low $V_{dd}$. This causes transient timing errors due to aging until the transient state disappears  in which generated defects, caused by BTI aging at the high $V_{dd}$, partially or fully recover. This is known as short-term aging~\cite{vanagingawarevoltagescaling}. Switching voltage from low $V_{dd}$ to high $V_{dd}$ does not cause aging effects as the degradation at low $V_{dd}$ is less and the circuits get more robust at high $V_{dd}$.

\subsection{Effects of Transistor Aging}
\label{sec:aging_effects}

Technology scaling is approaching its limits  displacing a few atoms in a transistor during operation is akin to aging and can endanger their key electrical properties. The key aging phenomena are Negative and Positive Bias Temperature Instabilities (NBTI and PBTI), with a potential to degrade the switching speed of pMOS and nMOS transistors. BTI occurs when the vertical electric field is applied to the transistor in which some of the minority carriers --that are being attracted to form the transistor's channel-- may combine with the available $Si$-$H$ bonds at the $Si$-$SiO_2$ interface layer resulting in \textit{interface traps}. Some of these carriers may move to the transistor's dielectric due to quantum tunneling and captured by the oxide vacancies resulting in \textit{oxide traps}. These defects interfere with the applied electric field and weaken it due to Coulomb scattering. As a result, the transistor can switch from OFF to ON state only at a higher gate voltage than the fresh device (i.e.~in the absence of aging). Hence, the threshold voltage ($V_{th}$) of the transistor increases. In addition, the generated interface traps reduce the mobility of carriers ($\mu$) as they move from source to drain due to Coulomb scattering. 

\noindent
\textbf{Aging-Induced Timing Errors:}
The delay of a transistor is proportional to its current in the ON state ($I_{ON}$).  $I_{ON}$ is a function of the threshold voltage and the carrier mobility as in Eq.~\ref{eq:I_d}~\cite{amrouch2016reliabilityaware}. An increase in $V_{th}$ plus a decrease in $\mu$ due to aging reduces $I_{ON}$ and increases transistor delay. 

\begin{equation}
\label{eq:I_d}
\text{Transistor delay } t_d = \frac{1}{I_{ON}} \text{ ; } 
I_{ON} \approx \frac{\mu}{2} \cdot C_{ox} \cdot \frac{W}{L}  \cdot  (V_{dd} - V_{th})^2
\end{equation}
where, $C_{ox}$, $W$ and $L$ are oxide capacitance, width, and length of transistor. $V_{dd}$, $V_{th}$, and $\mu$ are operating voltage, threshold voltage, and carrier mobility, respectively. 
 
Aged transistors slowdown increasing the likelihood of timing violations and errors in circuits, as shown in Eq.~\ref{eq:timing_errors}, if no (or in-sufficient) timing guard band is included. This is because the switching frequency is unsustainable causing timing violations in critical paths and these.propagate to outputs manifesting as errors. 

\begin{equation}
\label{eq:timing_errors}
t_{CP} =  \sum\limits_{d_i \in CP}^{} t_{d_i}, \quad t_{CP} (aging) > t_{clock} \Rightarrow \text{\textit{timing errors }} \textbf{\textbf{!}} \nonumber
\end{equation}

\noindent
\textbf{The Hidden Impact of Voltage Scaling:}  From Eq.~\ref{eq:I_d}, the impact of aging-induced $\Delta V_{th}$ largely depends on the operating voltage ($V_{dd}$). \textit{Therefore, the same increase in $V_{th}$ due to aging (e.g.,~$25$mV) can result in a much larger degradation in $I_{ON}$ and thus in the transistor speed when $V_{dd}$ is scaled down}. 
The timing errors that the circuit will exhibit, under aging effects, are subject to \textit{aging-induced degradation} ($V_{th}$, $\mu$) and \textit{operating voltage} ($V_{dd}$). The combination magnifies the impact of aging and shifts the aging problem from a sole long-term degradation (i.e.~a degradation that may take months to cause timing errors in circuits) to a \textit{short-term degradation} (i.e.~a degradation that might need merely hours to cause timing errors). Such a magnification in the impact of aging can point to using this as  
a knob to detect Trojans. In this study, we consider the relatively longer-term aging (at multiple controllable levels of aging) that can be induced by the various aging effects discussed above.

\subsection{Combining Circuit Aging With  Over-Clocking}
Aging alone does not create sufficient delay in a circuit path for the timing errors to propagate to the output. To check this hypothesis, we generated the outputs for the clean and Trojan-inserted IC using a nominal clock (i.e., without over-clocking) with different states of aging  (i.e with different aging-induced $\Delta V_{th}$). The difference in outputs for both the cases is insufficient to detect a Trojan. 

Similarly, over-clocking alone does not create significant bit-error patterns at the output. When the Trojan is placed far from the critical path, over-clocking does not generate a distinguishable signature of errors. We apply an ML classifier on over-clocking only data to see if it can detect a Trojan on the critical path. The results show a false negative rate $\geq$50\%. Windowing over multiple input vectors and majority voting during testing does not help. False negative rate increases when multiple input vectors are used to test.

\textit{Combining transistor aging with over-clocking generates a distinguishable signature of timing errors at the IC output and this can be used to detect Trojans with a high accuracy. Implementing and demonstrating this idea is the key novel contribution of this paper. }

In a nutshell, the delay caused due to aging alone is insufficient to produce bit errors at the output of an IC. 
Also, ICs include a timing guard band to ensure reliability. By over-clocking, we either narrow or remove the timing guard band. Over-clocking in combination with aging creates a robust \textit{signature} of bit-error patterns at the output. 

\section{AGING-AWARE CELL LIBRARIES: BRIDGING THE \\GAP BETWEEN PHYSICS AND SYSTEM LEVEL}
\label{sec:cell_lib}

\begin{figure*} [h]
  \includegraphics[width=\textwidth]{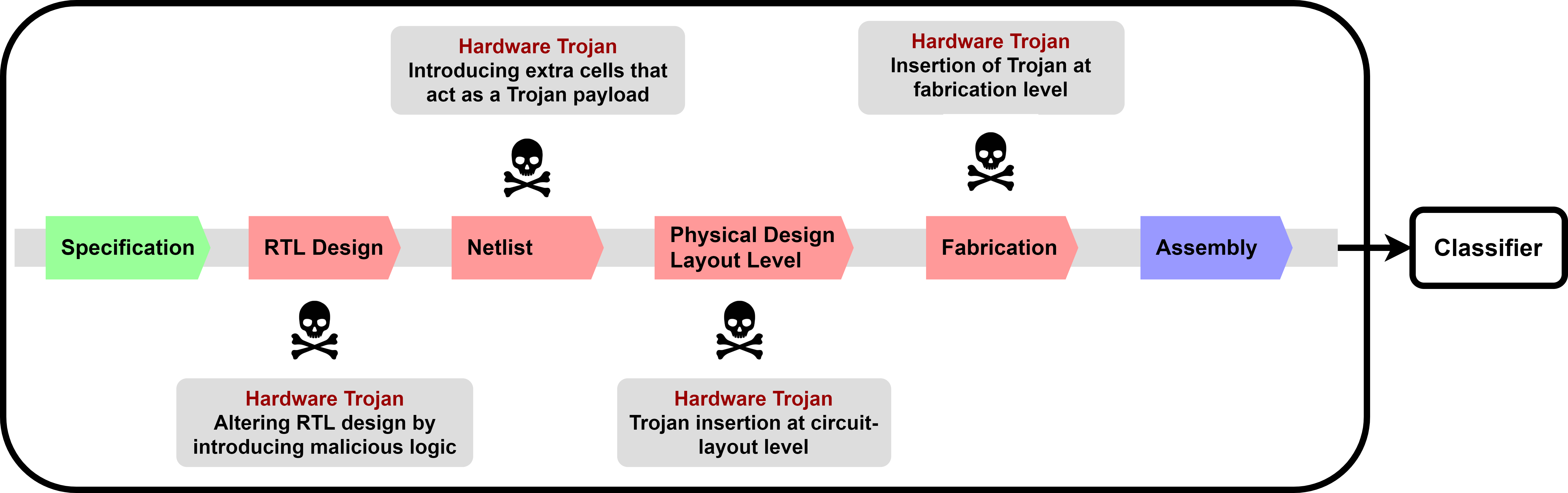}
  \caption{Stages in the IC design flow where a Hardware Trojan can be inserted.}
  \label{fig:trojan_attack_stages}
\end{figure*}

Aging is ``driven'' by different underlying mechanisms of defect generation that occur at the atomic level. In order to investigate and quantify how such defects may propagate all the way up to the system level, where they manifest as timing errors, the intervening abstraction layers must be carefully traversed. In addition, real digital circuits typically consist of numerous paths, which are all similar to each other with respect to overall delay. Therefore, when aging-induced degradation takes place, it is challenging to accurately quantify how the timing paths will be violated and how such violations will translate into errors at the circuit outputs. This necessitates an accurate modeling of how standard cells will behave in the presence of aging degradations. Any investigation in this direction requires that we use  commercial tool flows for static timing analysis in order to rely on their underlying mature algorithms evolved over decades. Otherwise, the impact of aging-induced degradation on  the delay of paths cannot be accurately captured  and, more importantly, any proposed technique would not be compilable  with the existing standard design flow of circuits.

To address these challenges, we create ``aging-aware cell libraries'' in which the delay of standard cells are characterized by considering the effects that aging-induced defects have on the electrical properties of pMOS and nMOS transistors, similar to~\cite{amrouch2016reliabilityaware}, \cite{amrouch2017impactofbti}. We start from the lowest level of abstraction where we employ state-of-the art physics-based BTI aging models to estimate the defects in pMOS and nMOS transistor and how they result in shifts in the transistor's parameters (i.e.~$V_{th}$ and $\mu$)~\cite{parihar2018btianalysistool}. Then, we employ SPICE simulation to estimate the delay and power of every standard cell considering the effects that $\Delta V_{th}$ and $\Delta \mu$  on the delay of the nMOS and pMOS transistors. We analyze every standard cell with $7 \times 7$ input signal slews and output load capacitances\footnote{This is how it is done in commercial standard cell libraries.}. All the generated data is stored using the ``liberty'' format, which is the standard format for existing commercial EDA tool flows (e.g., Synopsys and Cadence). To cover a wide range of aging effects, we create the aging-aware cell libraries for a various aging stress conditions (i.e.~various duty cycles~\footnote{Duty cycle is the \% of operation time for which the transistor is ON.}). We  start from $0$\% representing no aging all the way to $100$\% representing the maximum aging that considers a continuous aging stress without any recovery, in steps of $10$\%. These standard cell libraries are compatible with EDA tool flows like Synopsys and Cadence. Hence, designers can plug them directly within the commercial static timing analysis tools for accurate timing analysis.

\noindent
\textbf{Implementation Details:} We target the $45$nm technology node. The methodology and implementation is not limited and applies to advance technology nodes. We employ state-of-the-art physics-based aging model~\cite{parihar2018btianalysistool}, which were validated against semiconductor measurements. They capture the defect generation of BTI aging under any stress condition.  The physics models support all technology nodes and different transistor structures such as FinFETs. To create the standard cell library, we used the open-source $45$nm nangate library~\cite{pdk45nm}. The library provides SPICE netlist for sequential and combinational standard cells. For SPICE simulations we use open-source Predictive Transistor Model (PTM)~\cite{ptm}.

\section{PROPOSED METHODOLOGY}
\label{sec:methodology}
Consider an IC that performs a function $f$ at a nominal operating clock. When an input $x$ is applied to the IC, it generates an expected output $f(x)$ as long as the IC is operating normally (i.e.,  operating without timing errors). If the IC is run on a faster clock (higher frequency),  it alters its functionality causing mis-matches (e.g., bit errors, longer settling times) in the observed outputs due to the induced timing violations. If the logic gates of the IC age and slow down, the results take more time to propagate and reach the output bus and the observed output behavior will be different from expected because of timing errors. Aging induces a delay increase in the IC, which then results in transient timing errors.
If the clock period is small enough to propagate the errors to the output, the output bits will differ from the expected outputs. This change in the output bit patterns helps detect extra circuits such as Hardware Trojan.  Timing guard band is not used in this study. It is used during the operation of a chip and is not a consideration for the detection mechanism. The maximum clock frequency is purposefully violated by overclocking the circuit.

To validate this hypothesis using simulations, we create standard cell libraries as described in Section~\ref{sec:cell_lib} that consider aging effects by using the degradation of threshold voltage ($V_{th}$) and carrier mobility ($\mu$) in nMOS and pMOS transistors. We created standard cell libraries for different stresses of duty cycle from 0\% - 100\% in steps of 10\%. A duty cycle of 100\% refers to 100\% aging stress and 0 refers to no aging. 

We consider attacks pre-synthesis (RTL) and post-synthesis (gate-level netlist). In the case of post-synthesis Trojan insertion, the attacker can be either the designer or the fabrication company. For the RTL attack, we assume that we have access to the genuine RTL that is corrupted before proceeding to synthesis. For Trojan insertion post synthesis, the RTL without the Trojan is synthesized to produce a gate-level netlist and the Trojan is inserted within the netlist to get a corrupted, Trojaned netlist. We use the aging-aware standard cell libraries and the (clean and corrupted) gate-level netlists as inputs to a static timing analysis tool to generate the standard delay format (SDF) files for the different aging states. 
 The SDF file reports the exact delay of every logic gate in the netlist. We generate a set of inputs and their outputs for the baseline IC using gate-level simulations. This set of inputs includes randomly generated inputs as well as input test patterns generated by Synopsys Tetramax  Automatic Test Pattern Generator (ATPG)~\cite{tetramax}  tool. We generate these input-output pairs for combinations of aging states and clock frequencies using gate-level simulations (with SDF annotations). We then develop a machine learning approach (discussed in Sections~\ref{sec:feature_extraction}, \ref{sec:anomaly_detection}) to compare the observed bit error patterns at the circuit output with the expected bit error patterns trained from a known-good device/simulation. Figure~\ref{fig:trojan_attack_stages} shows the stages where a hardware Trojan can be inserted. Our ML classifier detects all the Trojans. We obtain the maximum clock period for an IC using static timing analysis and generate input/output dataset for smaller clock periods that create pronounced output bit-error patterns.

\subsection{Feature Extraction for Machine Learning}
\label{sec:feature_extraction}

\begin{figure}
\includegraphics[width=0.48\textwidth]{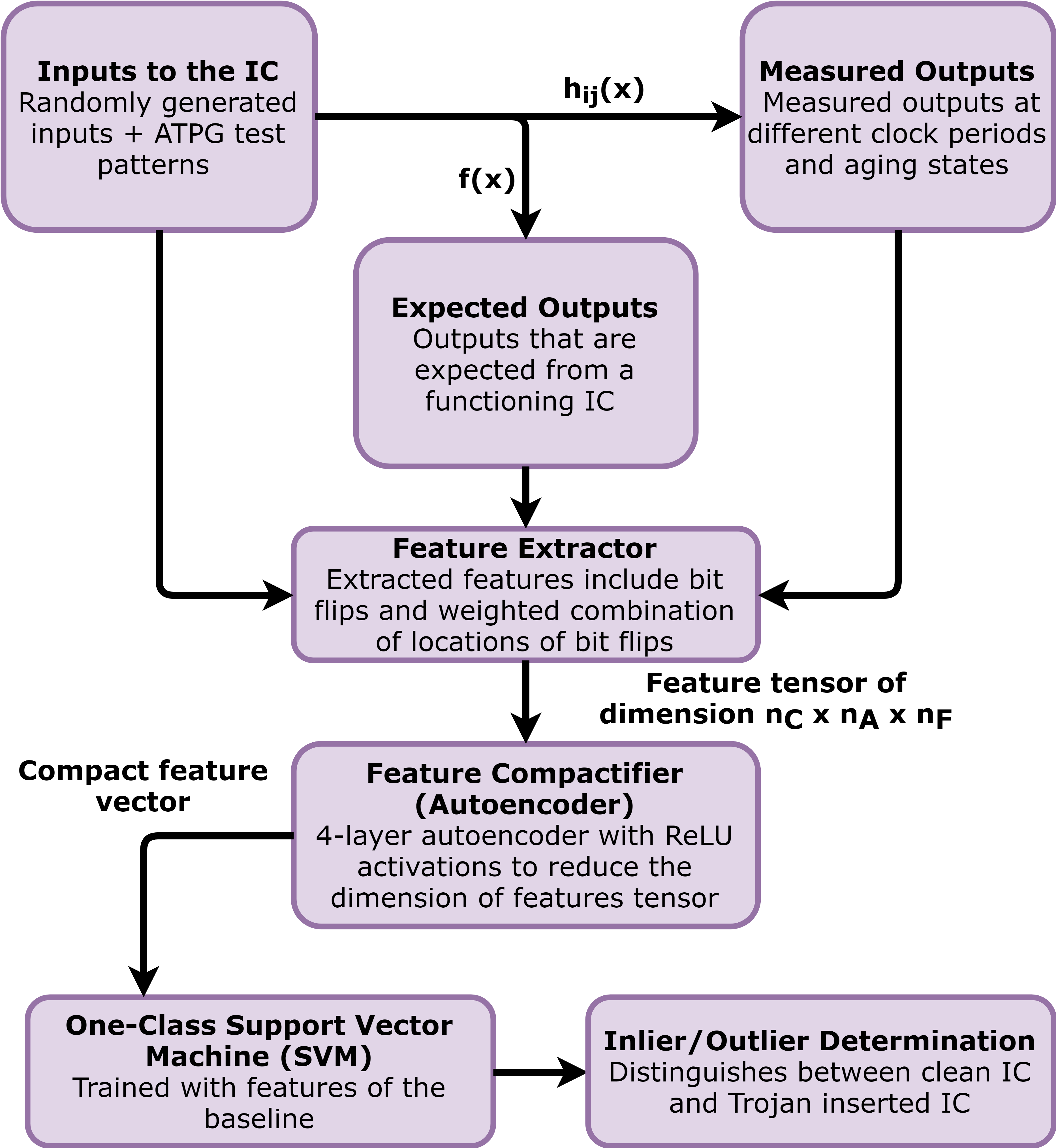}
\caption{ML-based detection of ICs with a Trojan.}
\label{fig:SVM}
\end{figure}

The Hardware Trojan detection methodology is based on applying inputs to the IC, observing the outputs for several clock periods and aging states, and comparing the observed output variations with ``baseline'' characteristics to detect anomalies that indicate presence of Trojans. A one-class classifier based on an auto-encoder and a one-class Support Vector Machine (SVM) is proposed as shown in Figure~\ref{fig:SVM}. 

To train a model using the baseline characteristics, we consider a set of inputs $x_T$. For each input $x\in x_T$, the output is recorded\footnote{After a number of cycles sufficient to read output in normal operation.} for a range of clock periods and aging states. Denoting clock periods by $t_1,\ldots,t_{n_C}$ and aging states by $a_1,\ldots,a_{n_A}$, the output for $i^{th}$ clock period and $j^{th}$ aging state is denoted as $h_{ij}(x)$. The clock periods $t_1,\ldots,t_{n_C}$, span significantly lower than the highest sustainable clock period under normal conditions\footnote{Nominal clock derived from slack analysis by Synopsys Primetime.} 
While the higher number of bit errors are expected at a higher clock, some bit errors are possible at higher than nominal clock in high-aging states. Given input $x$, the expected output is denoted by $y_0=f(x)$.

Given an input $x$, the expected output $f(x)$ is deterministic, while the measured outputs $h_{ij}(x)$ are stochastic,  when the IC is over-clocked and aged. IC-to-IC variations add to the  variability of $h_{ij}(x)$. To account for variability and achieve robust anomaly detection, we do not simply learn mappings from inputs to expected outputs. Rather, we learn a deeper model of how the observed outputs (i.e., observed bit error patterns) vary with over-clocking and aging. The trained classifier does not compare measured with expected outputs. It uses the variation in patterns of bit errors as a signature for the IC and learns a model of the characteristics of these variations (independent of the applied inputs). Hence, inputs used during testing of an IC are independent of inputs used in training (see Section~\ref{sec:experimental}).   

Given an input $x$, the expected output $y_0=f(x)$, and a measured output $y=h_{ij}(x)$ for the $i^{th}$ clock period and the $j^{th}$ aging state, the mismatch between $y$ and $y_0$ is measured by a set of  four features:  number of 0$\rightarrow$1 bit flips, 1$\rightarrow$0  bit flips and weighted combinations of 0$\rightarrow$1 and 1$\rightarrow$0 bit flips considering their bit locations. Denote the bit length of the output (i.e., $y$ or $y_0$) as $m$ and given any binary number $a$ of bit length $m$, define the ``bit indicator functions'' $\underline{1}(a)$ and $\underline{0}(a)$ as the subsets of ${\cal M}=\{0,\ldots,m-1\}$, given by  
\begin{align}
  \underline{1}(a) = \{r\in {\cal M} | a\&2^r > 0\} \\
  \underline{0}(a) = \{r\in {\cal M} | a\&2^r = 0\}
\end{align}
where $\&$ is the bit-wise AND. $\underline{1}(a)$ and $\underline{0}(a)$ capture subsets of bit locations in $\{0,\ldots,m-1\}$ corresponding to 1 or 0 bits in $a$. Given $y$,  $y_0$, the feature vector is defined as
\begin{align}
  f_1(y,y_0) &= \sum_{r\in(\underline{0}(y_0)\cap \underline{1}(y))}1
  \\
  f_2(y,y_0) &= \sum_{r\in(\underline{1}(y_0)\cap \underline{0}(y))}1
  \\
  f_3(y,y_0) &= \sum_{r\in(\underline{0}(y_0)\cap \underline{1}(y))}2^r
  \\
  f_4(y,y_0) &= \sum_{r\in(\underline{1}(y_0)\cap \underline{0}(y))}2^r
  \\
  F(y,y_0) &= [f_1(y,y_0),f_2(y,y_0),
  \nonumber\\
  &\quad f_3(y,y_0),f_4(y,y_0)]^T \in {\cal R}^{n_F};n_F=4
\end{align}

Using the feature vector $F(.,.)$, a three-dimension feature tensor is defined to characterize variations of bit errors over the set of clock periods and aging states. Given input $x$, expected output $y_0=f(x)$, and measured outputs $\{h_{ij}(x),i=1,\ldots,n_C;j=1,\ldots,n_A\}$, the three-dimensional tensor $\overline F(x)$ of dimension $n_C\times n_A\times n_F$ is defined where its $(i,j,k)^{th}$ element is the $k^{th}$ element of $F(h_{ij}(x),y_0)$.

\begin{figure*}[t]
\includegraphics[width=\textwidth]{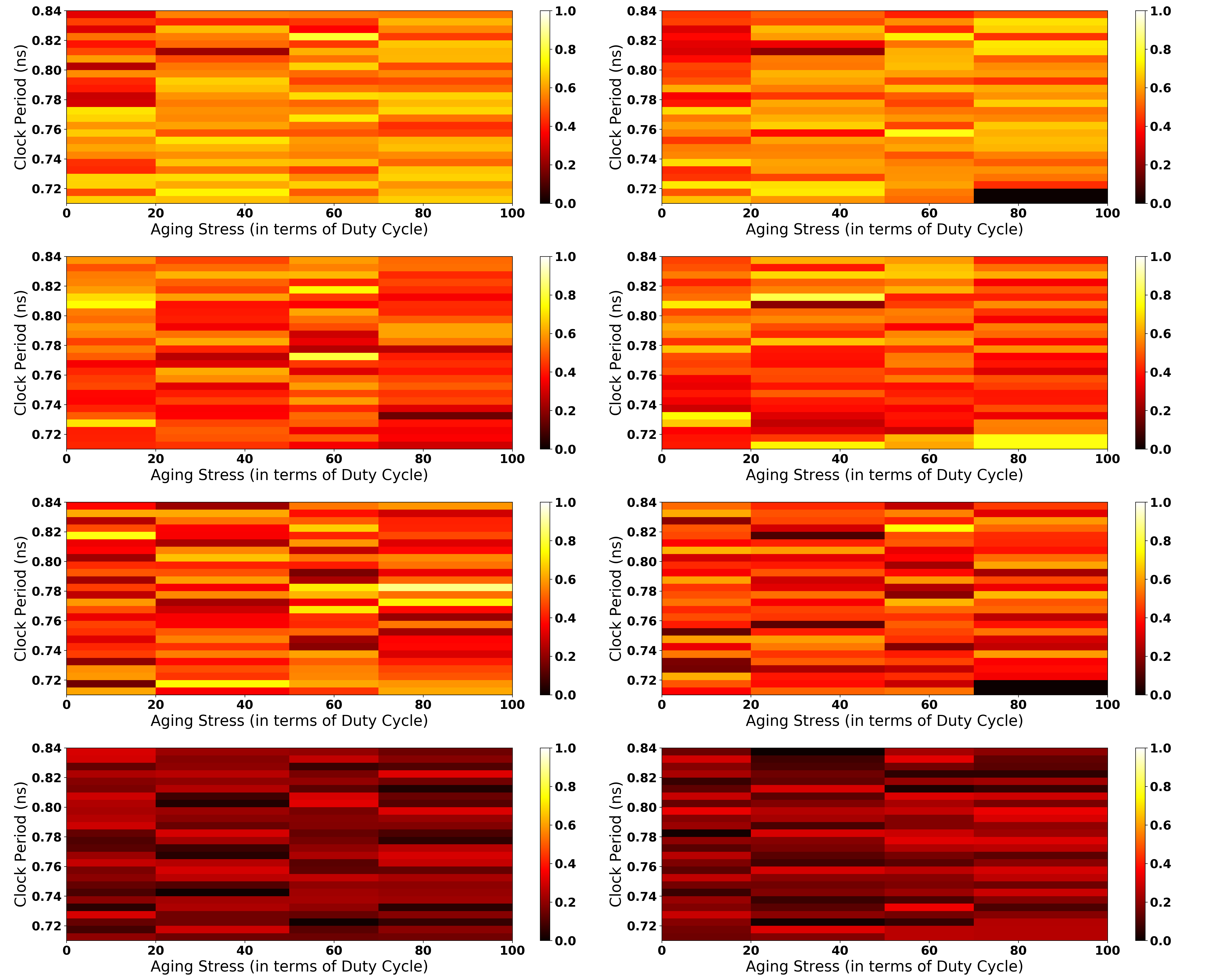}
\caption{Visualization of features for different aging states (duty cycle 0\% $\implies$ no aging and a duty cycle 100\% $\implies$ maximum aging) and clock periods for clean IC (left) and Trojaned IC (right) for RSA (when Trojan is inserted into the netlist).}
\label{features}
\end{figure*}

Figure~\ref{features} visualizes the feature tensor computation across clock periods and aging states for RSA circuit (Trojan inserted in the nelist). The figure shows the features extracted at the different clock periods and aging states for an input as a heat map. Although the feature values for clean and Trojaned ICs look similar, the classifier can distinguish a clean IC from a Trojaned one (see  Section~\ref{sec:experimental}).
From left to right in each of the plots, the aging stress increases from 0\% to 100\% and the Y-axis represents different clock periods and the over-clocking is more towards the bottom. The four plots in each column show the four types of features as discussed above. They correspond to (from the top) the number of 0$\rightarrow$1 bit flips, number of 1$\rightarrow$0 bit flips, weighted combinations of 0$\rightarrow$1 and 1$\rightarrow$0 bit flips considering their bit locations, respectively. The heat map shows how the features are varying with aging stress and clock periods as well as the
difference between a clean IC and Trojaned IC. Lighter colour indicates that the particular kind of feature at the given clock period and aging stress is more pronounced.

\subsection{Machine Learning for Trojan Detection}
\label{sec:anomaly_detection}
To train the anomaly detector, the feature tensors $\overline F(x)$ are generated for each input $x$ in the training set $x_T$. We combine features from multiple inputs by taking a window of size 1 during training for better accuracy in detecting Trojans. From the set $\{\overline F(x),x\in x_T\}$, we train a one-class classifier for outlier detection. Given a feature tensor computed from outputs measured from an IC under test, the  trained classifier  determines if the feature tensor is ``different (outlier)'' from that for a Trojan-free IC. A one-class classifier based on an autoencoder and a one-class SVM is used (Figure~\ref{fig:SVM}).  From the feature tensor of dimension $n_C\times n_A\times n_F$, a compact feature representation is computed using a four-layer autoencoder having a two-layer encoder and a two-layer decoder with ReLU activations. The feature vector from the hidden layer of  autoencoder is input to a one-class SVM.

When testing an IC, inputs $\tilde x$ in a test set $\tilde x_T$ are applied to the IC and the outputs $y_{ij}=h_{ij}(\tilde x)$ are measured for the $i^{th}$ clock period and $j^{th}$ aging state. Feature vectors $F(y_{ij},f(\tilde x))$ and the feature tensor $\overline F(\tilde x)$ are extracted. The feature tensor is passed through the trained autoencoder to extract a low-dimensional feature vector which is then passed through a one-class SVM. The test set $\tilde x_T$ can be disjoint from or overlap with the training set $x_T$. In either case, the anomaly detector is input-independent and does not rely on matching actual values of outputs, but on feature patterns of bit error variations across clock periods and aging states. Since the detector operates on a feature tensor extracted from outputs measured for a single input, applying one input $\tilde x$ is sufficient for inlier/outlier determination. However, to improve accuracy, multiple inputs $\tilde x$ plus majority voting is used to generate the inlier/outlier estimate. Section~\ref{sec:experimental} shows that $\geq$ 95\% accuracy of inlier/outlier determinations of clean vs. Trojan-inserted ICs) are obtained with a single input. Majority voting with multiple inputs improves it to 100\%.

In summary, the approach implemented is as follows:
\begin{itemize}
    \item Training: A simulated model of the baseline (Trojan-free) IC is used to train using a set of inputs which is a combination of random and ATPG-generated inputs and a predefined set of operating condition  tuples (e.g. clock frequency, aging state) are applied to the IC. For each of the conditions, inputs are applied and outputs are measured. The feature vectors are generated from each of the measured output to train the SVM. The feature vectors for challenging Trojans are enhanced by collecting the features over a set of inputs (called a bin).
    \item The testing procedure is similar to the training. We collect input-output measurements at different clock frequencies and aging states. Testing is done on the baseline and Trojan ICs to measure accuracy of the classifier. To improve accuracy for the challenging Trojans, the number of bins (called a batch) used for testing at a time is increased. We use a batch of 5 bins during training.
    \item Implementation: In the deployment site, same procedure as during testing is used. Different clock frequencies and aging states (i.e., operating condition) can be applied to the IC under test and input-output measurements with a batch of inputs at each operating condition can be collected. The extracted features are  run through the classifier to determine if the IC is Trojaned or not.
\end{itemize}

\section{EXPERIMENTAL SETUP}
\label{sec:setup}
\subsection{Trojan Benchmarks Used in the Study}
\begin{figure}[h]
\includegraphics[width=8.5cm]{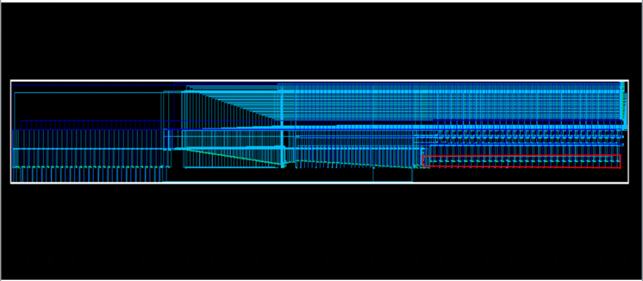} 
\caption{Gate-level circuit of synthesized Trojan-infected RSA circuit. The Trojan is contained in the red area of Figure.}
\label{rsacircuit}
\end{figure}

We use the following 32-bit RSA and the 128-bit AES circuits from the Trust-Hub~\cite{trusthub}. 
\begin{itemize}
\setlength\itemsep{0em}
\item  
The Basic RSA-T100 benchmark  implements a Shift-and-Add algorithm for modular multiplication. 
The trigger checks for an input and  activates the Trojan when this is found on the input bus. 
The Trojan leaks the secret key (private exponent) through output bus. 
Figure \ref{rsacircuit} shows the gate-level netlist of the synthesized RSA circuit with a Trojan (the Trojan is inserted at the RTL). The Trojan occupies 0.75\% of the IC area.

\item The AES-T100 uses a 128-bit key. In the baseline AES, the plaintext goes through 10 rounds of substitution, mix column, and shift rows. The AES-T100 Trojan is always on. The Trojan leaks bits from the secret key. The eight least significant bits are leaked through power side channel before which they are XORed with the bits generated from Linear-Feedback Shift Register (LFSR). This modification to the key obfuscates the power readings which allows only the adversary to recover the key.

\item The AES-T1000 benchmark uses a 128-bit key and has a trigger similar to Basic RSA-T100. The Trojan leaks the key using the technique similar to AES-T100. The difference is that the AES-T1000 has a Trojan trigger. 
\end{itemize}

\subsection{Synthesis and Simulation}
\begin{figure}[t]
\includegraphics[width=0.48\textwidth]{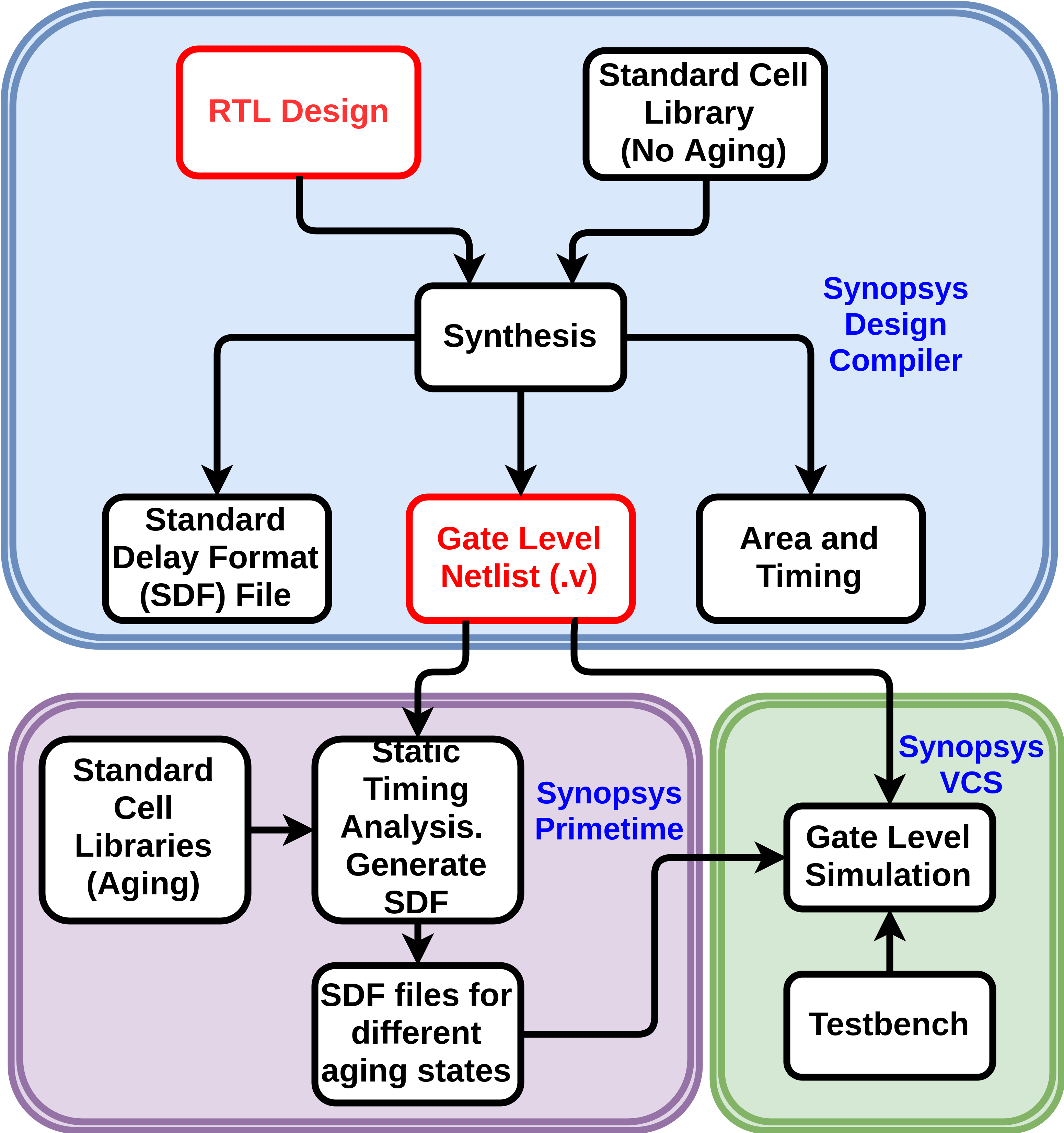}
\caption{The CAD tool flow. The CAD tools used are in blue background. The RTL design stage where the attack can happen is in red. The steps used to create aging-aware SDF files are in purple. The simulation steps are in green.}
\label{fig:EDA_tool_flow}
\end{figure}

We use Synopsys Design Compiler~\cite{dccompiler} and a 45 nm technology library operating at 1.1V, 25 $^o$C (see Section~\ref{sec:cell_lib}) without aging to synthesize the baseline and the Trojan circuit (RTL from Trust-Hub) to produce gate-level netlists. The netlist of the baseline circuit and the Trojan are combined to produce a netlist with a Trojan so as to keep the original circuit unchanged. Synopsys Primetime~\cite{primetime} is used along with 45 nm technology with different percentages of aging stress to create SDF files, each for a case of aging stress, using the netlists created for Trojan-free and Trojaned circuits. 
The gate-level simulations at different clock periods is performed using Synopsys VCS~\cite{vcs}. 
Figure~\ref{fig:EDA_tool_flow} shows the tool flow.

\subsection{SVM-based Machine Learning}
A one-class SVM is trained to learn bit error patterns produced at the outputs (details in Section~\ref{fig:SVM}); therefore, detecting Trojans.
  Features are extracted for one input at a time to train the classifier in Experiment 1. The number of input vectors used for windowing over features is denoted by $k$ (size of bin) and  we will empirically show that  $k$ = 5 provides improved accuracy to detect the more challenging Trojans.

\subsection{Modeling IC-to-IC Variations}
\label{sec:modeling_variations}
IC-to-IC delay variations due to temperature, pressure differences during manufacturing of ICs is also considered and addressed. These variations are modeled  by altering the delay parameters of the gates in the SDF files that are caused by each of the aging states. There are two types of variations in ICs - 1) IC-to-IC variation and 2) on-chip variation (die-to-die). The variations in delay from IC to IC can be typically 5\% or more~\cite{pufdevadas}. The change that we applied to the SDF files is as follows: 1) 5\% change in each parameter to model IC-to-IC variation, and 2) Gaussian random variations of zero mean and 4\% standard deviation ($\sigma$) to model on-chip variations. The combined effect corresponds to variations of up to 17\% (considering 3$\sigma$ plus 5\% as in the first case).
At advanced technology nodes, the percentage variation in delay due to manufacturing process increases and aging effects become pronounced, as has been demonstrated in Intel measurements for the 14nm FinFET technology~\cite{Natarajan2014A1L}.

 \section{EXPERIMENTAL RESULTS}
 \label{sec:experimental}
 
 We show the efficacy of our approach on three Trojaned circuits from the Trust-Hub~\cite{trusthub}: 1) BasicRSA-T100 (Experiments 1 and 2), 2) AES-T100 (Experiment 3) , and 3) AES-T1000 (Experiment 4). The Trojan sits on the critical path for BasicRSA-T100 and about 4209 paths (rank of the path) off of the critical path for the AES-T100 and 2753 paths (rank of the path) off the critical path for the AES-T1000. Therefore, the detection of the Trojan in the AES circuit is much more challenging. To train and test the classifiers, we build a corpus of datasets with overclocking and aging. To show that overclocking is insufficient on its own even on the simplest circuit with Trojan on the critical path, we collect data for the BasicRSA-T100 using overclocking only as well.  We also show results for BasicRSA-T100 when the Trojan is inserted at the RTL and netlist. This is due to the fact that the synthesized circuit changes considerably  when insertion is performed at the RTL level; therefore, the Trojan detection will be easier. However, in AES-T100 and AES-T1000, only a side channel Trojan (no feedback to the original circuit) is included and therefore injection of the Trojan into RTL will not change the original circuit once synthesized.  In all these cases, we use the data from Trojan-free ICs to train and the data from the Trojaned ICs to test. The inputs used during training are not re-used for testing to make sure that the classifier indeed learns useful patterns. Trojans in the training and testing dataset are kept dormant (except in Experiment 3 but the Trojan does not affect the output bits in Experiment 3) to make the problem challenging. The training and testing (which are offline) phases require larger dataset than during deployment (i.e., inference). 
 The number of input-output pairs of data collected for each of the circuits is between 4000 and 4300 which takes a few kilo bytes to store. The features collected over the data also is not more than a few hundred kilo bytes. The training is offline and needs to be done only once per circuit. It takes roughly five minutes to train each of the circuits. For testing and inference phases, the computations are not complex and does not take more than a few seconds.

  \begin{figure*}[!htb]
\centering
\begin{tabular}{cc}
\subfloat[Clean IC]{\includegraphics[width=0.49\textwidth]{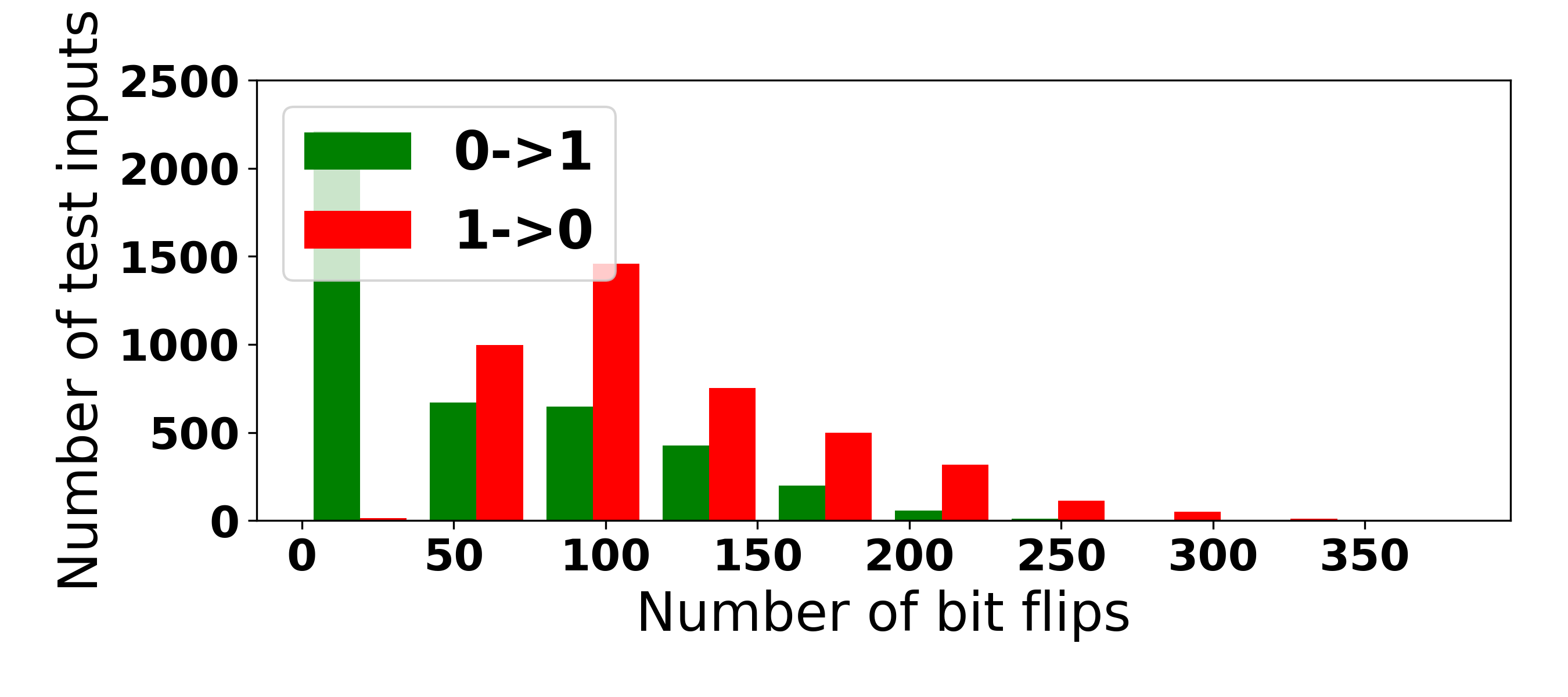}\label{bfcl0}}&
\subfloat[IC with Trojan inserted in RTL]{\includegraphics[width=0.49\textwidth]{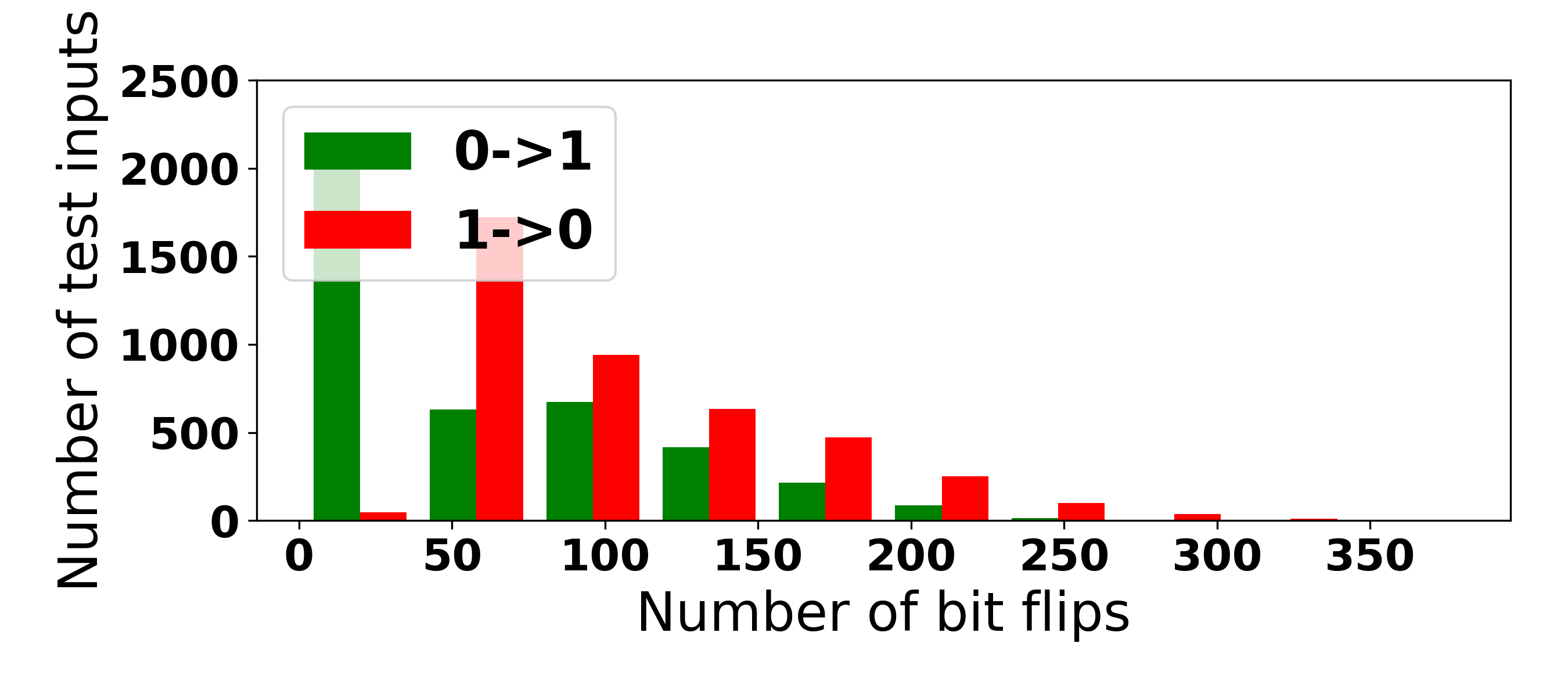}\label{bftr0}}
\end{tabular}
\caption{Bit flips induced by over-clocking for BasicRSA-T100 without the aging effects.}
 \end{figure*}

 \begin{figure*}[!htb]
\centering
\begin{tabular}{cc}
\subfloat[Clean IC]{\includegraphics[width=0.49\textwidth]{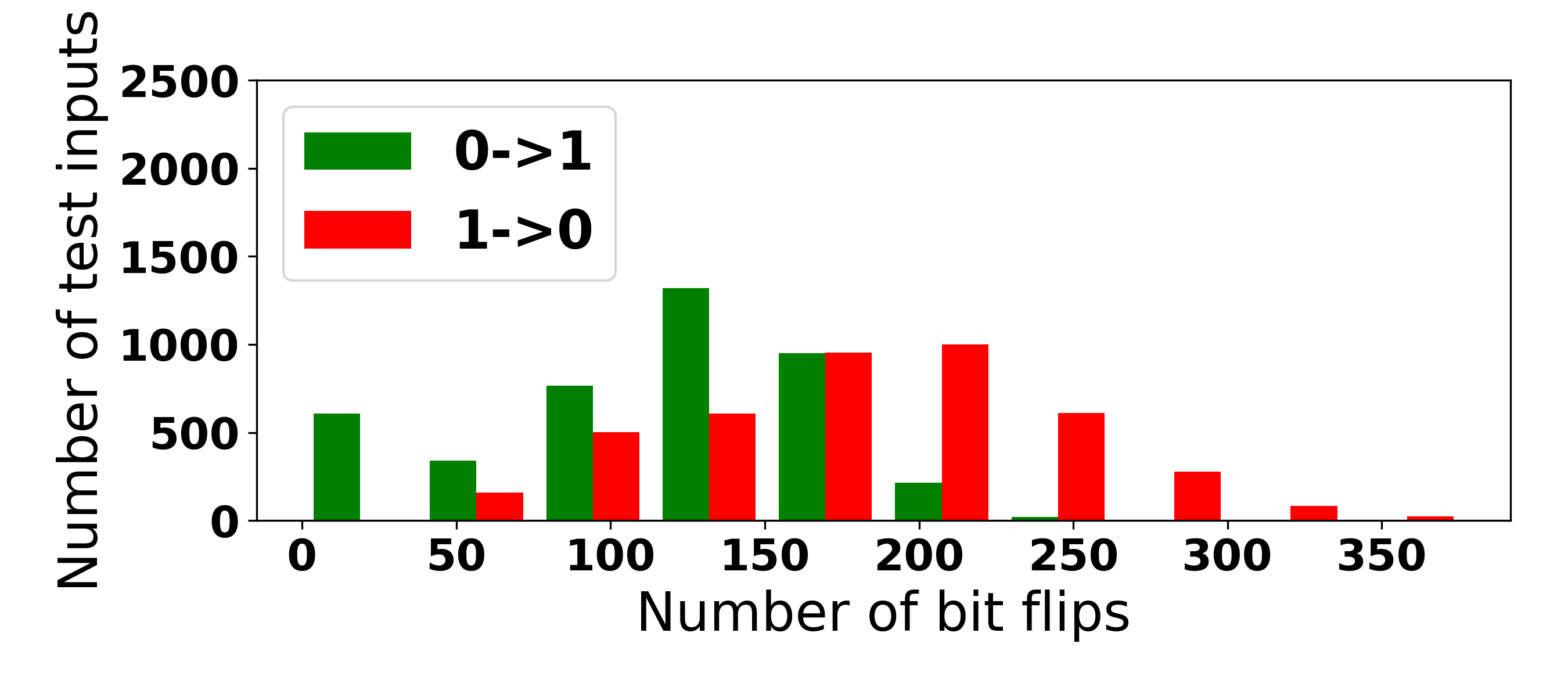}\label{bfcl10}}&
\subfloat[IC with Trojan inserted in RTL]{\includegraphics[width=0.49\textwidth]{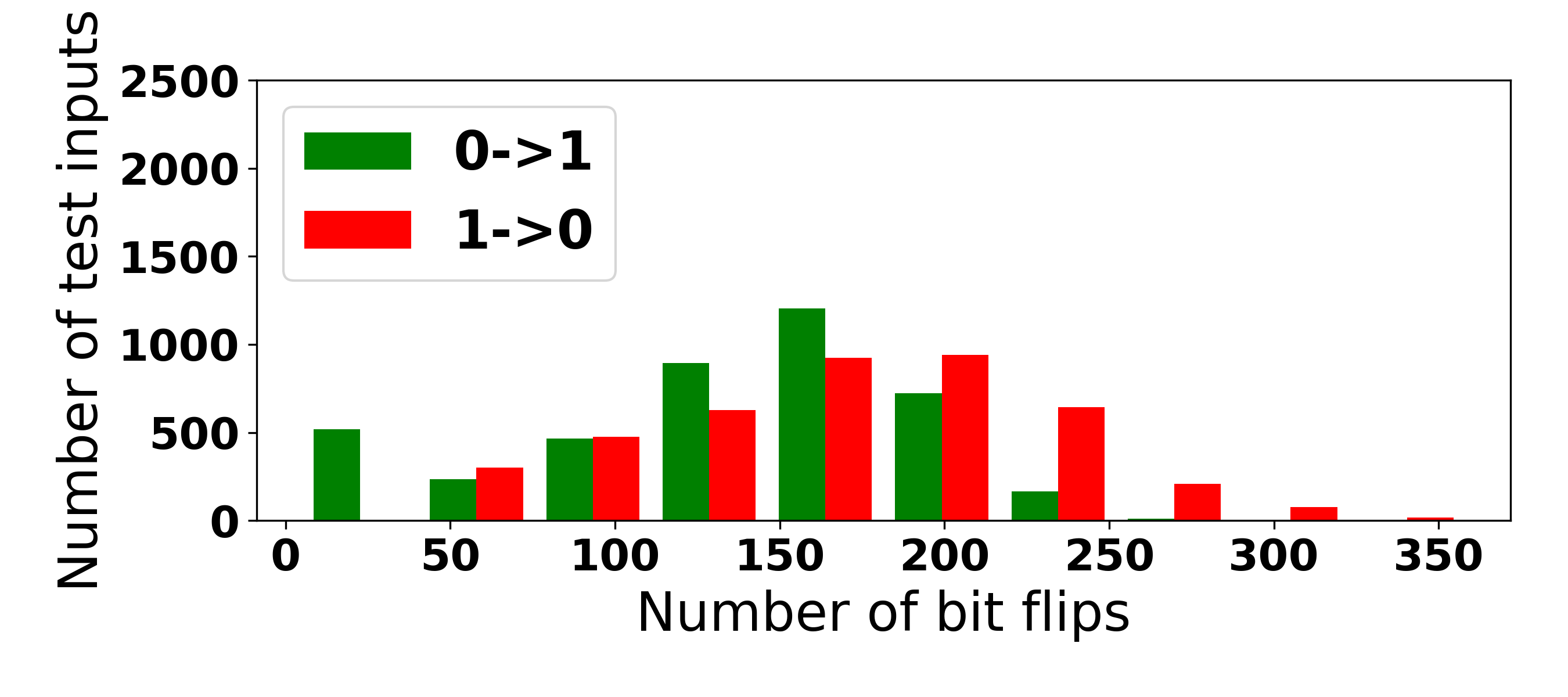}\label{bftr10}}
\end{tabular}
\caption{Bit flips induced by over-clocking for BasicRSA-T100 when aging is maximum.}
 \end{figure*}
 
 \begin{figure*}[!htb]
\centering
\begin{tabular}{cc}
\subfloat[Clean IC]{\includegraphics[width=0.49\textwidth]{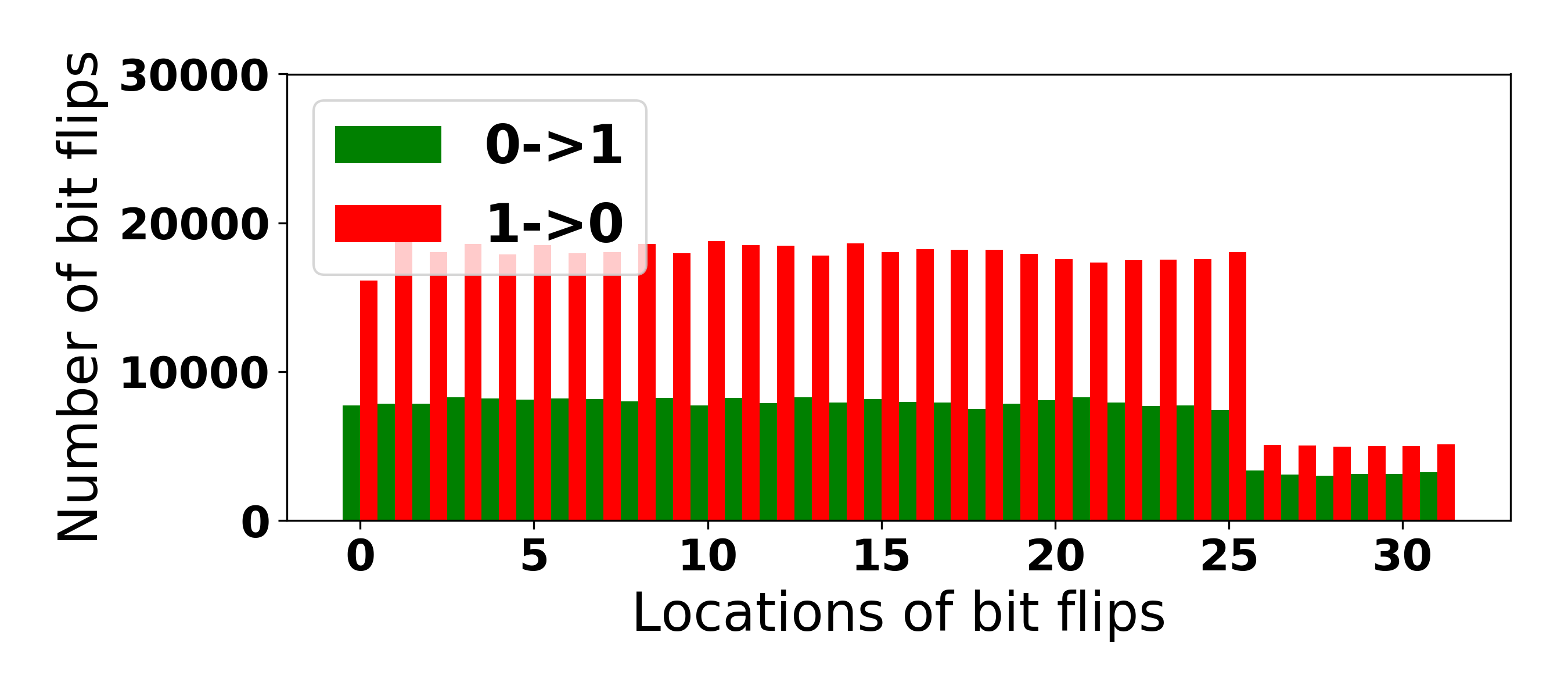}\label{bfloccl0}}&
\subfloat[IC with Trojan inserted in RTL]{\includegraphics[width=0.49\textwidth]{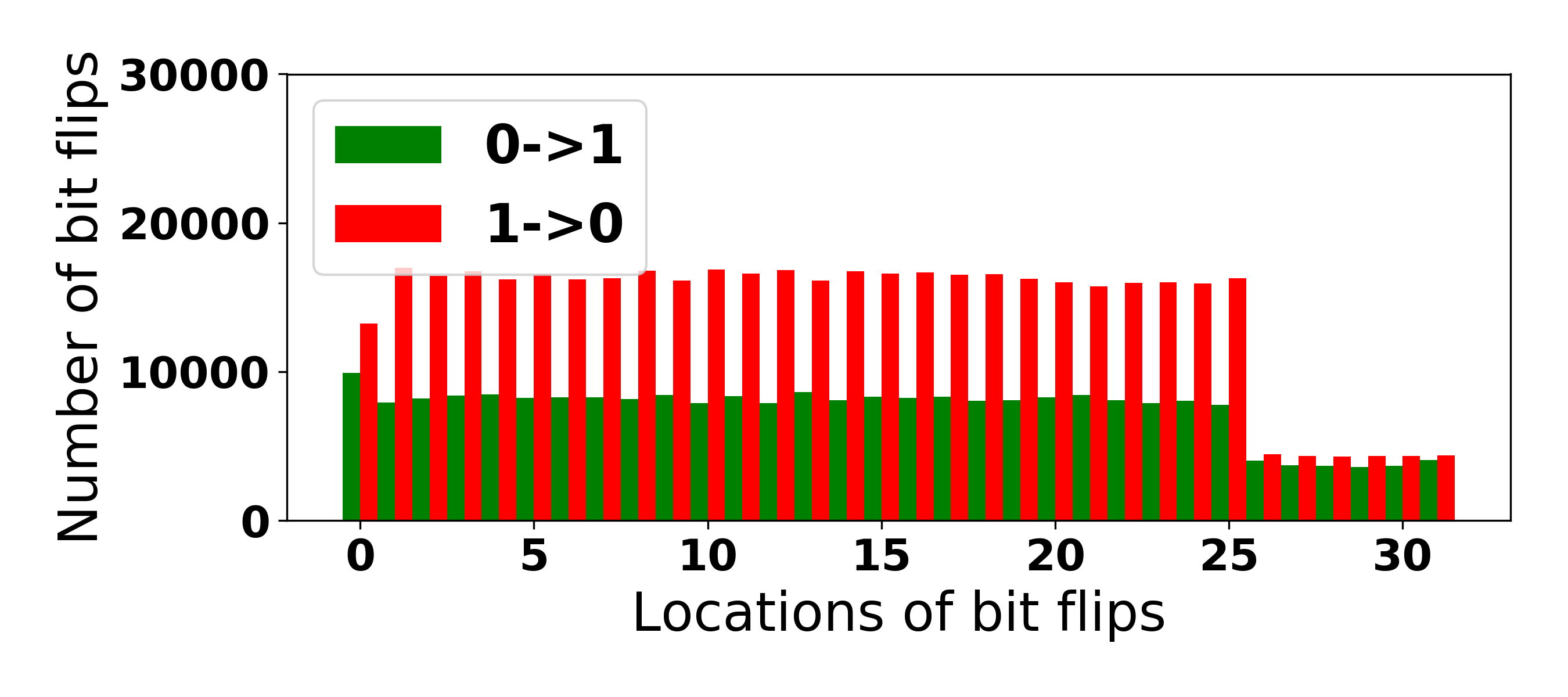}\label{bfloctr0}}
\end{tabular}
\caption{Weighted locations of bit flips without the aging effects in RSA circuit.}
 \end{figure*}
 
 \begin{figure*}[!htb]
\centering
\begin{tabular}{cc}
\subfloat[Clean IC]{\includegraphics[width=0.49\textwidth]{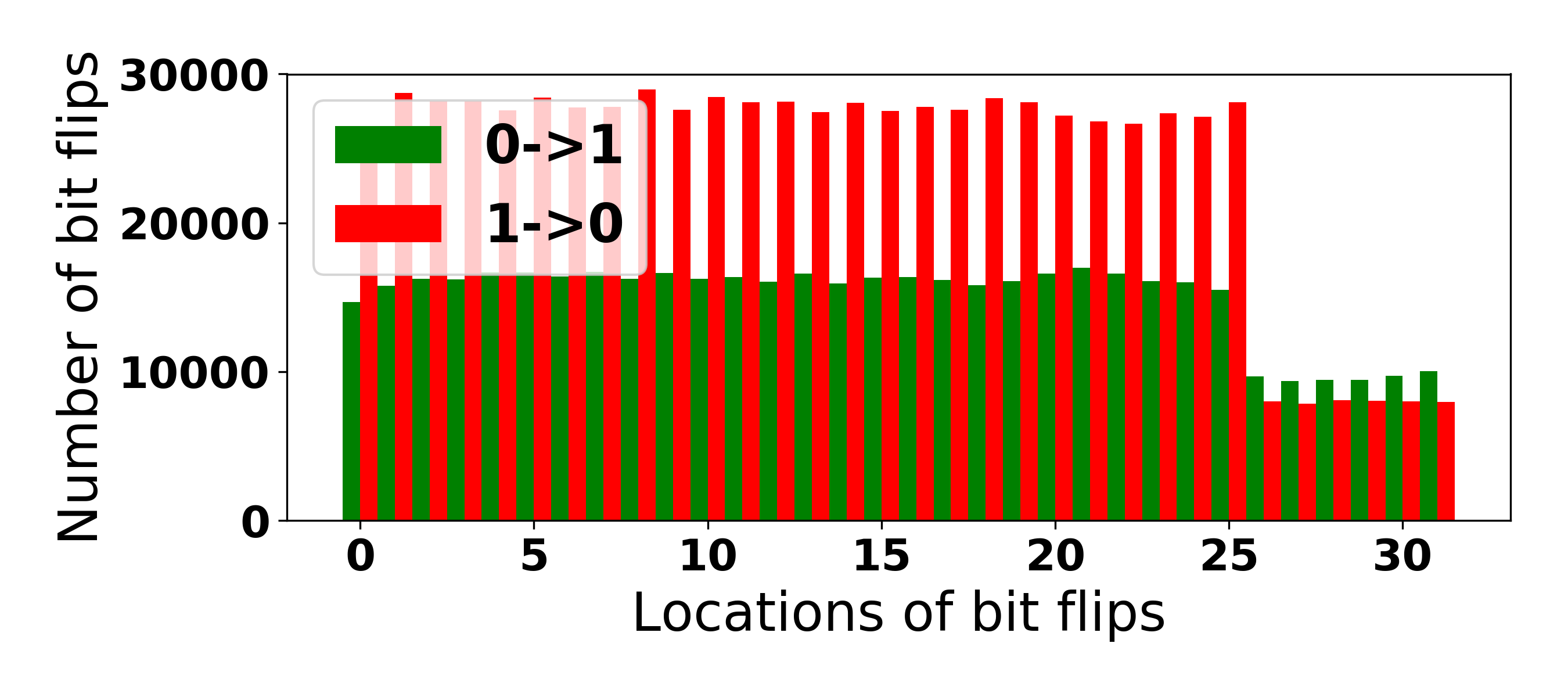}\label{bfloccl10}}&
\subfloat[IC with Trojan inserted in RTL]{\includegraphics[width=0.49\textwidth]{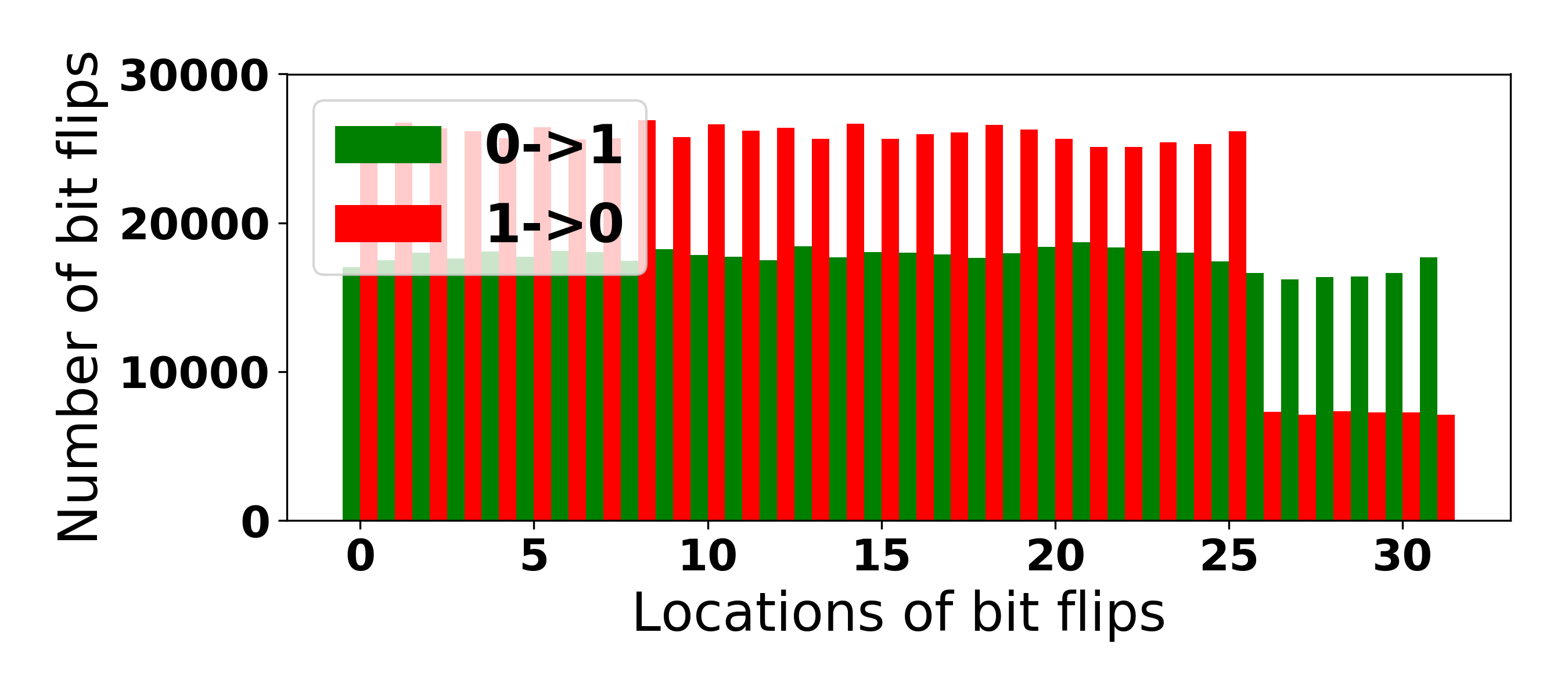}\label{bfloctr10}}
\end{tabular}
\caption{Weighted locations of bit flips when the aging is maximum in RSA circuit.}
 \end{figure*}

\begin{figure*}[!htb]
\centering
\begin{tabular}{cc}
\subfloat[Clean IC]{\includegraphics[width=0.5\textwidth]{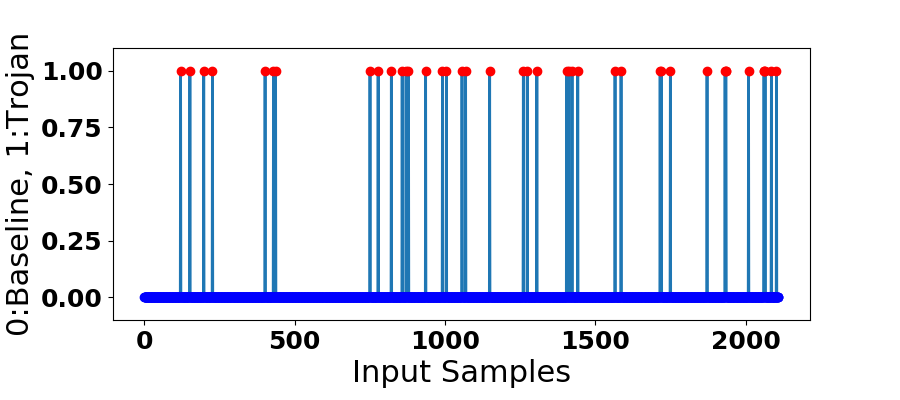}\label{clsing}}&
\subfloat[IC with Trojan inserted in RTL]{\includegraphics[width=0.5\textwidth]{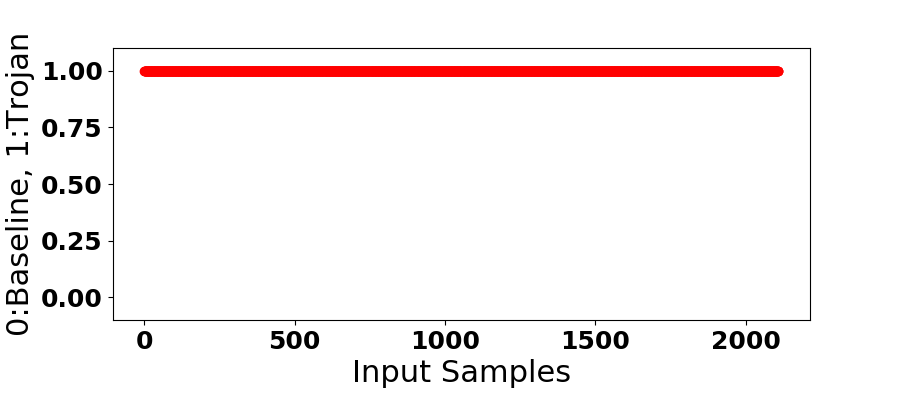}\label{trsing}}
\end{tabular}
\caption{Time series of anomaly detection using a single input for RSA circuit.}
 \end{figure*}
 

\begin{figure*}[!htb]
\centering
\begin{tabular}{cc}

\subfloat[Clean IC]{\includegraphics[width=0.5\textwidth]{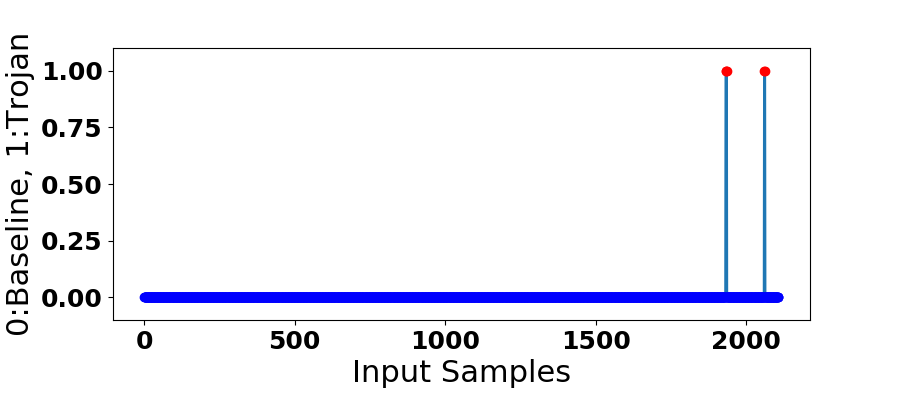}\label{cl3}}&

\subfloat[IC with Trojan inserted in RTL]{\includegraphics[width=0.5\textwidth]{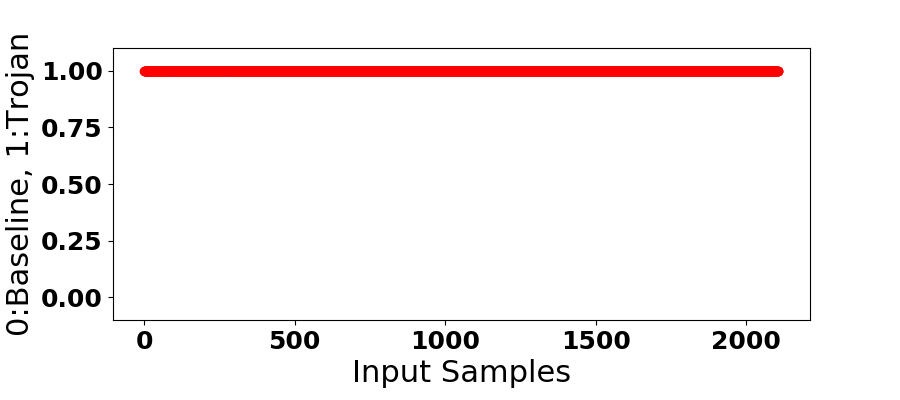}\label{tr3}}
\end{tabular}
\caption{Time series of anomaly detection using a batch of 3 inputs in RSA circuit.}
 \end{figure*}


  \begin{figure*}[!htb]
\centering
\begin{tabular}{cc}

\subfloat[Clean IC]{\includegraphics[width=0.5\textwidth]{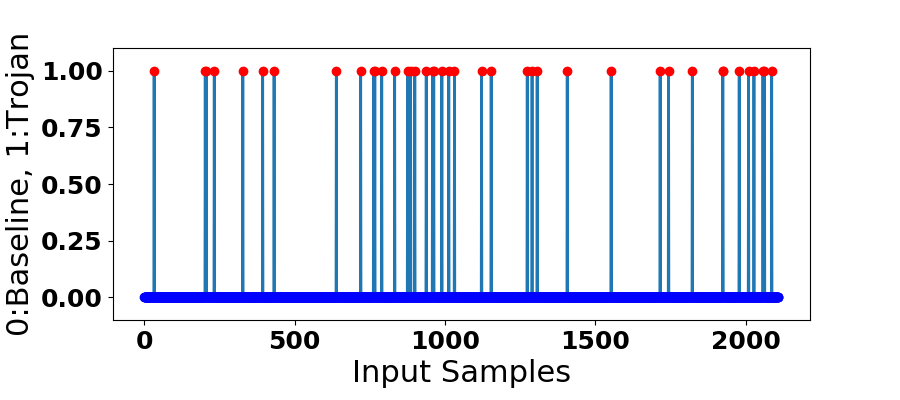}\label{clchvar}}&

\subfloat[IC with Trojan inserted in RTL]{\includegraphics[width=0.5\textwidth]{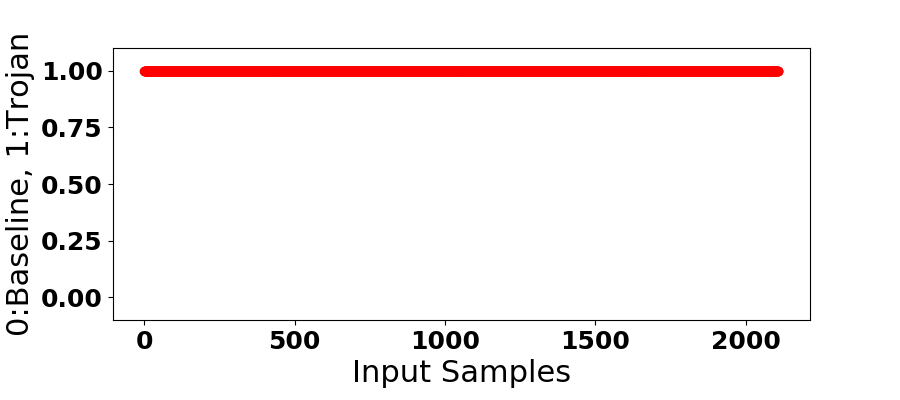}\label{trchipvar}}
\end{tabular}
\caption{Anomaly detection for IC-to-IC variation using single input in RSA circuit.}
 \end{figure*}

 \begin{figure*}[!htb]
\centering
\begin{tabular}{cc}

\subfloat[Clean IC]{\includegraphics[width=0.5\textwidth]{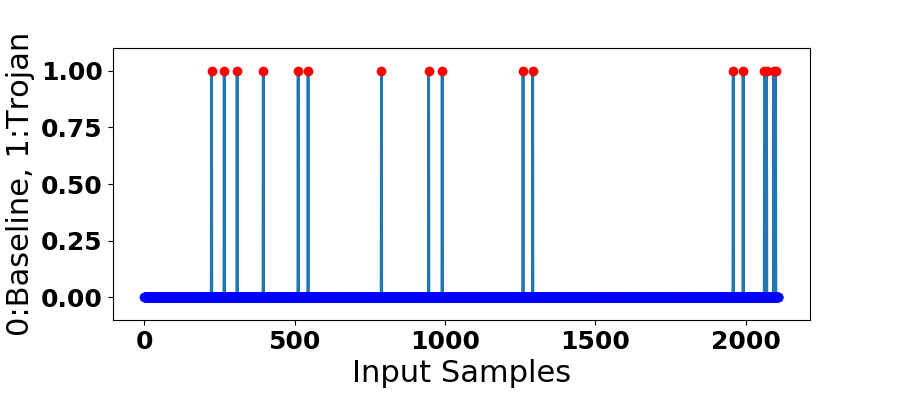}\label{pureoverclockcl}}&

\subfloat[IC with Trojan inserted in RTL]{\includegraphics[width=0.5\textwidth]{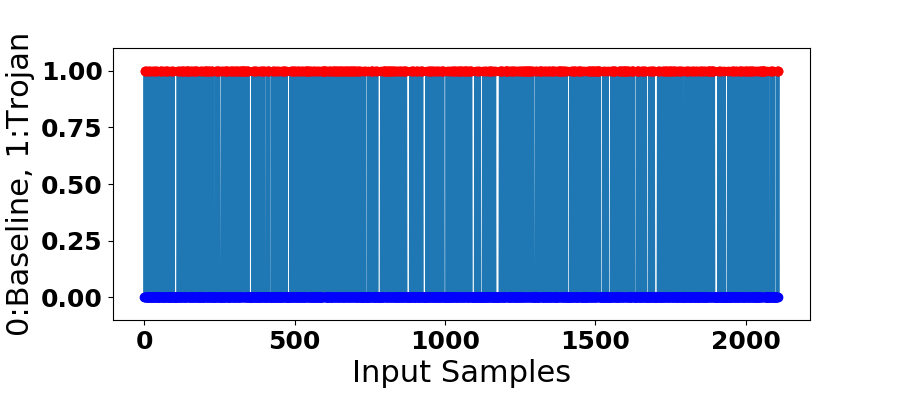}\label{pureoverclocktr}}
\end{tabular}
\caption{Anomaly detection for pure over-clocking using single input in RSA circuit.}
 \end{figure*}
 
\subsection{Experiment 1: Basic RSA-T100 Trojan Inserted into RTL}
\label{sec:RSARTL}

The baseline RSA circuit when synthesized has a  clock period of 2.17 ns. The Trojan occupies 0.75\% of the circuit. For the Trojaned and baseline ICs, we collect two sets of data:
\begin{enumerate}
\setlength\itemsep{0em}
\item {\bf Overclocking with aging:} We over-clock the RSA circuit and collect data by sweeping the overclocking in the range 1.125 ns - 1.4 ns with a step of 0.005 ns for Trojan and Trojan-Free cases and for all aging states.
\item {\bf Overclocking without aging:} We over-clock the RSA circuit and collect input/output data in the range 0.9 ns -- 1.4 ns in steps of 0.005 ns without considering the aging effects for clean and Trojaned ICs. 
\end{enumerate}

The data set is generated for 4226 inputs. Of these, 4096 are random patterns and 130 are ATPG patterns for the clean RSA circuit. We extracted four types of features (details in Section~\ref{sec:feature_extraction}). 
 Figures \subref*{bfcl0}, \subref*{bftr0} show the histograms of bit flips at the outputs for the no-aging case of RSA circuit with Trojan inserted at RTL.
Figures \subref*{bfcl10}, \subref*{bftr10} show the histograms for the maximum aging case. The bit flips largely depend on the inputs and the key used in RSA algorithm. Depending on the positions of bits in the inputs and the key used by the algorithm, each bit will have a different path from the input port to the output port. Additionally, when aging is applied to the circuit, it induces more delay in several paths that leads to an increase in number of bit flips. This is evident from the figures in which the histograms shifted towards right when aging is applied.
Figures \subref*{bfloccl0}, \subref*{bfloctr0} and Figures \subref*{bfloccl10}, \subref*{bfloctr10} show the bar charts of the weighted location of bit flips for no and maximum aging, respectively. The locations of bit flips has changed from no aging case to worst aging case and from no Trojan case to the Trojan inserted case as discussed. These figures show that the chosen features provide a discernible difference when a Trojan is inserted.

In case of overclocking with aging, we use half of 4226 inputs for training and the other half for testing. When a single input is used for training, the accuracy is 99.3\% on a clean IC and 100\% for a Trojaned IC as shown in Figures \subref*{clsing}, \subref*{trsing}. False positive rate is 1.4\% and false negative rate is 0\%. When a 3-input batch is used, the accuracy improves to 99.95\% for the clean IC and 100\% for IC with Trojan as seen in Figures \subref*{cl3}, \subref*{tr3}. The false positive rate is 0.09\% and the false negative rate is 0\%. A larger batch size is required for challenging Trojans. During deployment, only 1 input or 3 inputs are sufficient.  When IC-to-IC variations are considered and one input is used to test, the false negative rate for data from the clean IC is 2.73\% and for Trojaned IC is 0\%. A 3-input batch yields 100\% accuracy.  Figures \subref*{clchvar}, \subref*{trchipvar} show the time series of anomaly detection for random variations in ICs for a single input. The accuracy decreases slightly when process variations are considered. The perturbations introduced to the IC to simulate are random and may fall more in the baseline increasing the accuracy.

We over-clocked the circuit by to 4$\times$ the frequency of the Trojan-free circuit without aging and tested the classifier. Figures~\subref*{pureoverclockcl}, \subref*{pureoverclocktr} show performance of classifier on over-clocked data. False positive rate from clean IC is 1.14\% and the false negative rate for the Trojaned IC is 73.69\%. Batch of 3 inputs increases the false negative rate to 81.42\%. The high accuracy obtained for Trojan detection can be attributed to aging and not just over-clocking. Table \ref{metricstablersa} provides accuracy, precision, recall, and F1 on BasicRSA-T100 (RTL Trojan) and BasicRSA-T100 (netlist Trojan). The definitions for the accuracy metrics Accuracy, Precision, Recall and F1 are given by Equations \ref{accuracy}, \ref{precision}, \ref{recall}, and \ref{f1}, respectively:
 \begin{align}
 \label{accuracy}
 Accuracy &=  \frac{TP + TN}{TP + TN + FP + FN} \\
 \label{precision}
 Precision &= \frac{TP}{TP + FP} \\
 \label{recall}
  Recall &= \frac{TP}{TP + FN} \\
  \label{f1}
  F1 &= \frac{2*(Recall*Precision)}{Recall + Precision}.
\end{align}
where $TP = True Positives$, $FP = False Positives$, $TN = True Negatives$, $FN = False Negatives$.

 \begin{table}
\centering
\caption{Classifier performance on Experiments 1 and 2.}
\begin{tabular}{lccccc}
\hline
& \multicolumn{2}{c}{RSA (RTL Trojan)} &  \multicolumn{2}{c}{RSA (netlist Trojan)}\\
\hline
& Single input & Three inputs &  Single bin& 16 bins\\
\hline
Accuracy& 0.9930 &  0.9995 & 1.0 & 1.0\\
\hline
Precision & 0.9862 & 0.9991 & 0.9504 & 0.9947\\
\hline
Recall & 1.0 & 1.0 & 1.0 & 1.0\\
\hline
F1 & 0.9931 & 0.9995 & 0.9745 & 0.9973\\
\hline
\end{tabular}
\label{metricstablersa}
\end{table}%

\subsection{Experiment 2: Basic RSA-T100 Trojan Inserted into Netlist}
\label{sec:RSAnetlist}
\begin{figure*}
\centering
\begin{tabular}{cc}
\subfloat[Clean IC]{\includegraphics[width=0.49\textwidth]{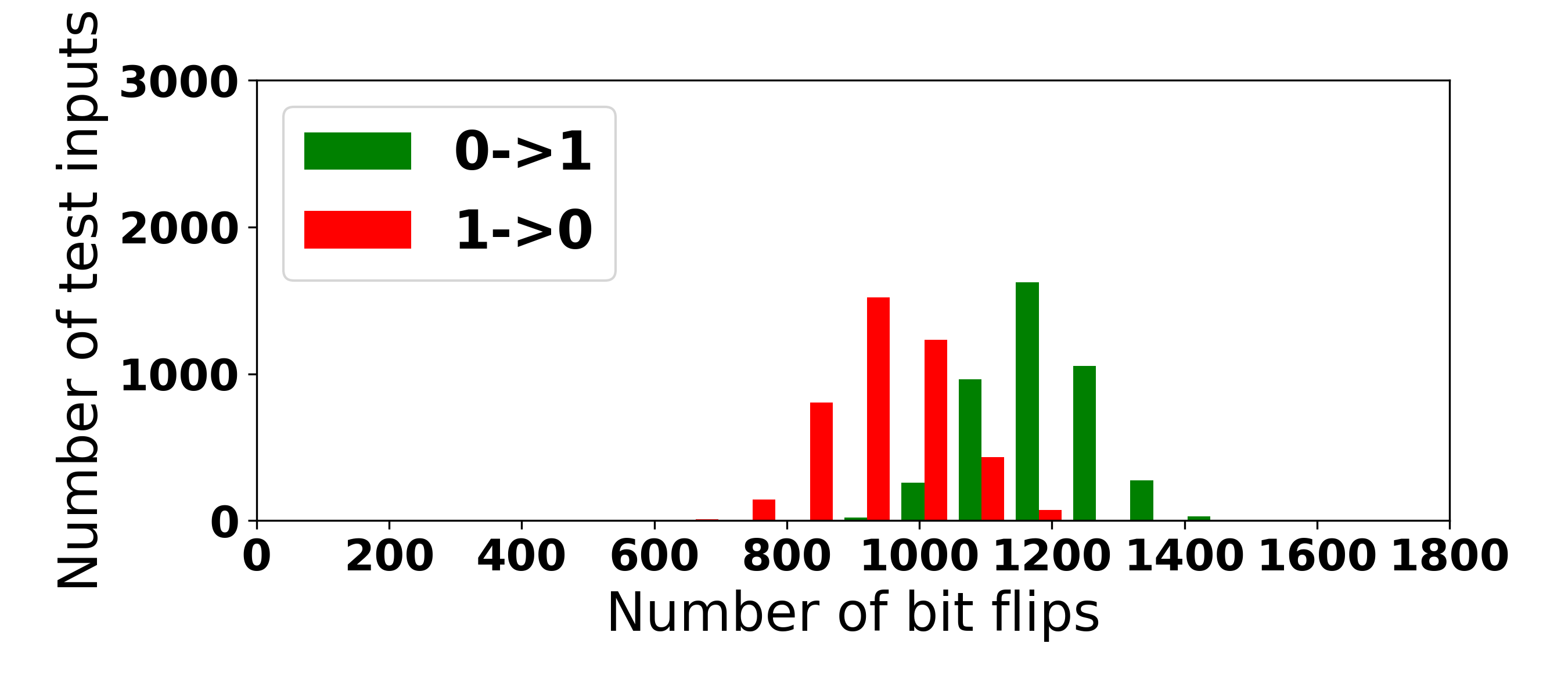}\label{fig:histbasenoAg}}&
\subfloat[IC with Trojan inserted into netlist]{\includegraphics[width=0.49\textwidth]{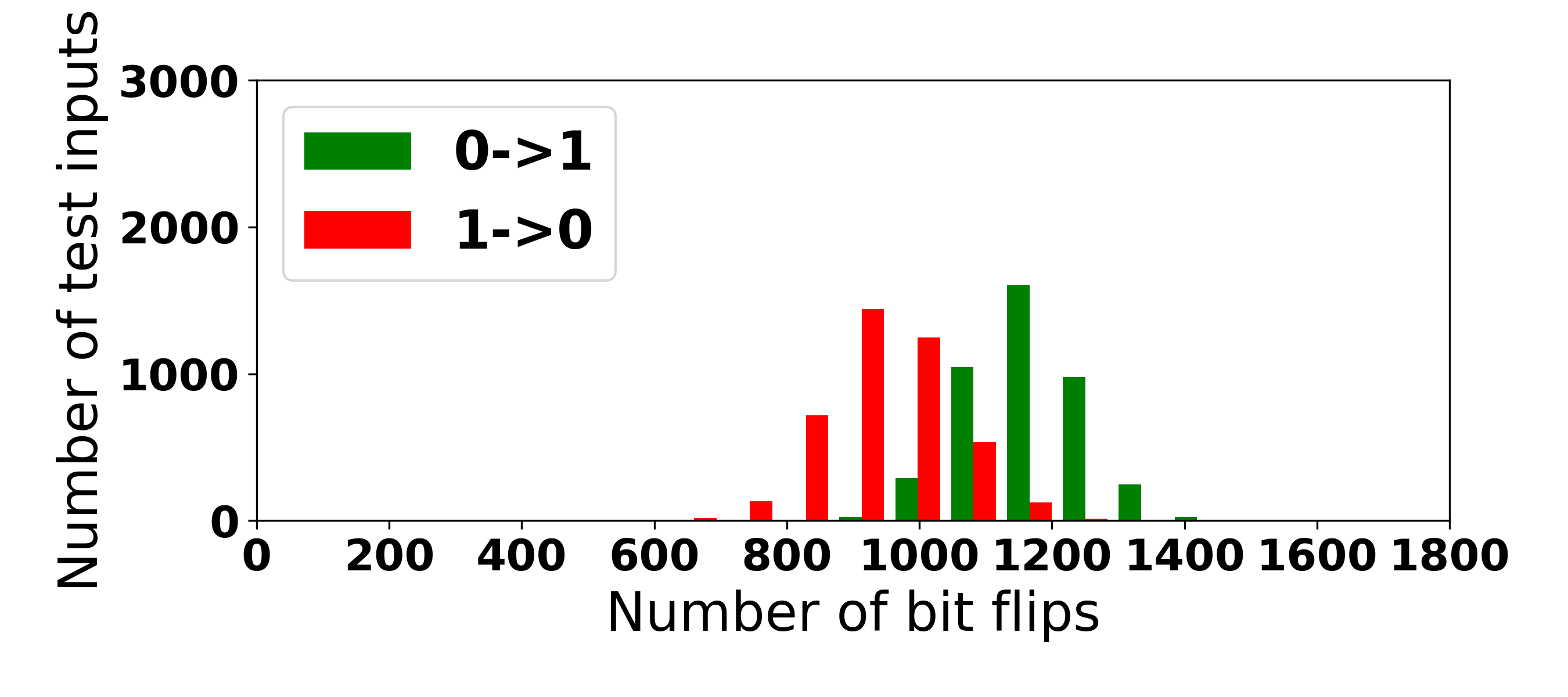}\label{fig:histtrnoAg}}
\end{tabular}
\caption{Bit flips induced by over clocking without the aging effects in RSA circuit.}
 \end{figure*}

\begin{figure*}
\centering
\begin{tabular}{cc}
\subfloat[Clean IC]{\includegraphics[width=0.49\textwidth]{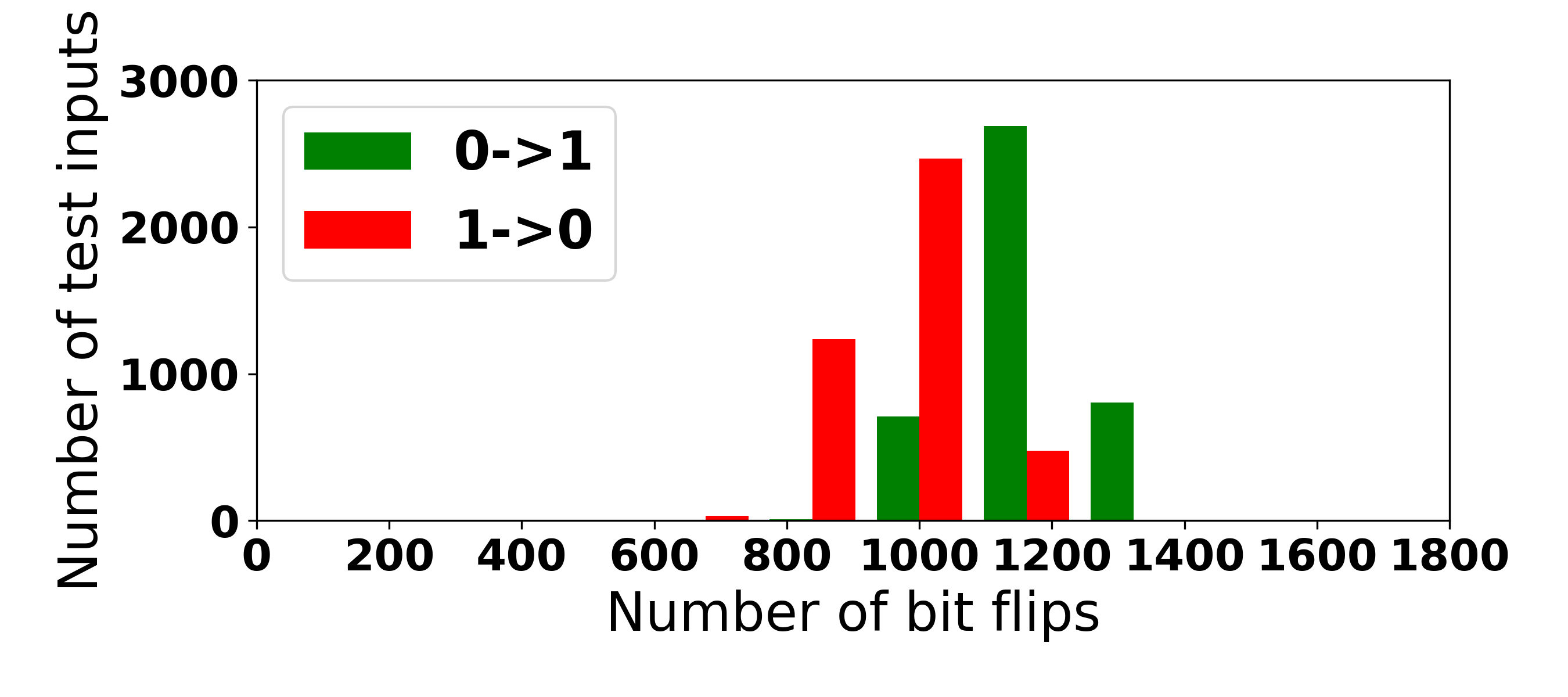}\label{fig:histbasemaxAg}}&
\subfloat[IC with Trojan inserted into netlist]{\includegraphics[width=0.49\textwidth]{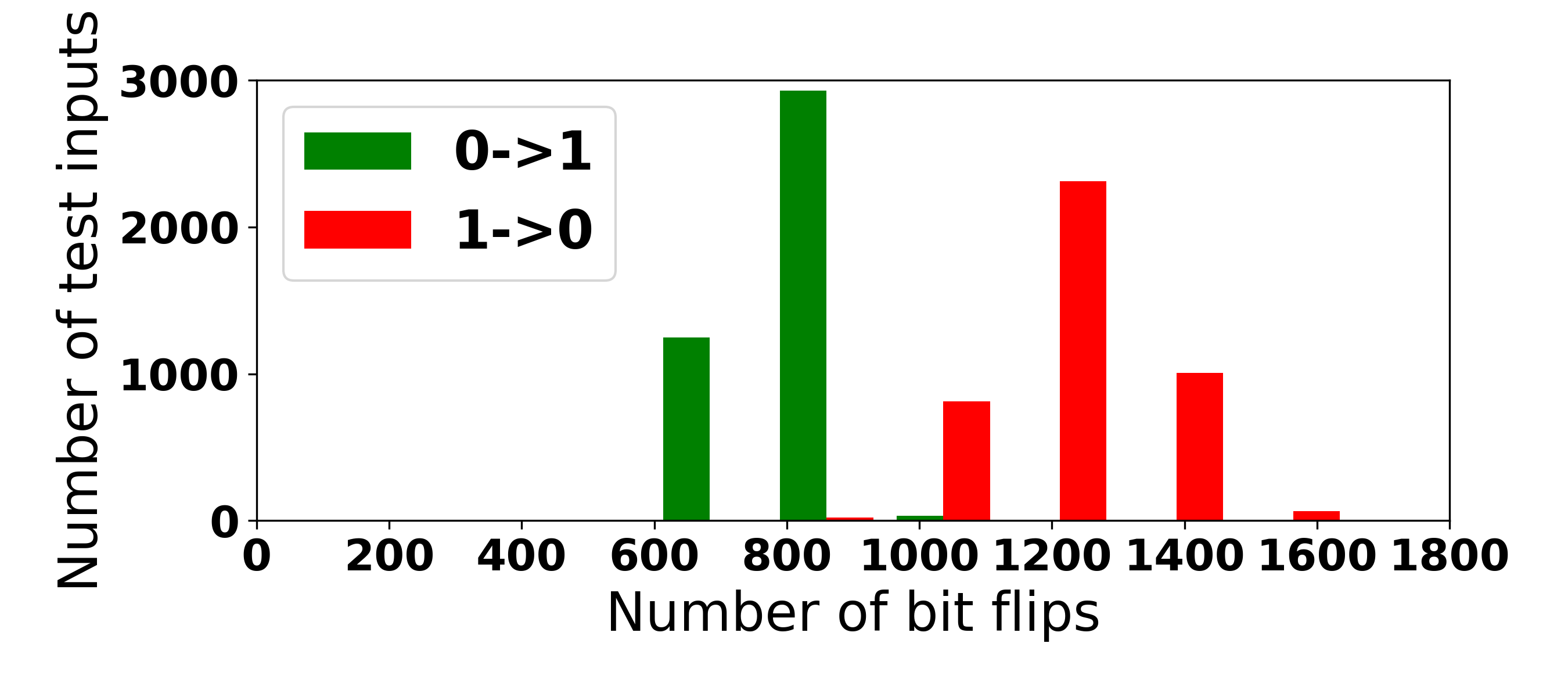}\label{fig:histtrmaxAg}}
\end{tabular}
\caption{Bit flips exacerbated by over clocking when the aging is maximum in the RSA circuit.}
 \end{figure*}

\begin{figure*}
\centering
\begin{tabular}{cc}
\subfloat[Clean IC]{\includegraphics[width=0.49\textwidth]{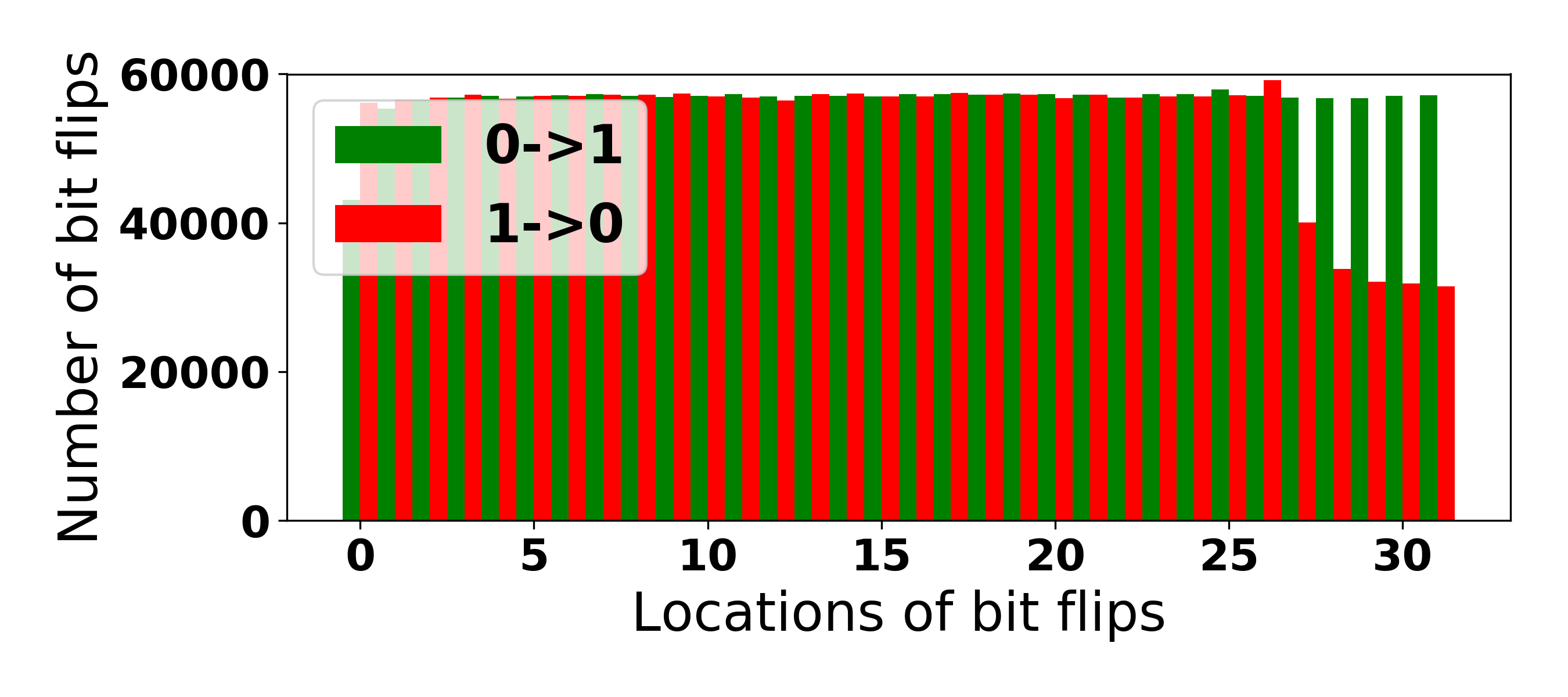}\label{fig:barbasenoAg}}&
\subfloat[IC with Trojan inserted into netlist]{\includegraphics[width=0.49\textwidth]{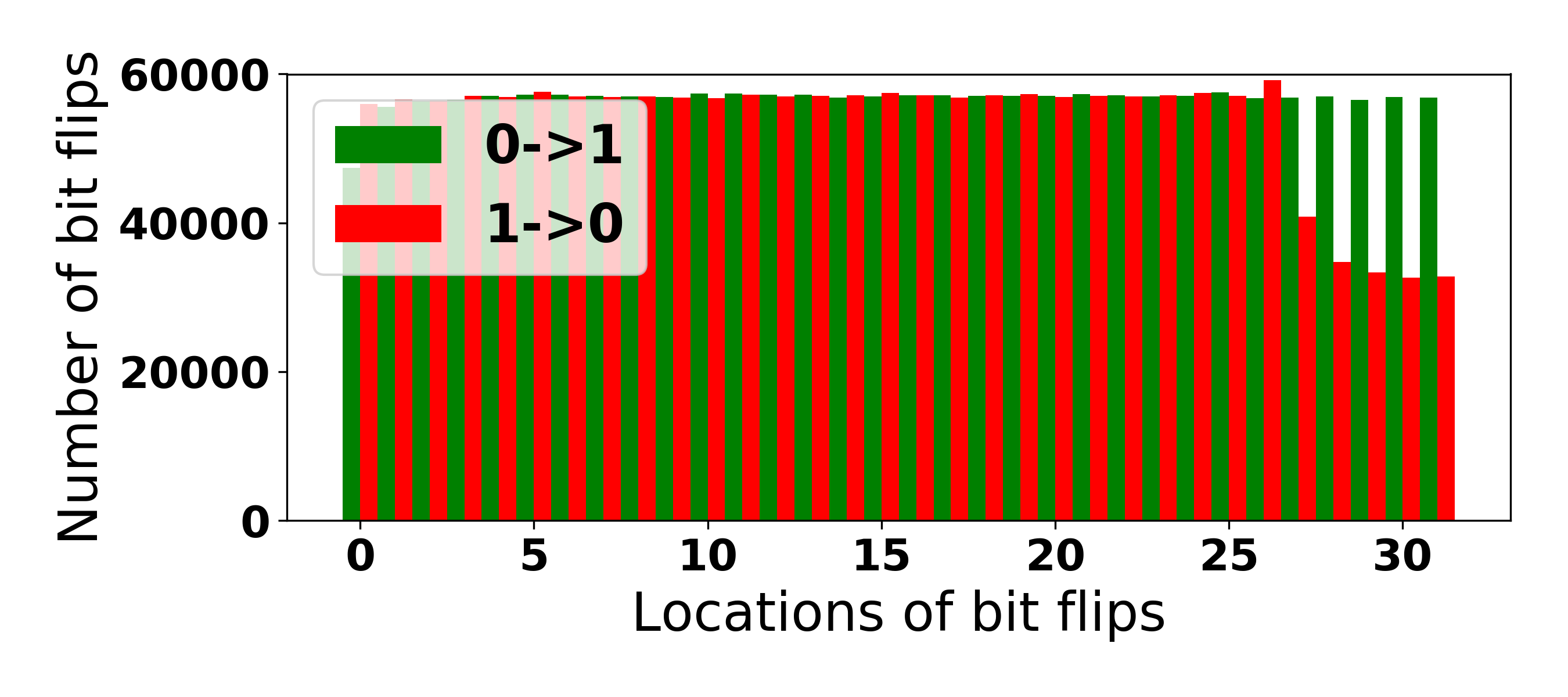}\label{fig:bartrnoAg}}
\end{tabular}
\caption{Weighted locations of bit flips without the aging effects in RSA circuit.}
 \end{figure*}

\begin{figure*}
\centering
\begin{tabular}{cc}
\subfloat[Clean IC]{\includegraphics[width=0.49\textwidth]{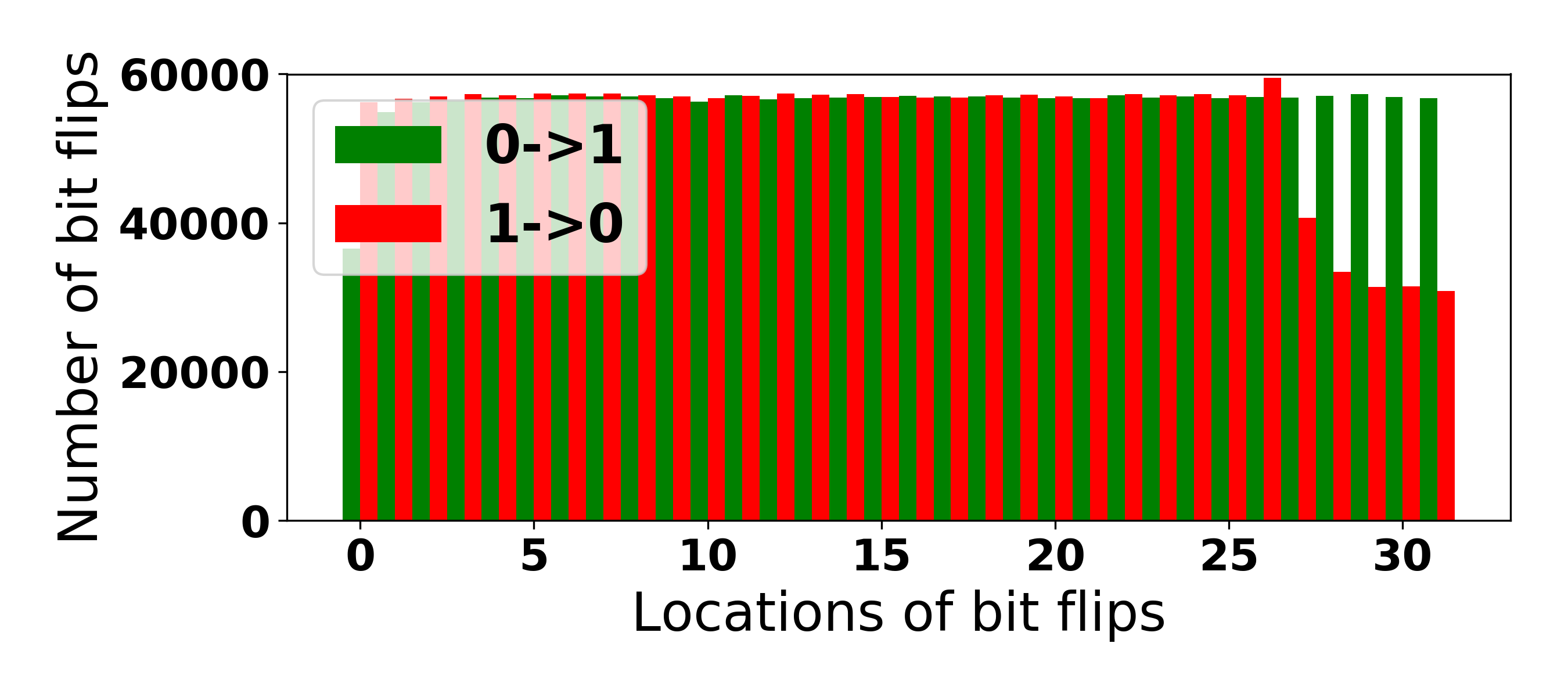}\label{fig:barbasemaxAg}}&
\subfloat[IC with Trojan inserted into netlist]{\includegraphics[width=0.49\textwidth]{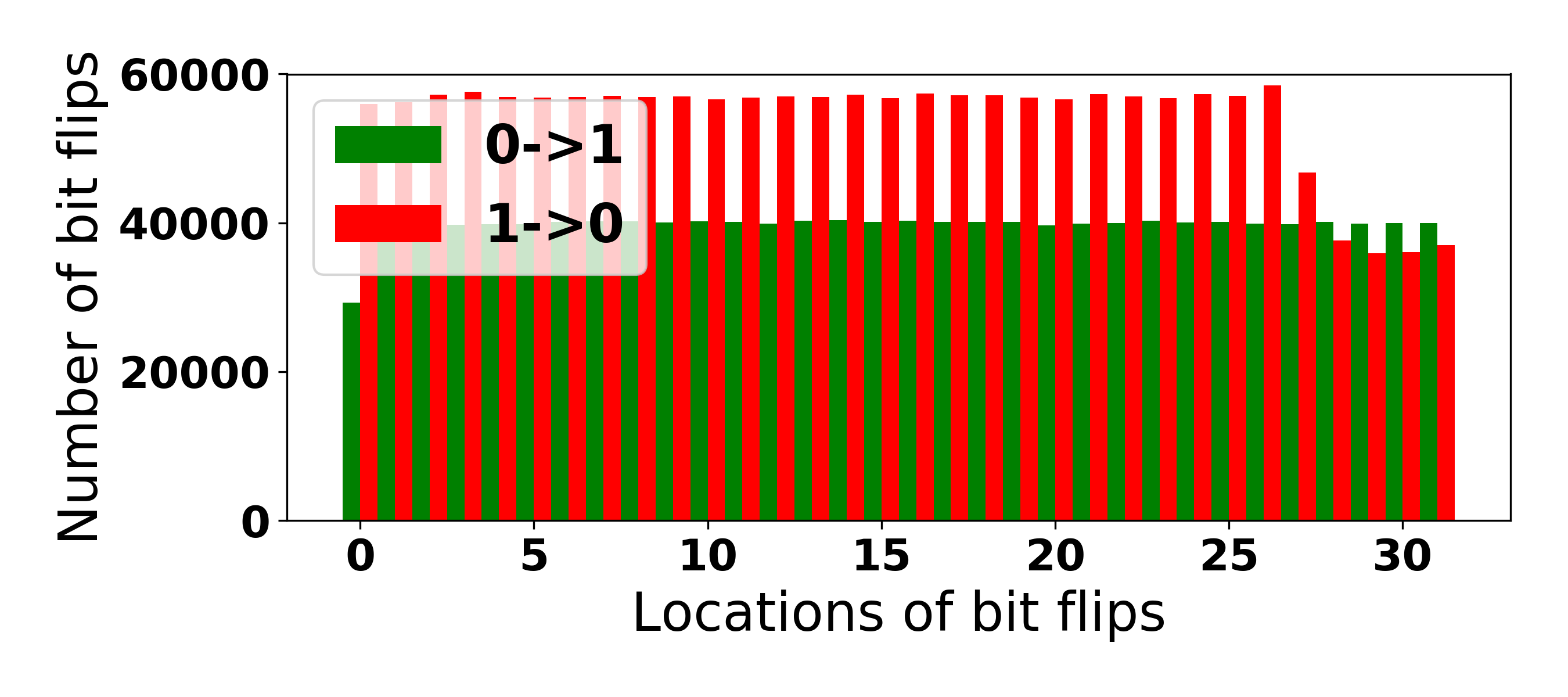}\label{fig:bartrmaxAg}}
\end{tabular}
\caption{Weighted locations of bit flips when the aging is maximum in RSA circuit.}
 \end{figure*}

We make Trojan detection challenging by inserting the Trojan into the synthesized netlist. This makes minimal alterations to the original design and hence difficult to detect. 
 The Trojan in this experiment occupies 0.22\% of the circuit area.
We collect data in the clock range 0.71 ns -- 0.84 ns with a step of 0.005 ns. This range is lower than the RTL inserted one since the synthesized circuit is similar to the original Trojan-free circuit. Aggressive over-clocking is  to get discernible patterns of bit errors at the outputs. 
 Figures \subref*{fig:histbasenoAg}, \subref*{fig:histtrnoAg} show the histograms of bit flips at output for no aging and Figures \subref*{fig:histbasemaxAg}, \subref*{fig:histtrmaxAg} show the histograms for maximum aging. Figures \subref*{fig:barbasenoAg}, \subref*{fig:bartrnoAg} show the bar charts of weighted location of bit flips for no aging and Figures \subref*{fig:barbasemaxAg} and \subref*{fig:bartrmaxAg} show the same for maximum aging.  These  figures show that the features provide a discernible difference when the Trojan is inserted in the circuit. 
 As the Trojan is inserted into the netlist, the overall netlist will now be different from the previous case (Experiment 1). The number of gates involved and the structure of the netlist changes. Thus, the bit flip distribution changes from the previous case.

\begin{figure*}
\centering
\begin{tabular}{cc}
\subfloat[Clean IC]{\includegraphics[width=0.5\textwidth]{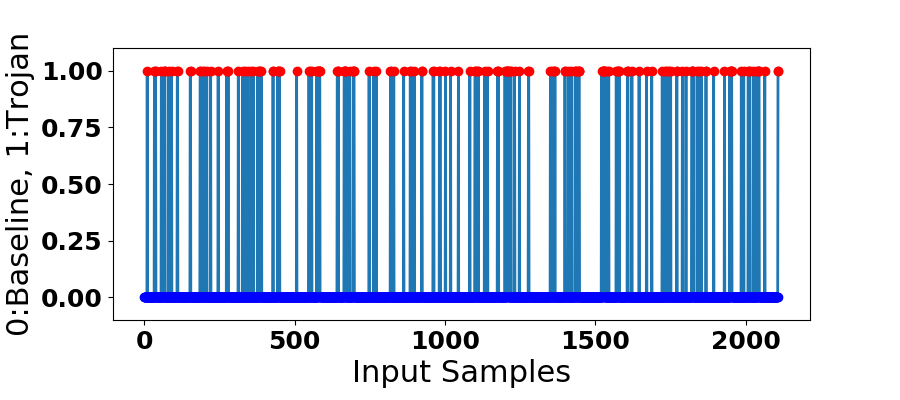}\label{fig:accb1}}&
\subfloat[IC with Trojan inserted into netlist]{\includegraphics[width=0.5\textwidth]{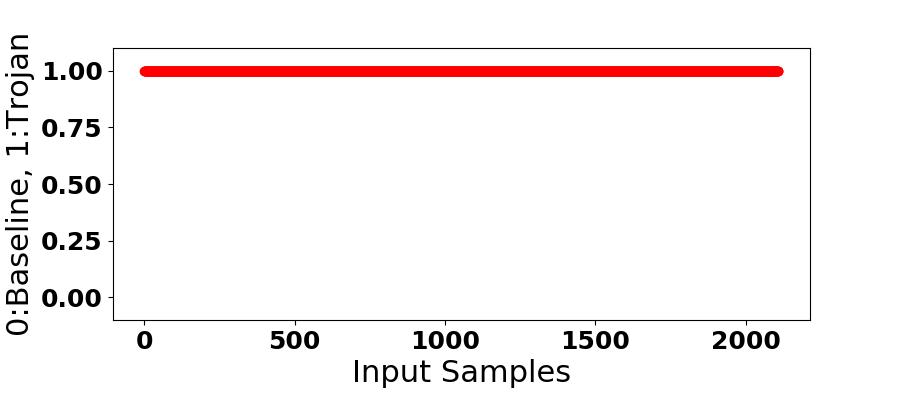}\label{fig:acctr1}}
\end{tabular}
\caption{Time series of anomaly detection using a single bin for RSA circuit.}
 \end{figure*}
 
 \begin{figure*}
\centering
\begin{tabular}{cc}
\subfloat[Clean IC]{\includegraphics[width=0.5\textwidth]{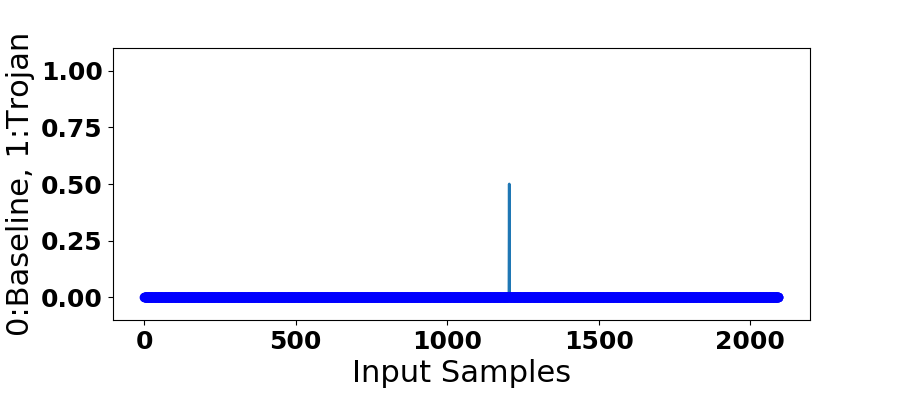}\label{fig:accb16}}&
\subfloat[IC with Trojan inserted into netlist]{\includegraphics[width=0.5\textwidth]{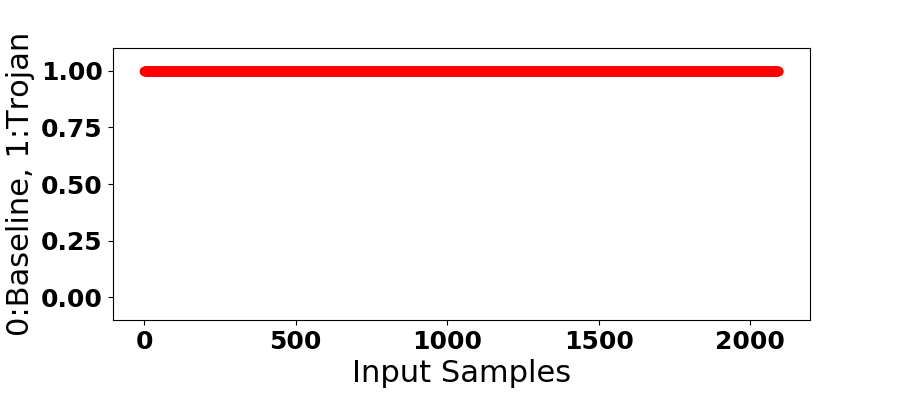}\label{fig:acctr16}}
\end{tabular}
\label{fig:anomaly_rsanetlist}
\caption{Time series of anomaly detection using a batch of 16-bins for RSA circuit.}
 \end{figure*}

\begin{figure*}[!htb]
\centering
\begin{tabular}{cc}

\subfloat[Clean IC]{\includegraphics[width=0.5\textwidth]{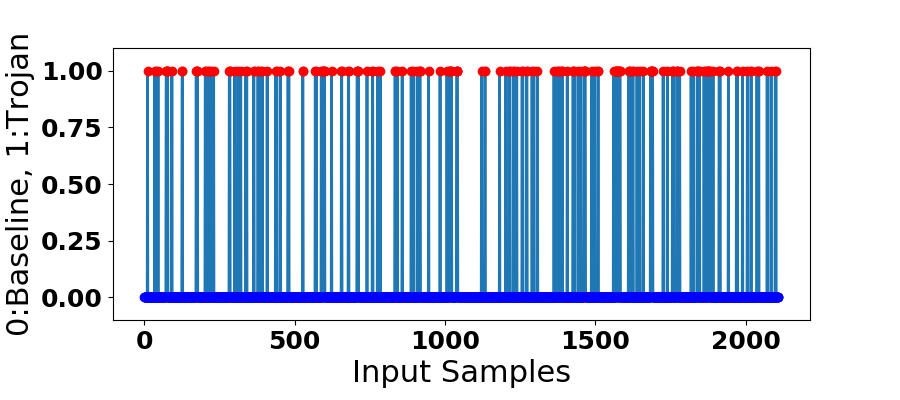}\label{fig:accbasesinglersachipvariation}}&

\subfloat[IC with Trojan inserted into netlist]{\includegraphics[width=0.5\textwidth]{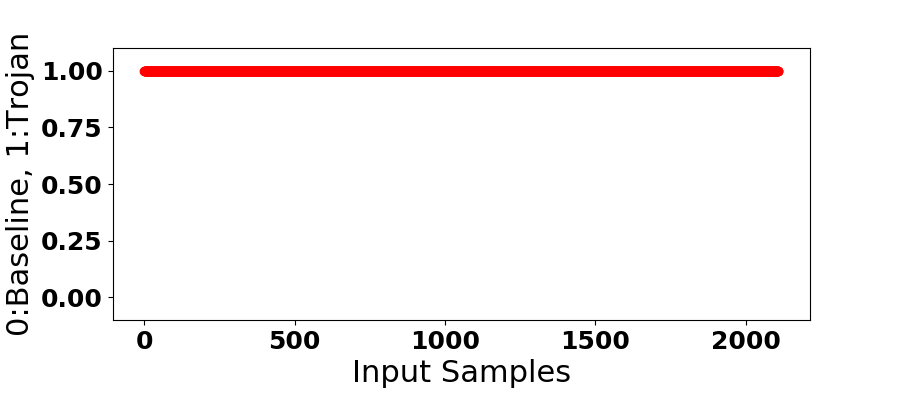}\label{fig:acctrsinglersachipvariation}}
\end{tabular}
\caption{Anomaly detection for IC-to-IC variation using single bin for RSA circuit.}
 \end{figure*}

The features are collected using a bin of 5 inputs at a time for training the classifier. When a single bin is used for testing, the accuracy of correctly classifying the Trojan-free case is 89.47\% and for the Trojan case is 100\%, as shown in Figure~\subref*{fig:accb1} and Figure~\subref*{fig:acctr1}, respectively. When 16 bins are used for testing, the accuracy for the clean IC increases to 99.47\% and that for Trojaned IC remains 100\%.
 When IC-to-IC variations are considered and single bin input is used to test, the false negative rate for data from clean IC is 13.45\% and for Trojan inserted IC, it is 0\%. A 16-bin input yields 99.57\% accuracy. For random variations in ICs, the number of inputs in the test data set is 4226. Figures \subref*{fig:accbasesinglersachipvariation}, \subref*{fig:acctrsinglersachipvariation} show the time series of anomaly detection for random variations in ICs for single input. Table \ref{metricstablersa} summarizes the results.  Using smaller number of bins yields more false positives.

\subsection{Experiment 3: AES-T100 Trojan Inserted into RTL and Netlist}
\label{sec:AEST100}
\begin{figure*}
\centering
\begin{tabular}{cc}
\subfloat[Clean IC]{\includegraphics[width=0.49\textwidth]{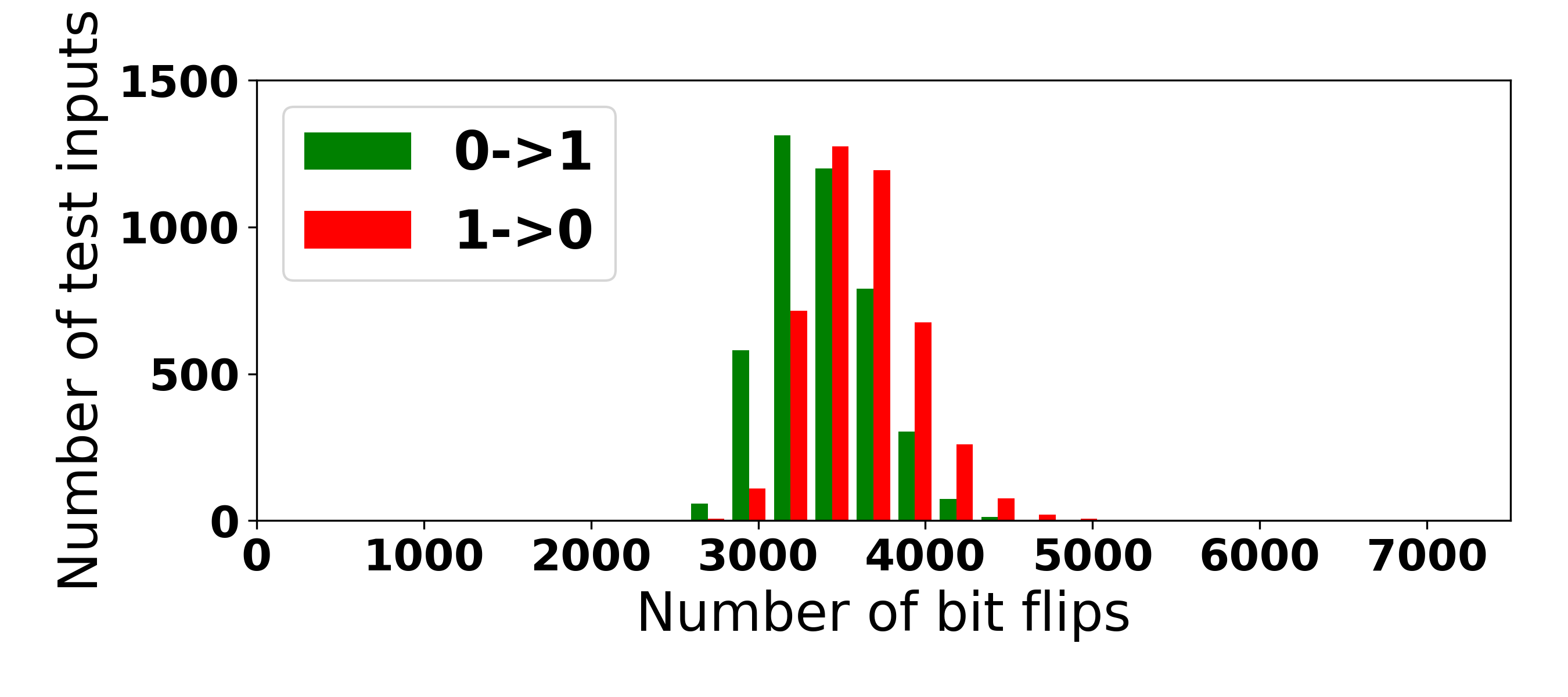}\label{fig:histbasenoAgaes100}}&
\subfloat[IC with Trojan]{\includegraphics[width=0.49\textwidth]{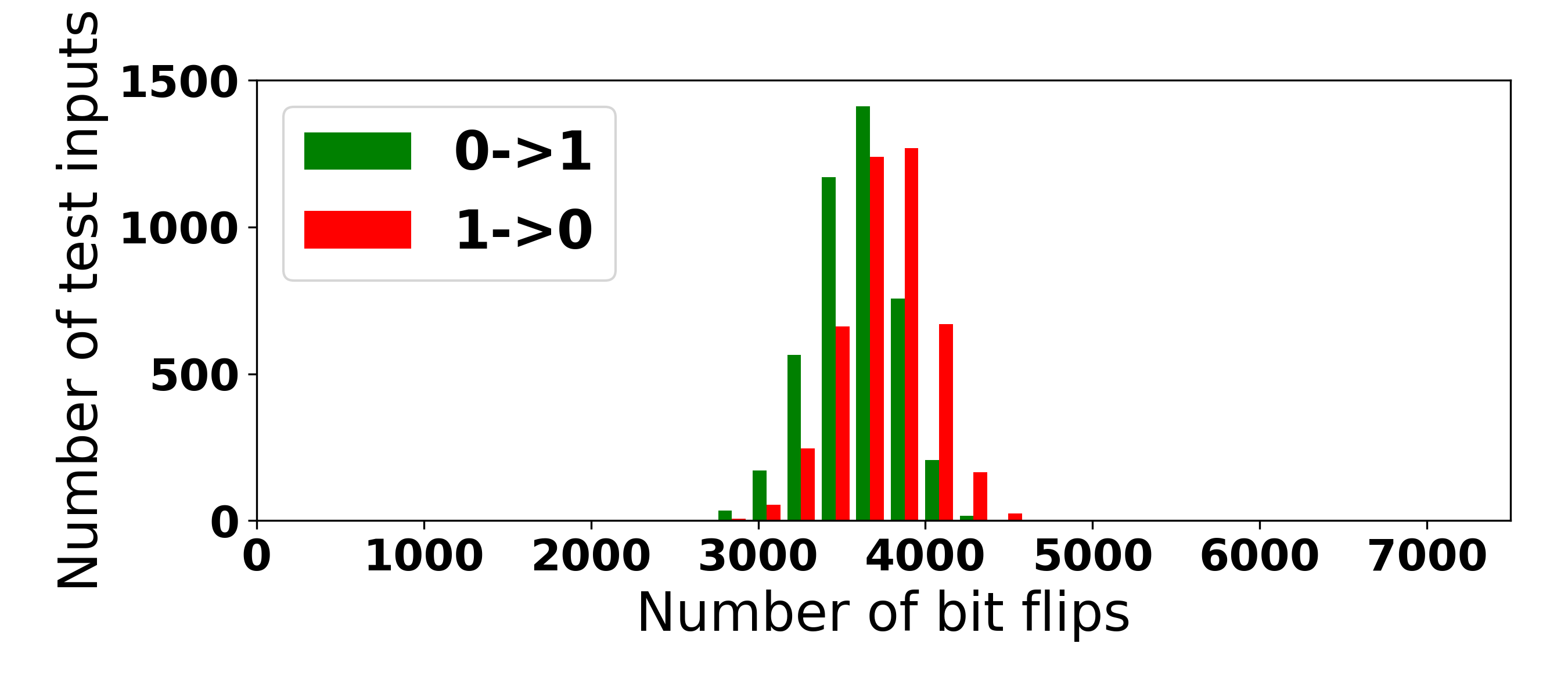}\label{fig:histtrnoAgaes100}}
\end{tabular}
\caption{Bit flips induced by over-clocking for AES-T100 with no aging.}
 \end{figure*}

\begin{figure*}
\centering
\begin{tabular}{cc}
\subfloat[Clean IC]{\includegraphics[width=0.49\textwidth]{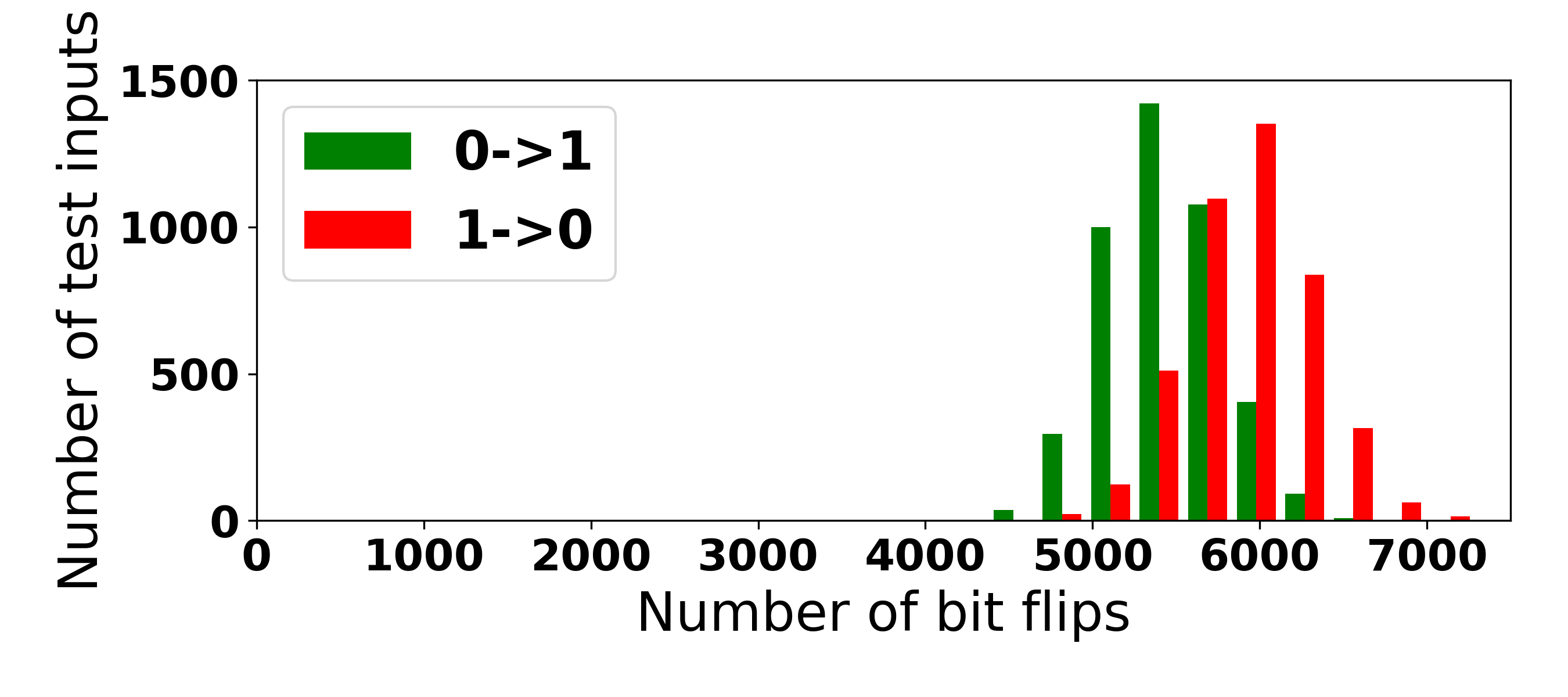}\label{fig:histbasemaxAgaes100}}&
\subfloat[IC with Trojan]{\includegraphics[width=0.49\textwidth]{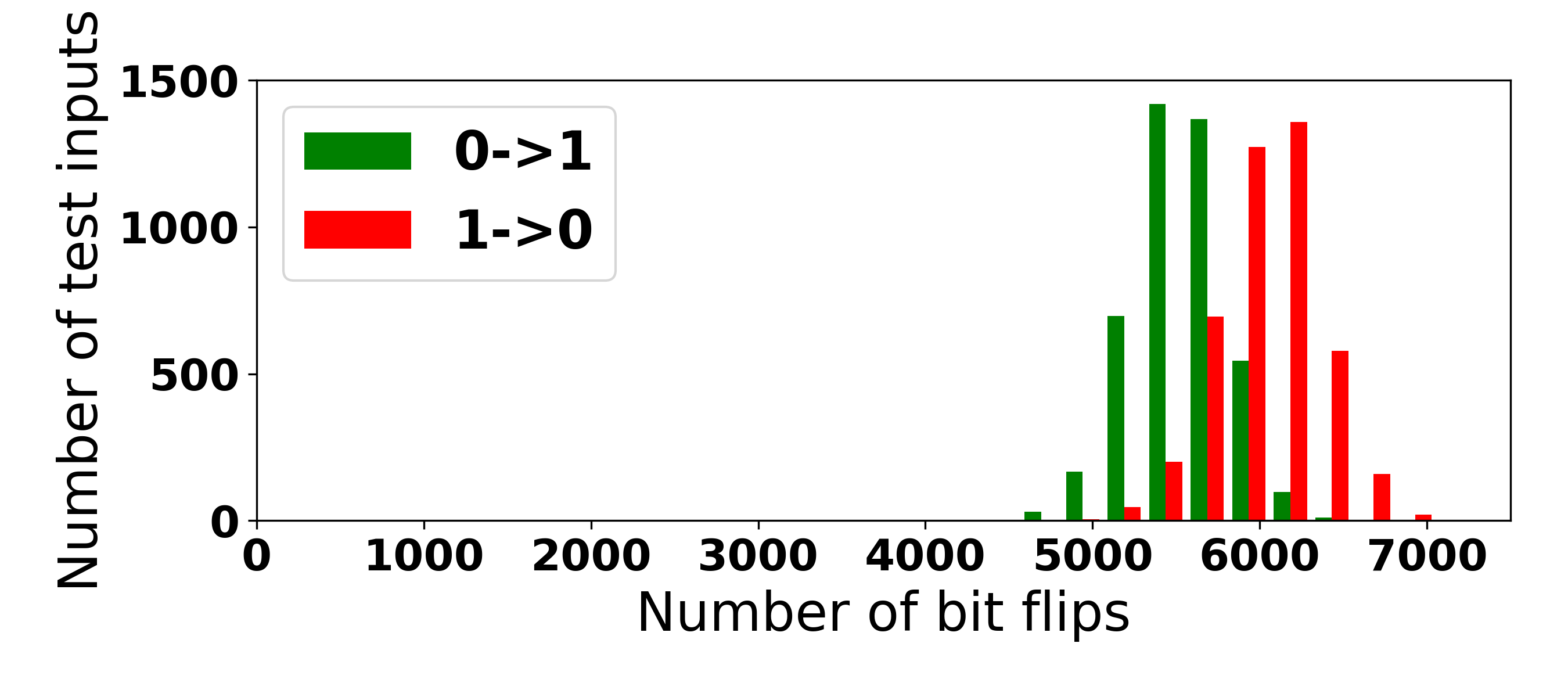}\label{fig:histtrmaxAgaes100}}
\end{tabular}
\caption{Bit flips induced by over-clocking for AES-T100 with maximum aging.}
 \end{figure*}

\begin{figure*}
\centering
\begin{tabular}{cc}
\subfloat[Clean IC]{\includegraphics[width=0.49\textwidth]{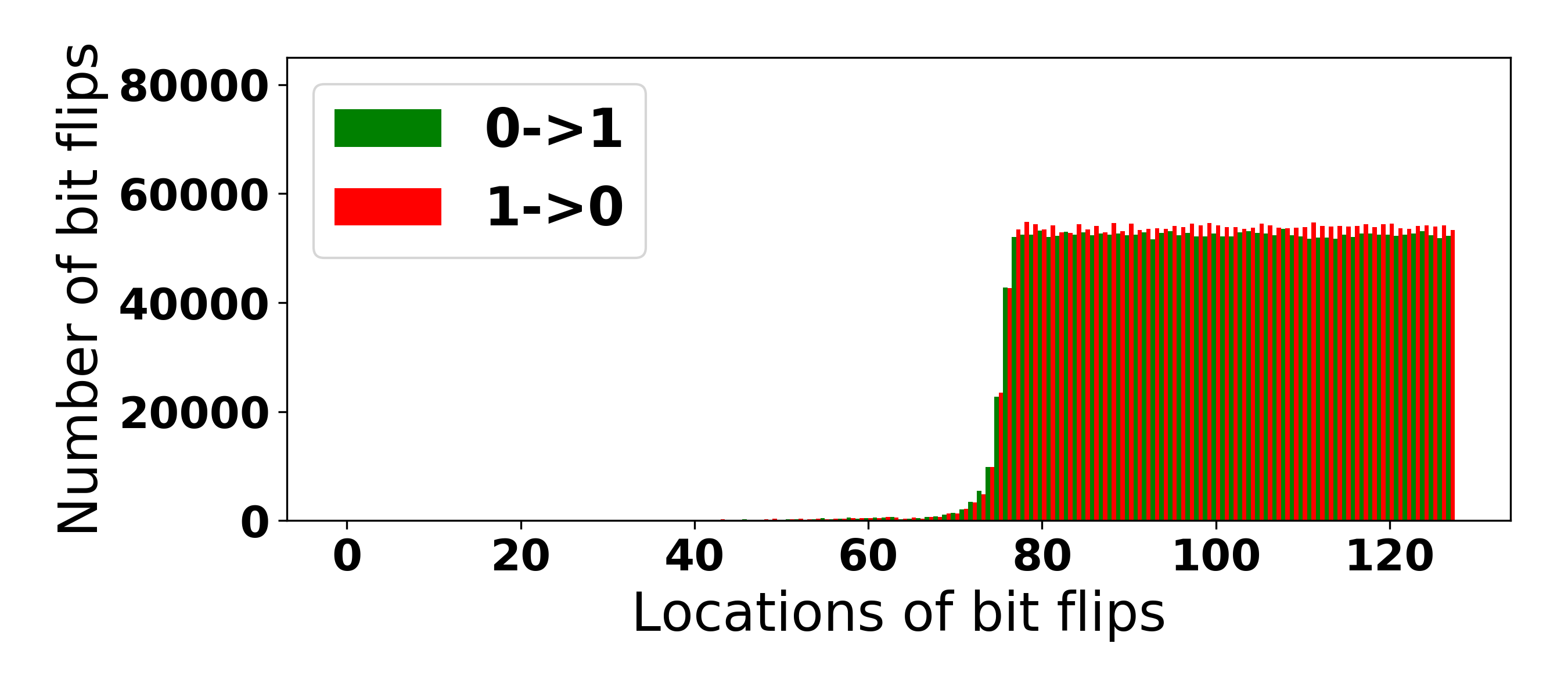}\label{fig:barbasenoAgaes100}}&
\subfloat[IC with Trojan]{\includegraphics[width=0.49\textwidth]{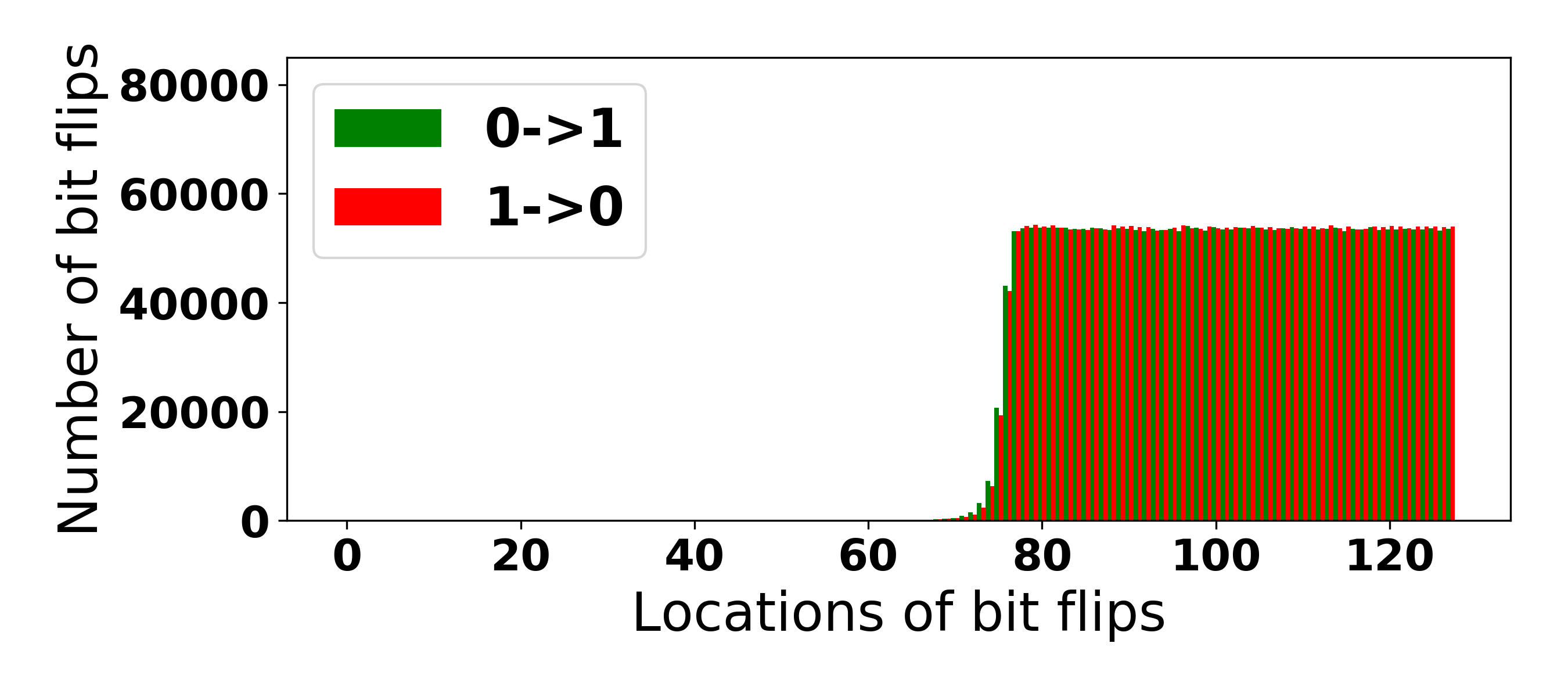}\label{fig:bartrnoAgaes100}}
\end{tabular}
\caption{Distribution of bit flip locations for AES-T100 without the aging effects.}
 \end{figure*}

\begin{figure*}
\centering
\begin{tabular}{cc}
\subfloat[Clean IC]{\includegraphics[width=0.49\textwidth]{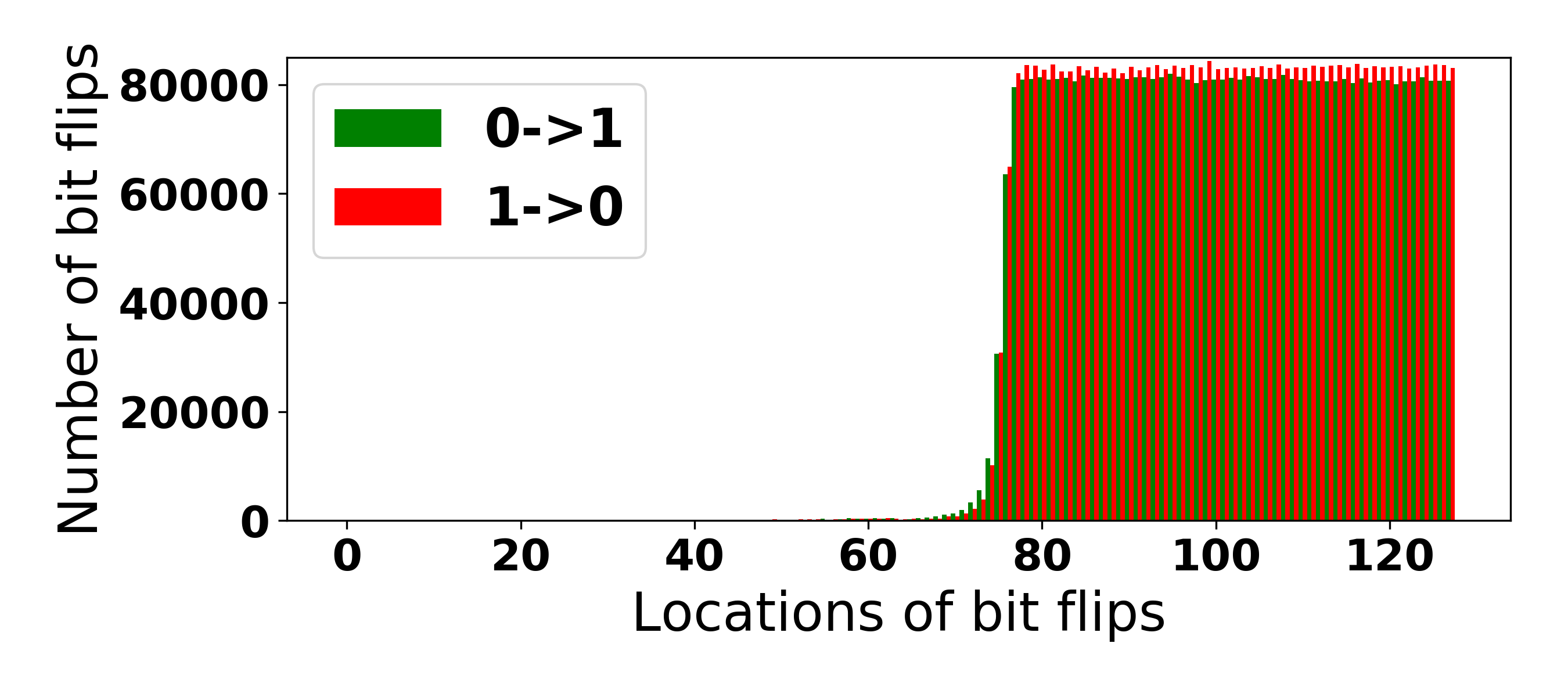}\label{fig:barbasemaxAgaes100}}&
\subfloat[IC with Trojan]{\includegraphics[width=0.49\textwidth]{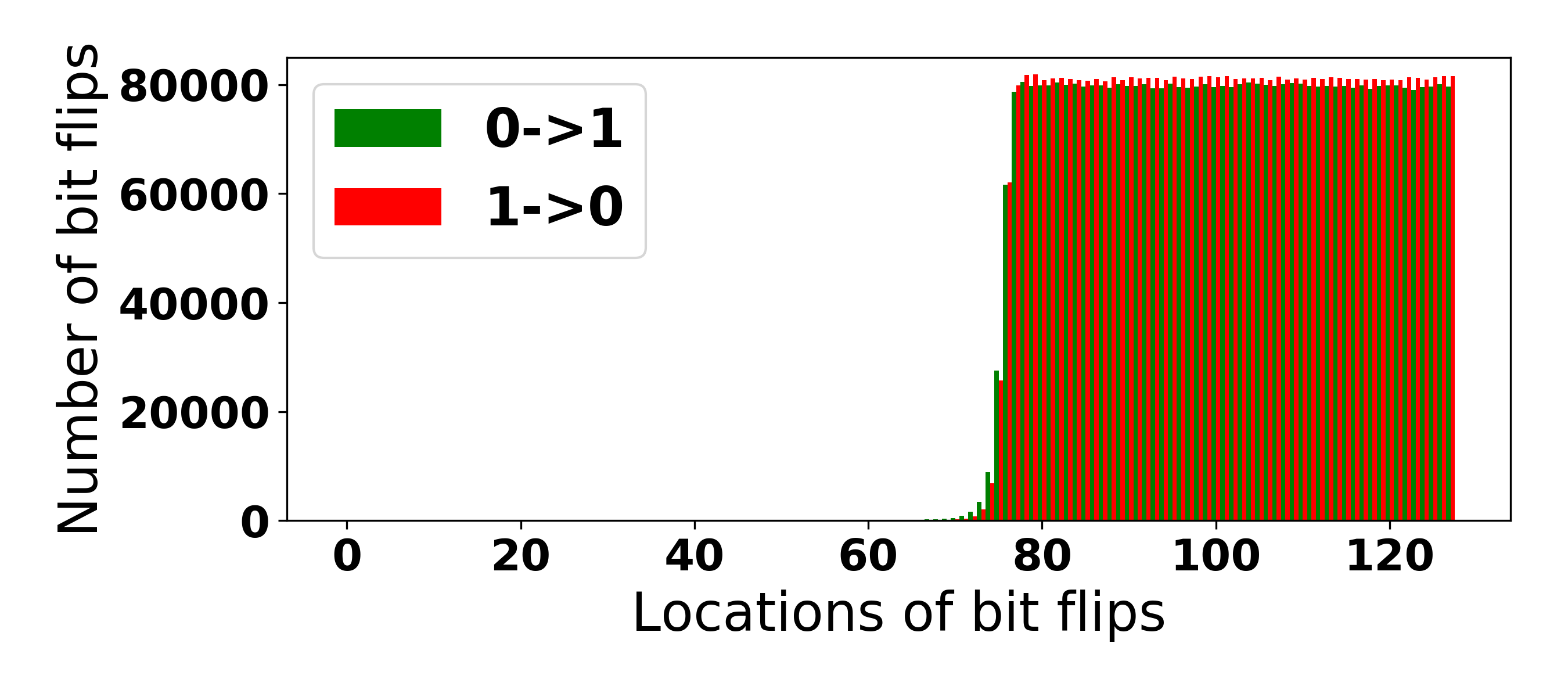}\label{fig:bartrmaxAgaes100}}
\end{tabular}
\caption{Distribution of bit flip locations for AES-T100 when maximum aging is considered.}
 \end{figure*}

 The Trojan in this experiment occupies 0.23\% of the circuit area.
The netlist generated by inserting Trojan at the RTL is same as the one obtained after inserting Trojans in the netlist. So we will show the results for one case. Since this Trojan is harder to detect (4209 paths away from the critical path) and the circuit is complex, the performance decreases when a single input is used for testing. Therefore, we enhance the features and use several bins containing different number of inputs at a time (during training and inference). For the AES circuit, the maximum clock that the circuit can be synthesized is 0.57 ns. We collect data in the clock range of 0.45 ns - 0.55 ns in steps of 0.005 ns. The data is generated for 4340 inputs.
 Figures \subref*{fig:histbasenoAgaes100} and \subref*{fig:histtrnoAgaes100} show the histograms of bit flips for no aging  and Figures \subref*{fig:histbasemaxAgaes100}, \subref*{fig:histtrmaxAgaes100} show the histograms for maximum aging. The number of bit flips have been increased from the no aging case to the worst aging case as can be seen by the shift in the histogram towards right. Figures \subref*{fig:barbasenoAgaes100}, \subref*{fig:bartrnoAgaes100} show the bar charts of weighted locations of bit flips at outputs for no aging. Whereas, Figures \subref*{fig:barbasemaxAgaes100} and \subref*{fig:bartrmaxAgaes100} show the bar charts for maximum aging.  The location of bit flips is concentrated on the least significant bits of the output. In each round of AES algorithm, ``Shift Rows'' operation rotates the bits by an amount that depends on their position. This induces more operations on most significant bits than that of least significant bits. Thus, the bit flips are more concentrated towards the end. Additionally, there is a change in bit-flip distribution and locations of bit flips from no aging to the worst aging case and the Trojan free circuit to the Trojaned circuit.
 These  figures show that these features provide a discernible difference when a Trojan is inserted.

\begin{figure*}
\centering
\begin{tabular}{cc}
\subfloat[Clean IC]{\includegraphics[width=0.5\textwidth]{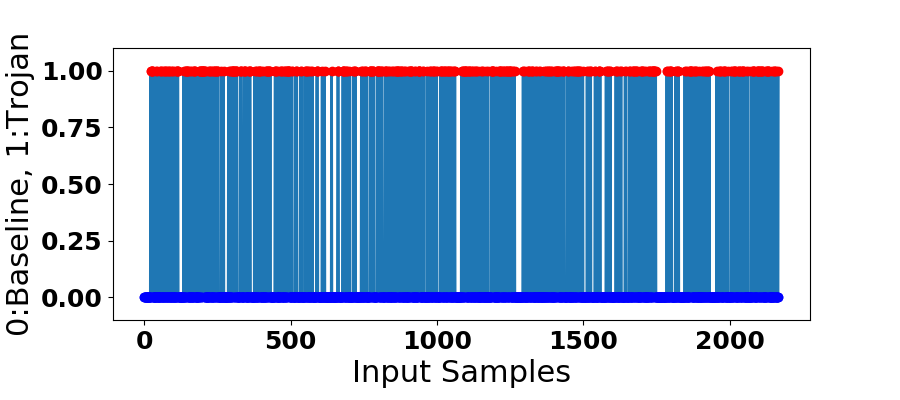}\label{fig:accb1aes}}&
\subfloat[IC with Trojan]{\includegraphics[width=0.5\textwidth]{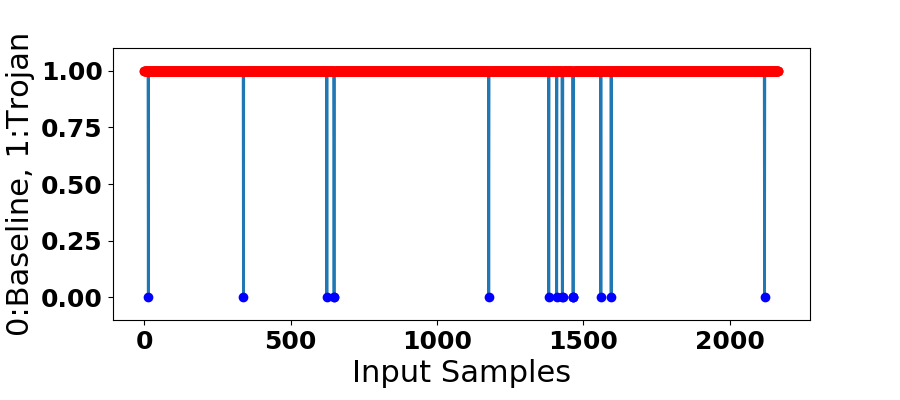}\label{fig:acctr1aes}}
\end{tabular}
\caption{Time series of anomaly detection using a single bin for AES-T100.}
 \end{figure*}

 \begin{figure*}
\centering
\begin{tabular}{cc}
\subfloat[Clean IC]{\includegraphics[width=0.5\textwidth]{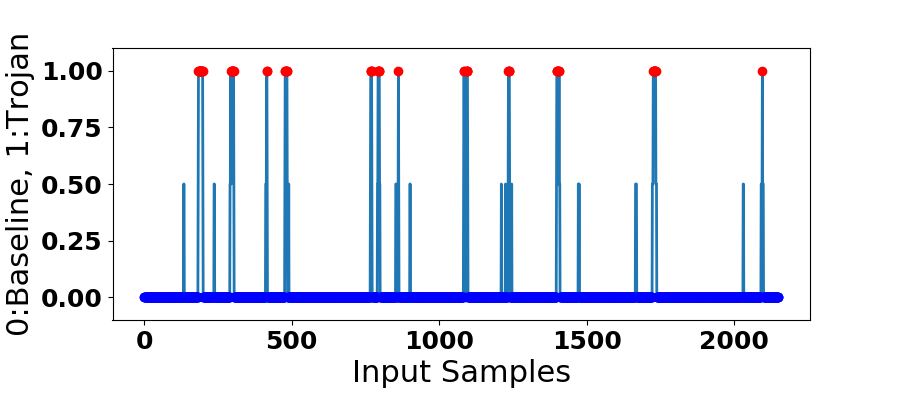}\label{fig:accb32aes}}&
\subfloat[IC with Trojan]{\includegraphics[width=0.5\textwidth]{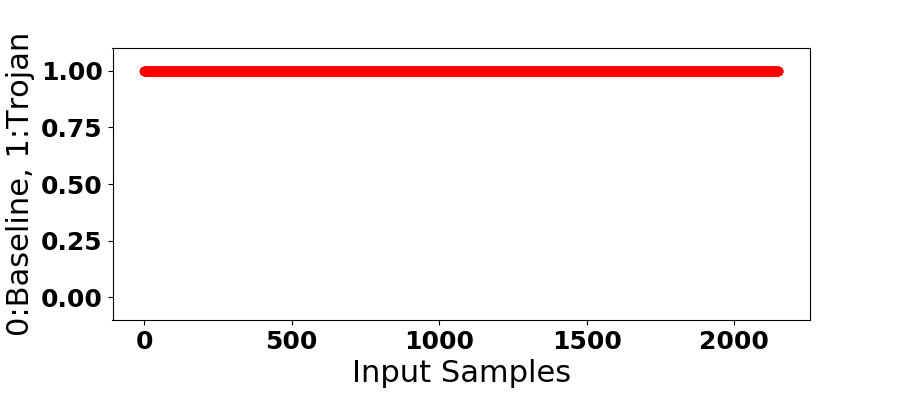}\label{fig:acctr32aes}}
\end{tabular}
\label{fig:anomaly_aes100}
\caption{Time series of anomaly detection using a batch of 32 bins for AES-T100.}
 \end{figure*}

\begin{figure*}
\centering
\begin{tabular}{cc}

\subfloat[Clean IC]{\includegraphics[width=0.5\textwidth]{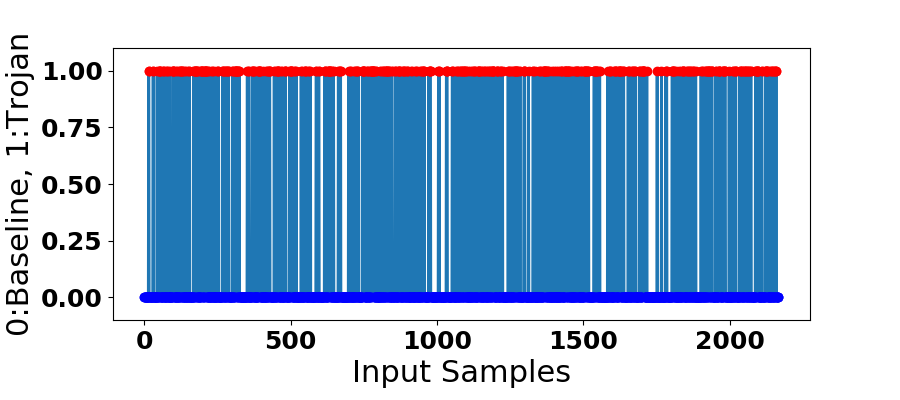}\label{fig:accbasesingleaes100chipvariation}}&

\subfloat[IC with Trojan]{\includegraphics[width=0.5\textwidth]{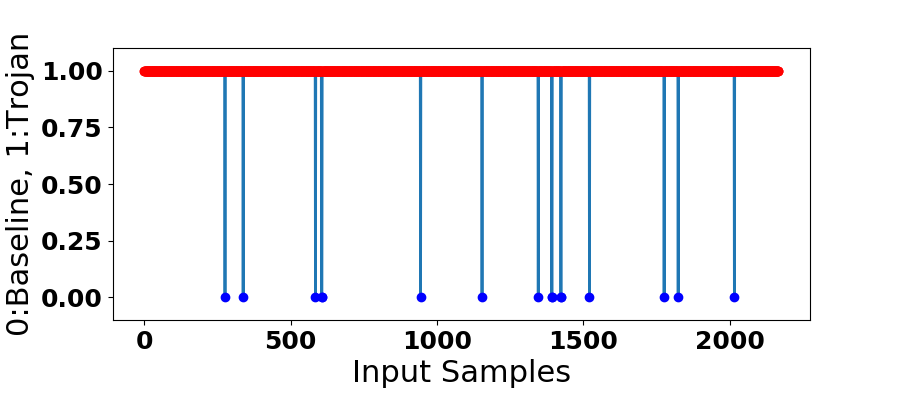}\label{fig:acctrsingleaes100chipvariation}}
\end{tabular}
\caption{Anomaly detection for IC-to-IC variation with a single bin for AES-T100.}
 \end{figure*}

With a single bin input (bin size = 5), the accuracy of correctly classifying a Trojan-free IC is 75.77\% and that for a Trojaned IC is 98.98\% (Figures \subref*{fig:accb1aes} and \subref*{fig:acctr1aes}, respectively). To reduce the false positive rate, we use a multiple-bin input. Using a batch of 32-bins increases the accuracy to 99.71\% for classifying a Trojan-free IC and 100\% accuracy when detecting a Trojaned IC (Figures~\subref*{fig:accb32aes} and~\subref*{fig:acctr32aes}).  When IC-to-IC variations are considered and a single bin input is used, the false negative rate for data from clean IC is 24.91\% and for Trojan-infected IC is 0.87\% (Figures \subref*{fig:accbasesingleaes100chipvariation}, \subref*{fig:acctrsingleaes100chipvariation}). A 32-bin input yields 99.29\% and 100\% accuracy for clean and Trojan-infected IC. The resulting precision for the classifier is presented in Table~\ref{precisiontableaes}.

\subsection{Experiment 4: AES-T1000 Trojan Inserted into RTL and Netlist}
\label{sec:AEST1000}
\begin{figure*}
\centering
\begin{tabular}{cc}
\subfloat[Clean IC]{\includegraphics[width=0.49\textwidth]{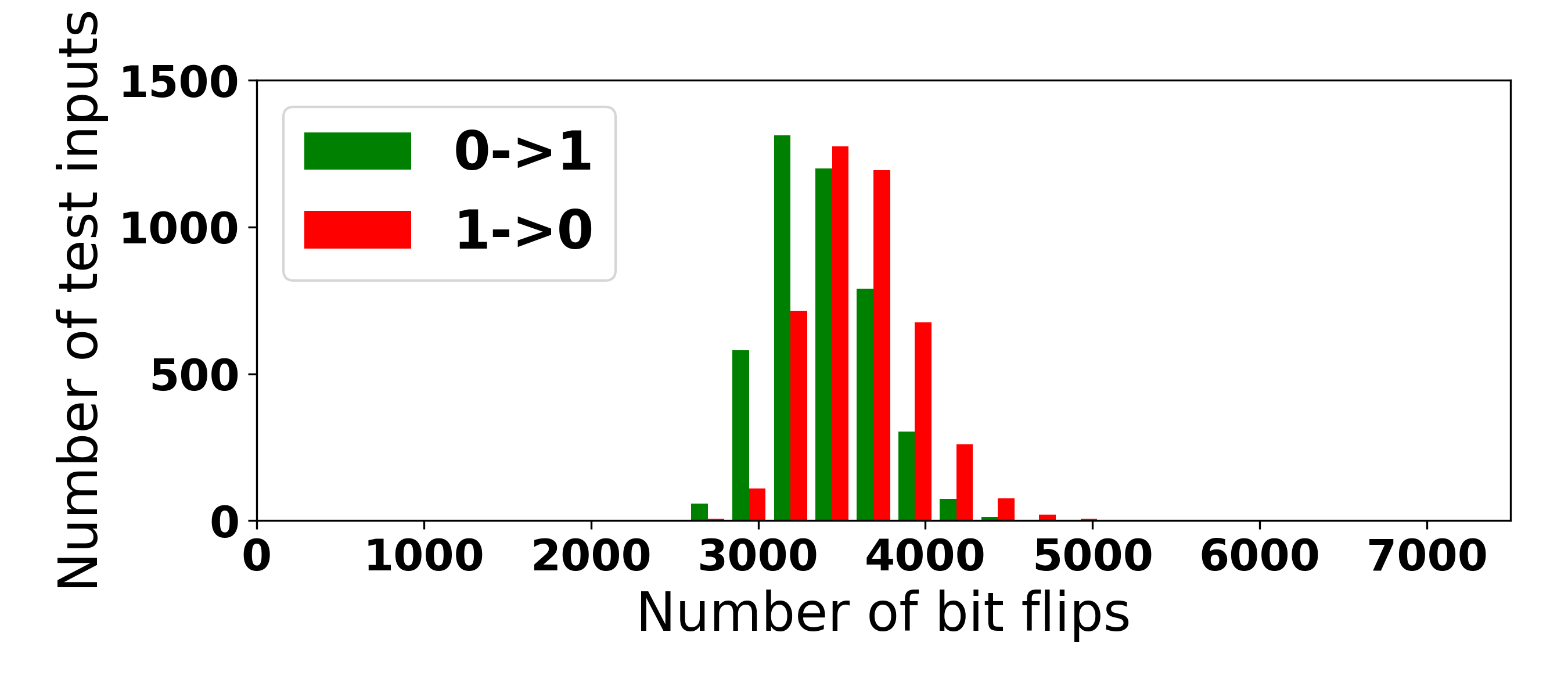}\label{fig:histbasenoAgaes1000}}&
\subfloat[IC with Trojan]{\includegraphics[width=0.49\textwidth]{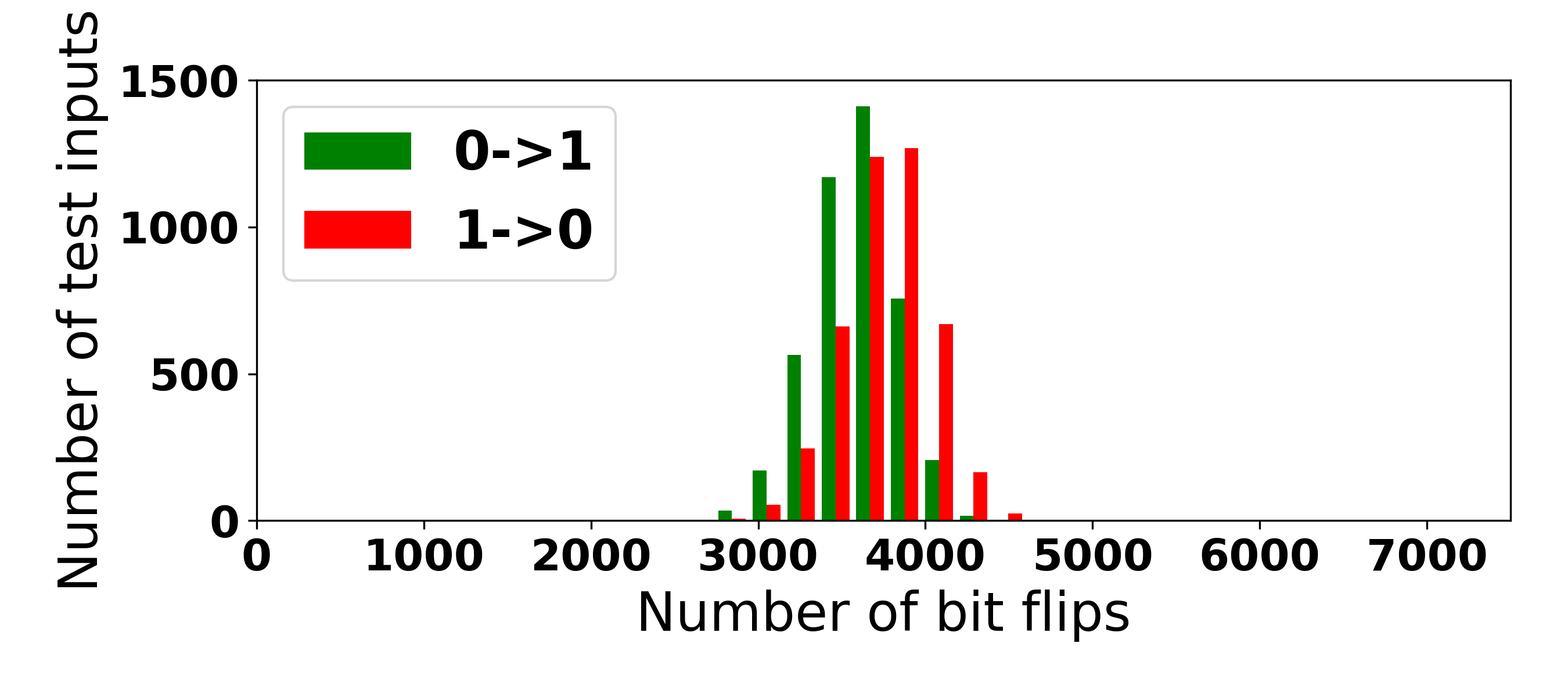}\label{fig:histtrnoAgaes1000}}
\end{tabular}
\caption{Bit flips induced by over-clocking for AES-T1000 without aging effects.}
 \end{figure*}

\begin{figure*}
\centering
\begin{tabular}{cc}
\subfloat[Clean IC]{\includegraphics[width=0.49\textwidth]{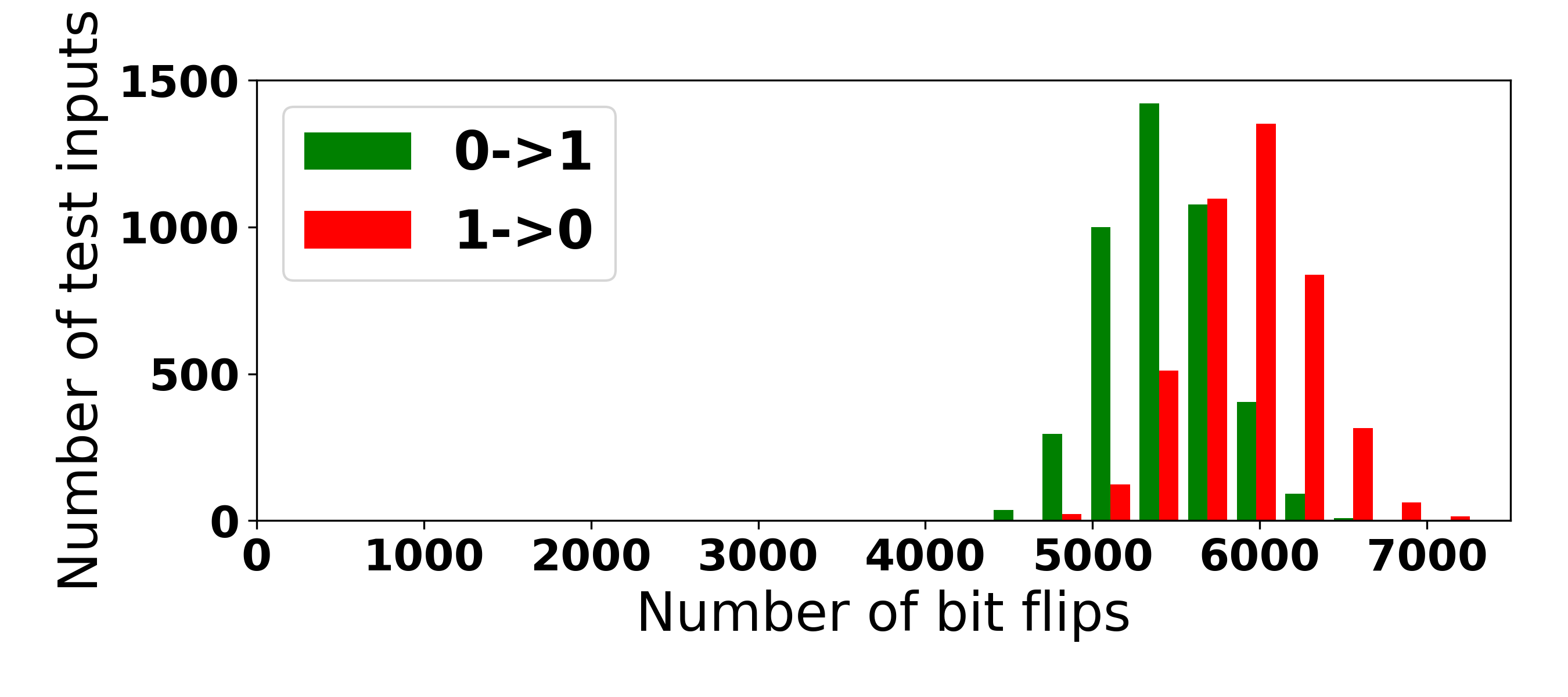}\label{fig:histbasemaxAgaes1000}}&
\subfloat[IC with Trojan]{\includegraphics[width=0.49\textwidth]{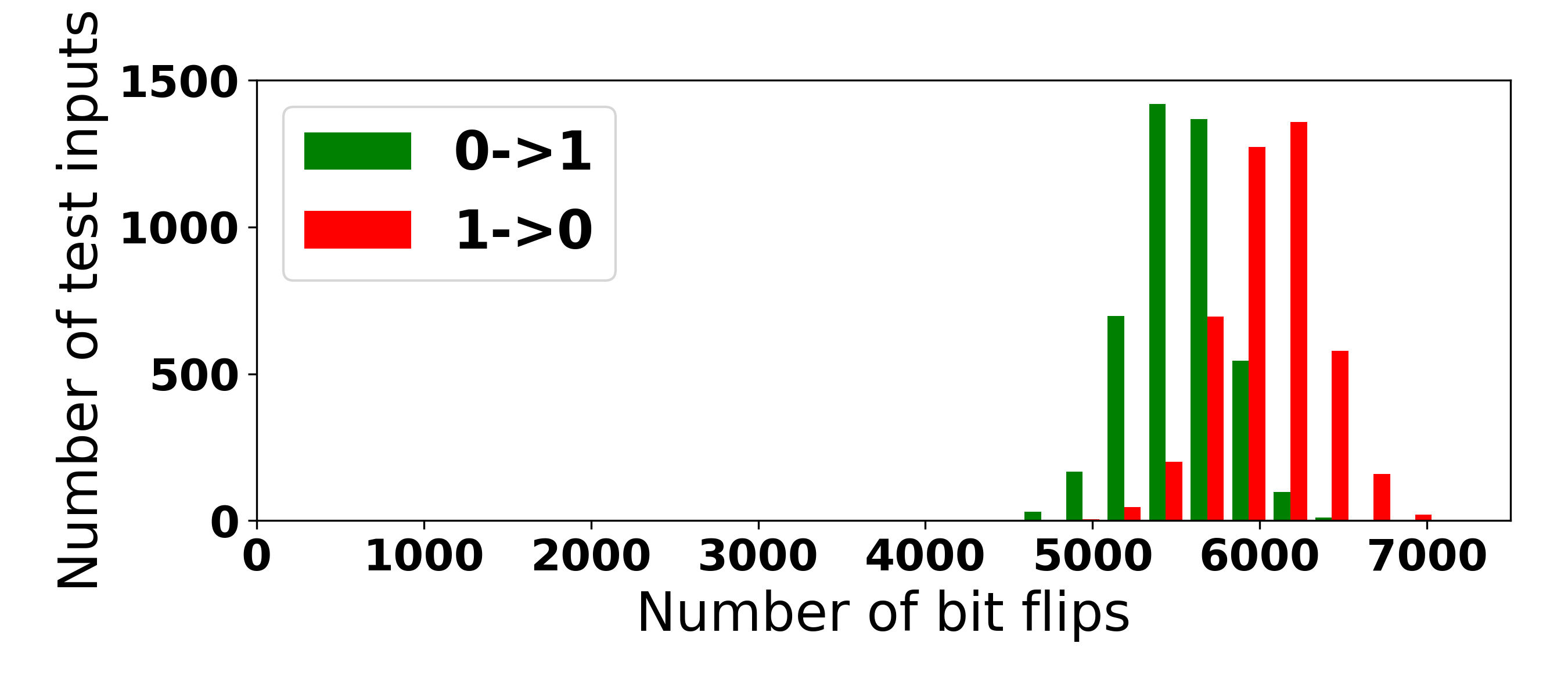}\label{fig:histtrmaxAgaes1000}}
\end{tabular}
\caption{Bit flips induced by over-clocking for AES-T1000 when aging is maximum.}
 \end{figure*}

\begin{figure*}
\centering
\begin{tabular}{cc}
\subfloat[Clean IC]{\includegraphics[width=0.49\textwidth]{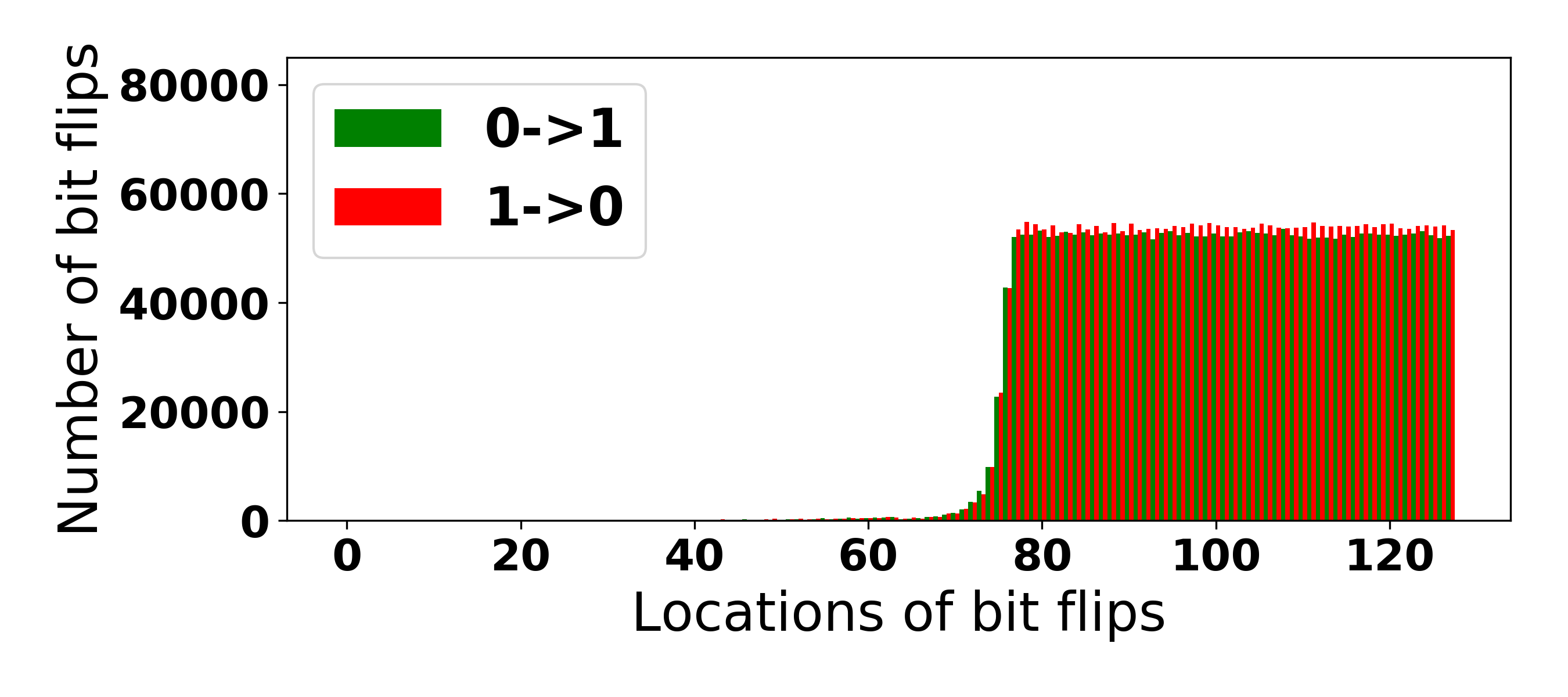}\label{fig:barbasenoAgaes1000}}&
\subfloat[IC with Trojan]{\includegraphics[width=0.49\textwidth]{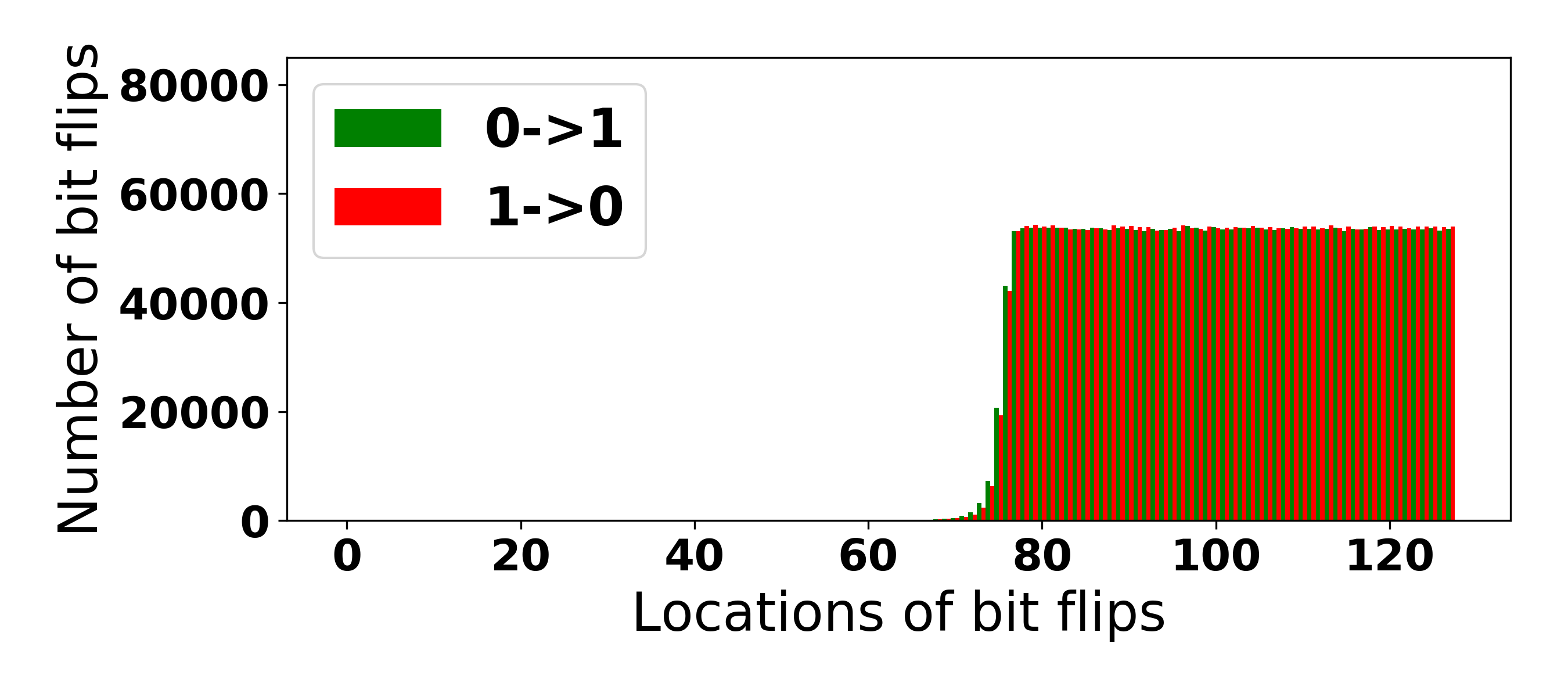}\label{fig:bartrnoAgaes1000}}
\end{tabular}
\centering
\begin{tabular}{cc}
\subfloat[Clean IC]{\includegraphics[width=0.49\textwidth]{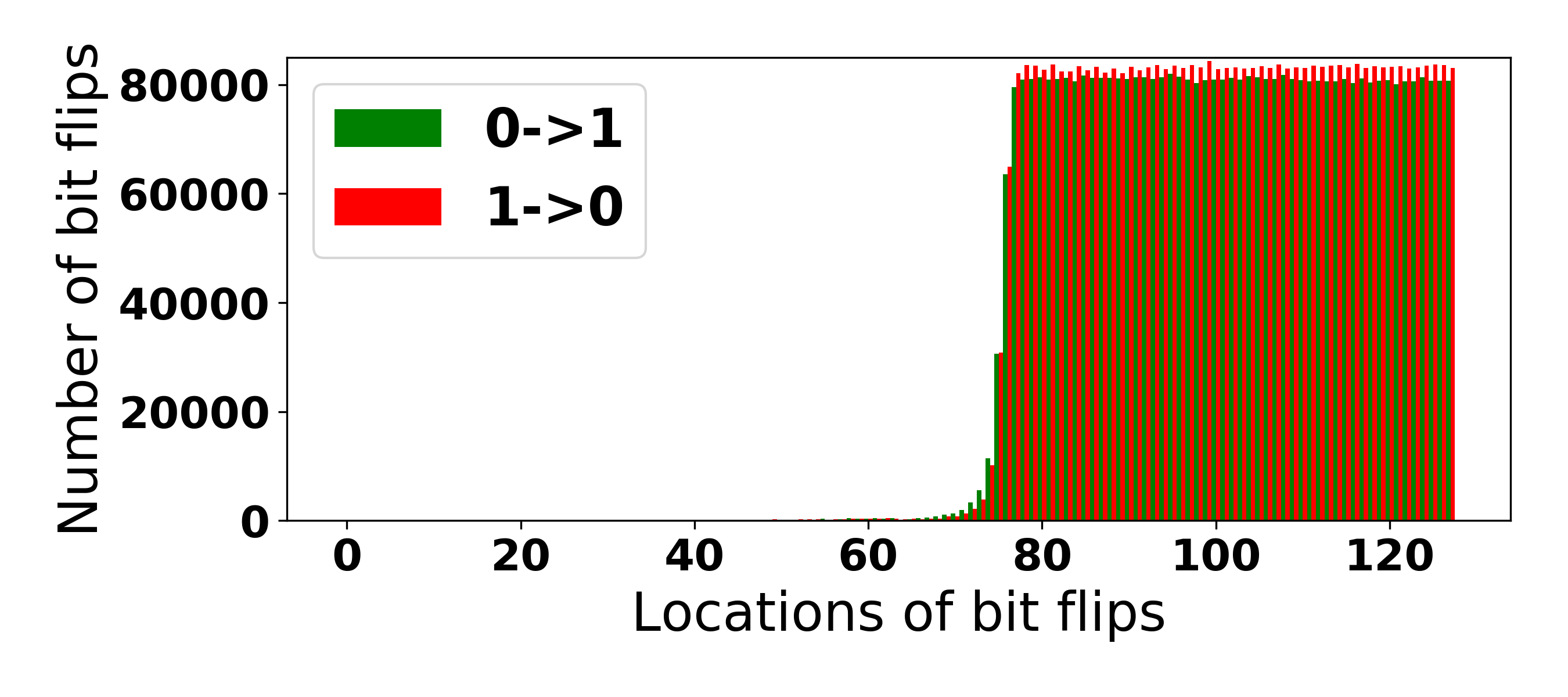}\label{fig:barbasemaxAgaes1000}}&
\subfloat[IC with Trojan]{\includegraphics[width=0.49\textwidth]{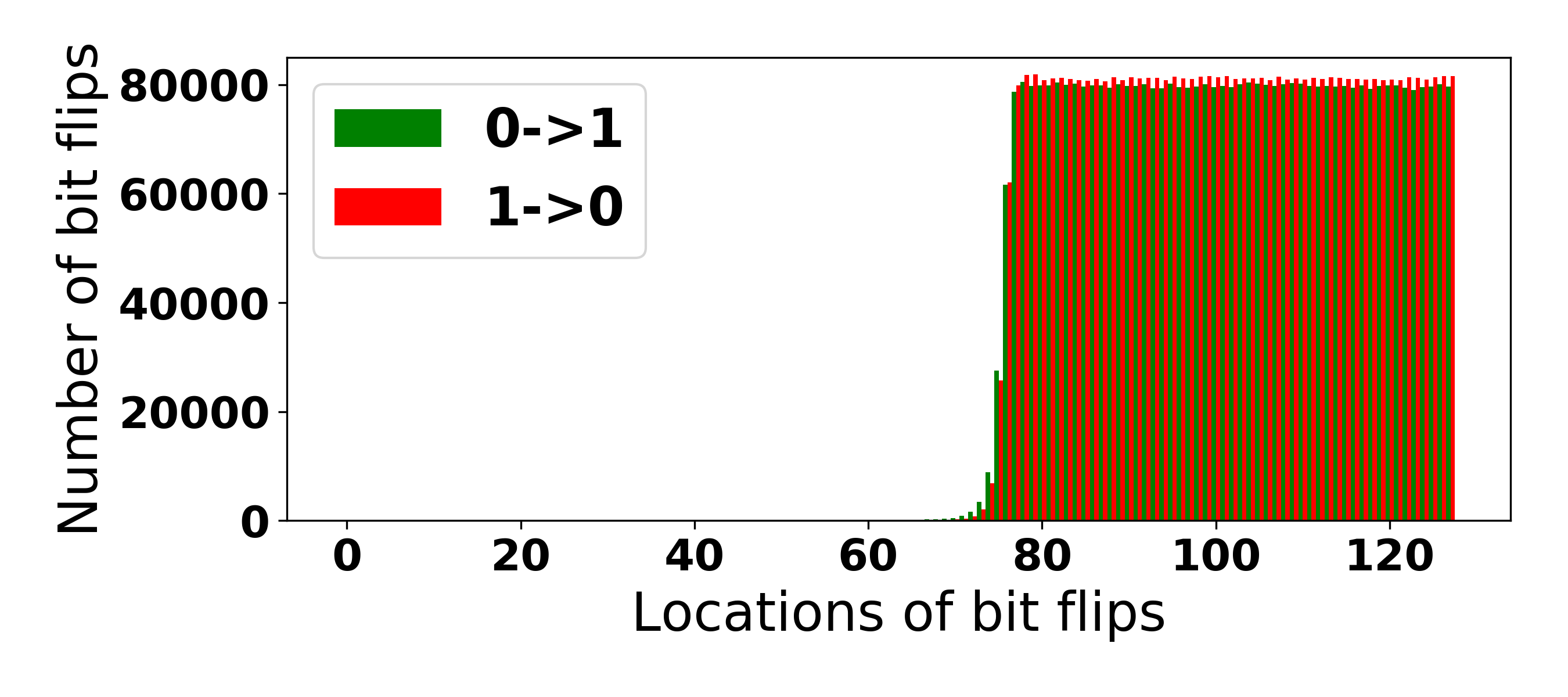}\label{fig:bartrmaxAgaes1000}}
\end{tabular}
\caption{AES-T1000 bit flip distribution for no aging (top) and max. aging (bottom).}
 \end{figure*}

\begin{figure*}
\centering
\begin{tabular}{cc}
\subfloat[Clean IC]{\includegraphics[width=0.5\textwidth]{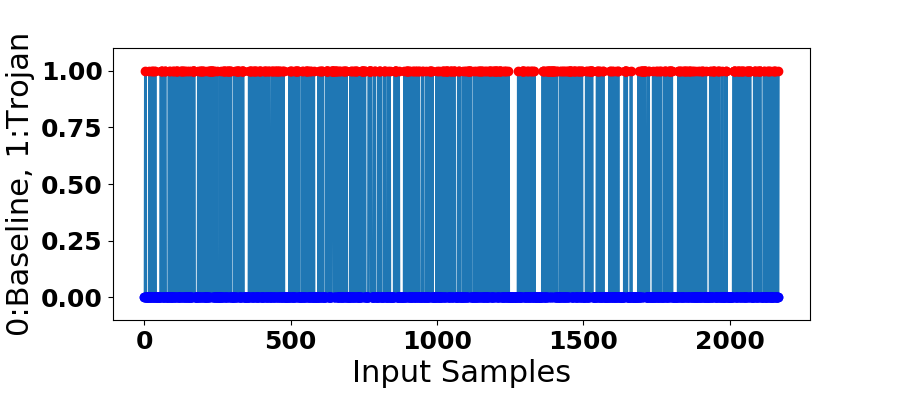}\label{fig:accb1aes1000}}&
\subfloat[IC with Trojan]{\includegraphics[width=0.5\textwidth]{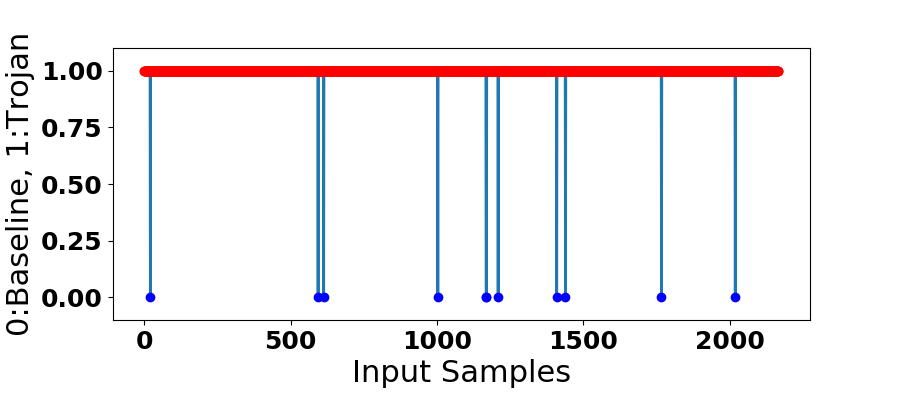}\label{fig:acctr1aes1000}}
\end{tabular}
\caption{Time series of anomaly detection using a single input for AES-T1000.}
 \end{figure*}

\begin{figure*}
\centering
\begin{tabular}{cc}
\subfloat[Clean IC]{\includegraphics[width=0.5\textwidth]{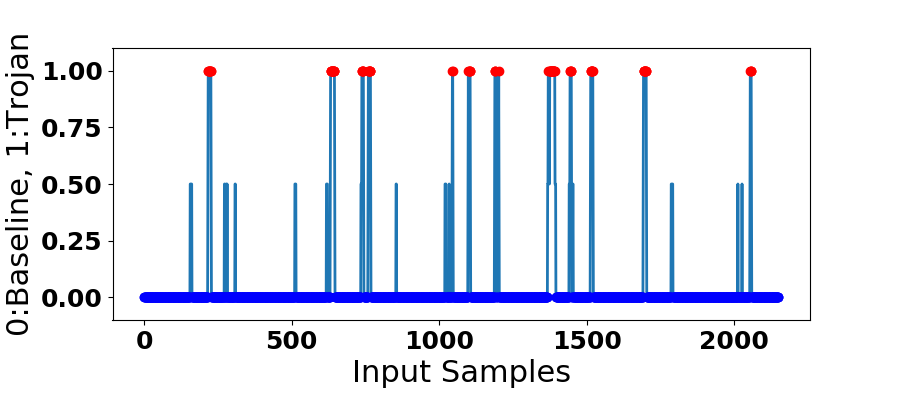}\label{fig:accb32aes1000}}&
\subfloat[IC with Trojan]{\includegraphics[width=0.5\textwidth]{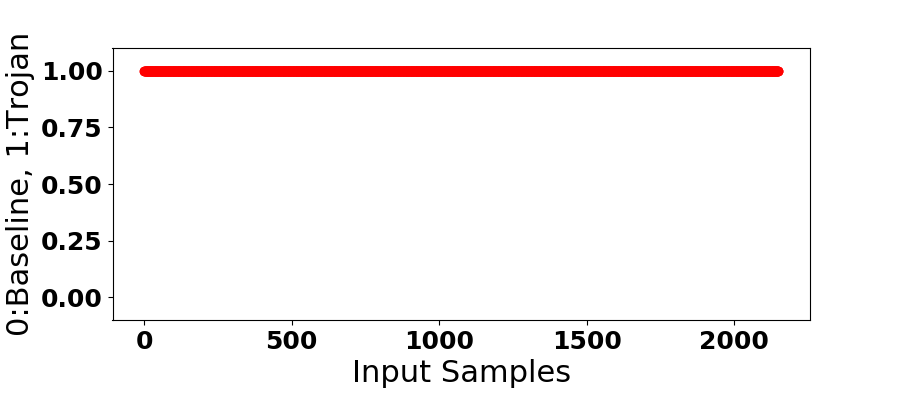}\label{fig:acctr32aes1000}}
\end{tabular}
\caption{Time series of anomaly detection using 32 bins input for AES-T1000.}
\label{fig:anomaly_aes1000}
\end{figure*}

\begin{figure*}
\centering
\begin{tabular}{cc}

\subfloat[AES-T100]{\includegraphics[width=0.49\textwidth]{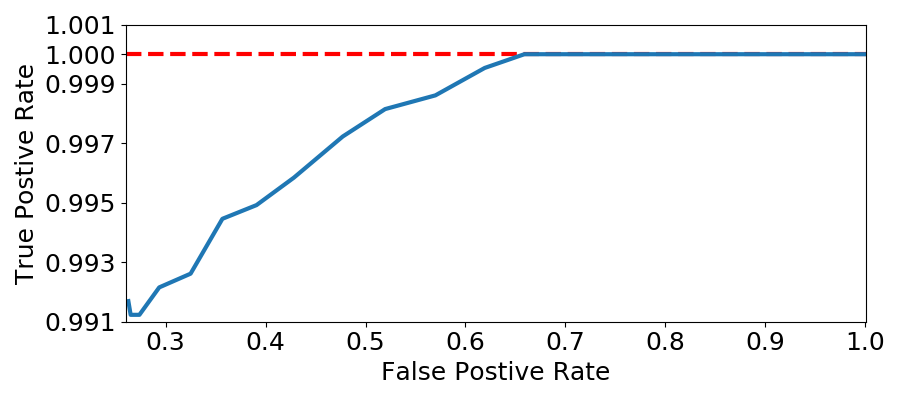}\label{fig:roc_T100}}&

\subfloat[AES-T1000]{\includegraphics[width=0.49\textwidth]{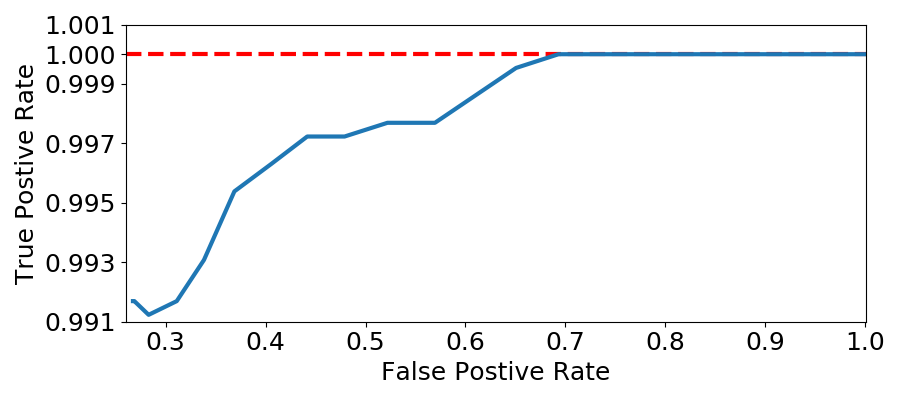}\label{fig:roc_T1000}}
\end{tabular}
\caption{ROC curves for different thresholds using a single input and with a bin size of k = 5.}
 \end{figure*}

\begin{figure*}
\centering
\begin{tabular}{cc}

\subfloat[Clean IC]{\includegraphics[width=0.5\textwidth]{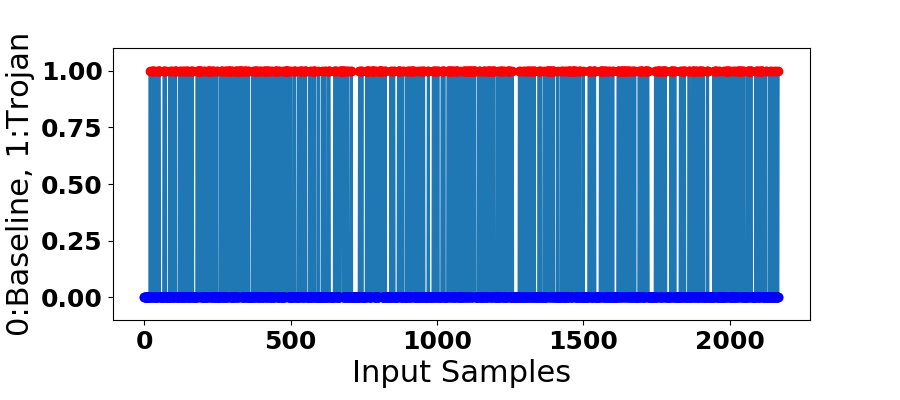}\label{fig:accbasesingleaes1000chipvariation}}&

\subfloat[IC with Trojan]{\includegraphics[width=0.5\textwidth]{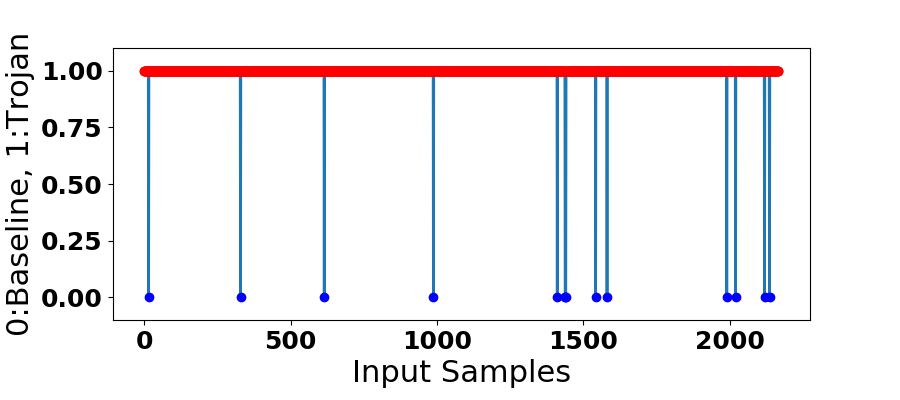}\label{fig:acctrsingleaes1000chipvariation}}
\end{tabular}
\caption{Anomaly detection for IC-to-IC variation with a single bin for AES-T1000.}
 \end{figure*}

 The Trojan in this experiment occupies roughly 0.3\% of the circuit area.
 Figures \subref*{fig:histbasenoAgaes1000},  \subref*{fig:histtrnoAgaes1000} show the histograms of bit flips for no aging case and Figures \subref*{fig:histbasemaxAgaes1000}, \subref*{fig:histtrmaxAgaes1000} show the histograms for maximum aging.  The number of bit flips have been increased from the no aging case to the worst aging case as can be seen by the shift in the histogram towards right. Figures \subref*{fig:barbasenoAgaes1000},  \subref*{fig:bartrnoAgaes1000} show the  bit flips of weighted location at outputs for no aging case. Whereas, Figures \subref*{fig:barbasemaxAgaes1000},  \subref*{fig:bartrmaxAgaes1000} show the bar charts for maximum aging. 
 As explained in the previous section (Experiment 3), the locations of bit flips are concentrated more towards the end. The bit flips and their locations vary from no aging case to worst aging case as well as Trojan free case to Trojan inserted case. These  figures show that the chosen features provide a discernible difference when the Trojan is inserted. With single bin as input (bin size of 5), the accuracy for correctly classifying Trojan-free IC is 75.72\% and that for Trojaned IC is 99.16\% as shown in Figures \subref*{fig:accb1aes1000},  \subref*{fig:acctr1aes1000}, respectively.
When a batch of 32-bins is used, accuracy increases to 99.85\%  for classifying Trojan-free IC and an accuracy of 100\% for detecting Trojan IC as in Figure~\subref*{fig:accb32aes1000}, Figure~\subref*{fig:acctr32aes1000}, respectively. Table \ref{precisiontableaes} summarizes the precision  of the model on AES-T100 and AES-T1000  when using single or multiple inputs. Figures \subref*{fig:roc_T100},  \subref*{fig:roc_T1000} show the ROC curves (true vs false positives at different thresholds). Considering IC-to-IC variations and a single bin as input, the false negative rate for data from clean IC is 20.33\% and 0.87\% for a Trojaned IC as in Figures \subref*{fig:accbasesingleaes1000chipvariation},  \subref*{fig:acctrsingleaes1000chipvariation}. A batch of 32-bins yields 99.15\% accuracy for clean and 100\% accuracy for Trojaned IC. 
 
 \begin{table*}
\centering
\caption{Precision on AES T100, T1000 for different values of k (size of bin).}
\begin{tabular}[t]{lccccccc}
\hline
& \multicolumn{3}{c}{AES-T100} &  \multicolumn{3}{c}{AES-T1000}\\
\hline
& Single bin & 16 bins & 32 bins &  Single bin& 16 bins & 32 bins\\
\hline
k = 1& 0.5680 &  0.5083 & 0.5047 & 0.7579 & 0.9217 & 0.9835\\
\hline
k = 3 & 0.7520 & 0.8918 & 0.9758 & 0.8840 & 0.9962 & 1.0\\
\hline
k = 5 & 0.7972 & 0.9256 & 0.9935 & 0.9103 & 0.9991 & 1.0\\

\hline
\end{tabular}
\label{precisiontableaes}
\end{table*}%

\section{CONCLUSION}
\label{sec:conclusions}
This study shows effectiveness of controlled  aging in detecting Trojans. 
A machine learning classifier and feature selection distinguishes genuine ICs from the ICs in which Trojans are far off the critical path. Over-clocking alone does not distinguish genuine ICs from Trojan-inserted ones. Over-clocking plus aging provides sufficient patterns of output bit errors to detect Trojans. A high detection accuracy is achieved with 32-bins as input with a bin size of 5 in case of AES circuits. We will study detection of the smallest Trojan on and off the critical path. We will consider cell libraries at different voltages so that aging is better approximated using fast voltage switching. While this is a detailed simulation study, it is interesting to demonstrate the method on complex circuits and Trojans and on real ICs.

\bibliographystyle{IEEEtran}
\bibliography{ref}

\ifCLASSOPTIONcaptionsoff
  \newpage
\fi

\begin{IEEEbiography}[{\includegraphics[width=1in,height=1.25in,clip,keepaspectratio]{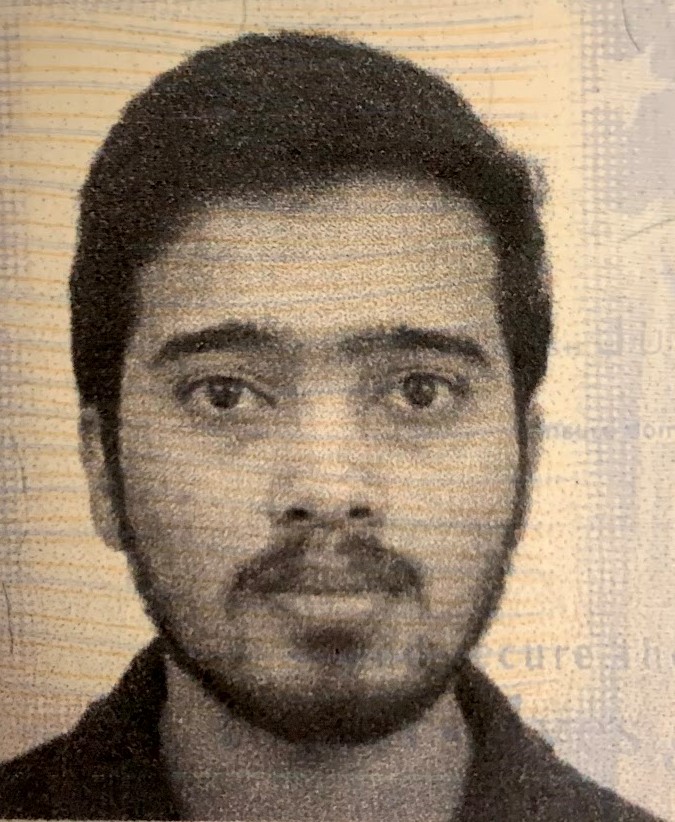}}]{Virinchi Roy Surabhi} received his B.Tech. degree in electrical engineering from the Indian Institute of Technology, Kanpur, India, in 2018. He is currently working toward the Ph.D. degree at NYU Tandon School of Engineering. His research interests include robotics, artificial intelligence and cyber-physical systems.
\end{IEEEbiography}

\begin{IEEEbiography}[{\includegraphics[width=1in,height=1.25in,clip,keepaspectratio]{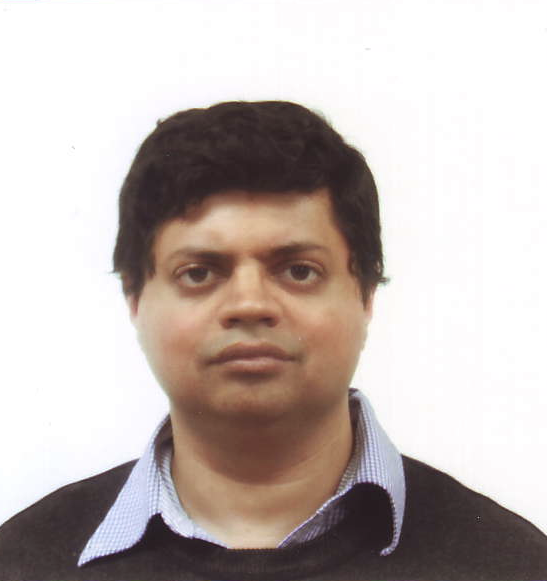}}]{Prashanth Krishnamurthy} received his B.Tech. degree in electrical engineering from Indian Institute of Technology, Chennai in 1999, and M.S. and Ph.D.  degrees in electrical engineering from Polytechnic University (now NYU), Brooklyn, NY in 2002 and 2006, respectively. He is currently a Research Scientist and Adjunct Faculty with the Department of Electrical and Computer Engineering at NYU Tandon School of Engineering, NY, and a Senior Researcher with FarCo Technologies, NY. He has co-authored over 120 journal and conference papers in the broad areas of autonomous systems, robotics, and control systems. He has also co-authored the book ``Modeling and Adaptive Nonlinear Control of Electric Motors'' published by Springer Verlag in 2003.  His research interests include autonomous vehicles and robotic systems, multi-agent systems, sensor data fusion, robust and adaptive nonlinear control, resilient control,  path planning and obstacle avoidance, machine learning, real-time embedded systems, electromechanical systems modeling and control, cyber-physical systems and cyber-security, decentralized and large-scale systems, high-fidelity and hardware-in-the-loop simulation, and real-time software implementations.
\end{IEEEbiography}

\begin{IEEEbiography}[{\includegraphics[width=1in,height=1.25in,clip,keepaspectratio]{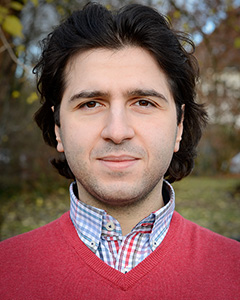}}] {Hussam Amrouch}  (S'11, M'15) is a Research Group Leader at the Chair for Embedded Systems (CES), Karlsruhe Institute of Technology (KIT), Germany. He is leading of the Dependable Hardware research group. He received his Ph.D. degree (Summa cum laude) degree from KIT in 2015. His main research interests are emerging technologies, design for reliability from physics to system level and machine learning. He holds seven HiPEAC Paper Awards. He has three best paper nominations at DAC'16, DAC'17 and DATE'17 for his work on
reliability. He currently serves as Associate Editor at Integration, the VLSI Journal. He is a member of the IEEE. ORCID 0000-0002-5649-3102
\end{IEEEbiography}

\begin{IEEEbiography}[{\includegraphics[width=1in,height=1.25in,clip,keepaspectratio]{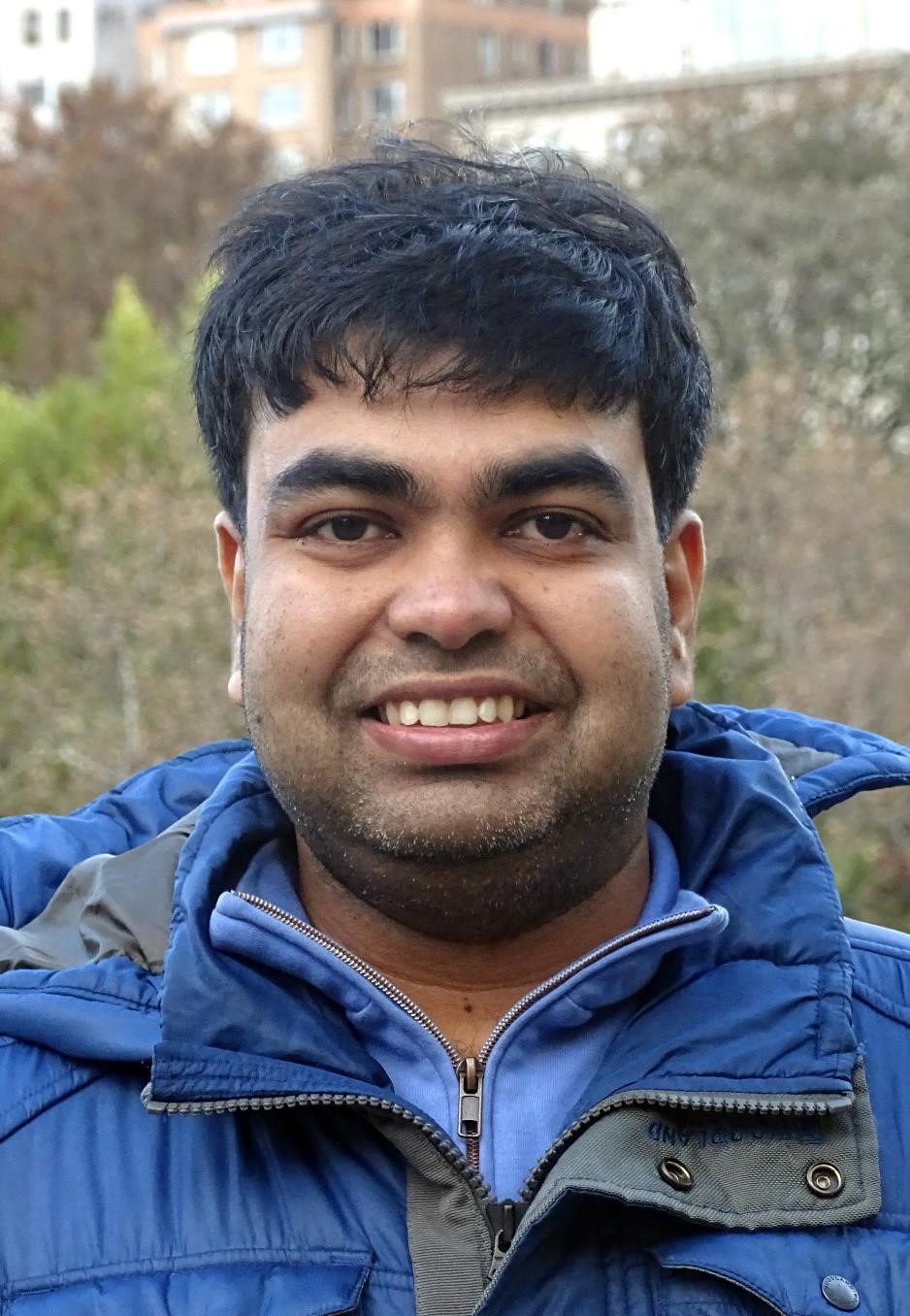}}]{Kanad Basu} received his Ph.D. from the department of computer and information science and engineering, University of Florida. His thesis was focused on improving signal observability for post-silicon validation. Post-Phd, he worked in various semiconductor companies like IBM and Synopsys. At IBM, he was responsible for the design on IBM Power and Z Processors. At Synopsys, he helped in development of DFTMAX Ultra, the state of the art low pin hardware test solution. During his PhD days, he did internships at Intel. Currently, he is an Assistant Research Professor at the Electrical and Computer Engineering Department of NYU. He has authored 2 US patents, 1 book chapter and several peer reviewed journal and conference articles. Dr. Basu was awarded the "Best Paper Award" at the International Conference on VLSI Design 2011. His current research interests are hardware and systems security. He is an Associate Editor for  IET  Computers  and Digital Technology Journal and a Guest Editor for Springer Journal of Electronic Testing. He has served as a Program Committee members for  various conferences including the VLSI Design Conference, Asian Test Symposium, etc. 
\end{IEEEbiography}

 \begin{IEEEbiography}[{\includegraphics[width=1in,height=1.25in,clip,keepaspectratio]{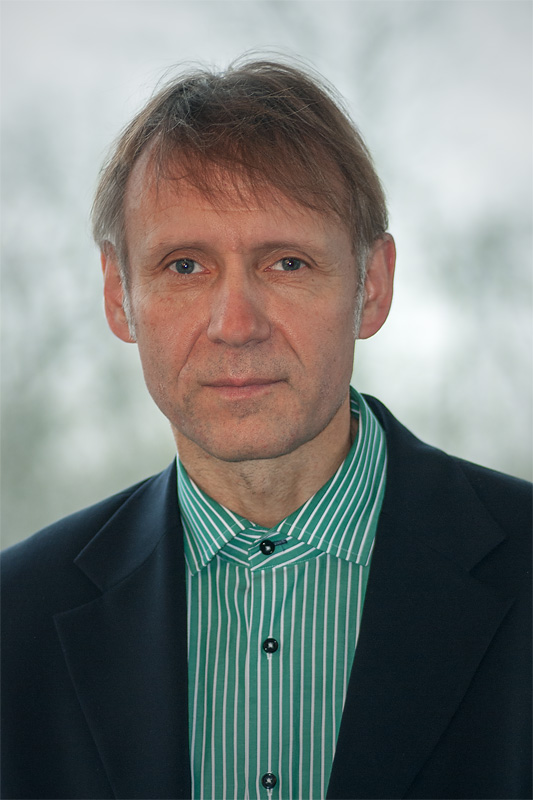}}]{J\"org Henkel}
(M'95-SM'01-F'15) is the Chair Professor for Embedded Systems at Karlsruhe Institute of Technology. Before that he was a research staff member at NEC Laboratories in Princeton, NJ. He received his diploma and Ph.D. (Summa cum laude) from the Technical University of Braunschweig. His research work is focused on co-design for embedded hardware/software systems with respect to power, thermal and reliability aspects. He has received six best paper awards throughout his career from, among others, ICCAD, ESWeek and DATE. For two consecutive terms he served as the Editor-in-Chief for the ACM Transactions on Embedded Computing Systems. He is currently the Editor-in-Chief of the IEEE Design\&Test Magazine and is/has been an Associate Editor for major ACM and IEEE Journals. He has led several conferences as a General Chair incl. ICCAD, ESWeek and serves as a Steering Committee chair/member for leading conferences and journals for embedded and cyber-physical systems. Prof. Henkel coordinates the DFG program SPP 1500 ``Dependable Embedded Systems'' and is a site coordinator of the DFG TR89 collaborative research center on ``Invasive Computing''. He is the chairman of the IEEE Computer Society, Germany Chapter, and a Fellow of the IEEE.
\end{IEEEbiography}

\begin{IEEEbiography}[{\includegraphics[width=1in,height=1.25in,clip,keepaspectratio]{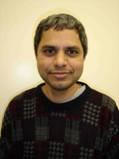}}]{Ramesh Karri} 
  is a Professor of ECE at New York University. He co-directs the NYU Center for Cyber Security (http://cyber.nyu.edu). He also leads the Cyber Security thrust of the NY State Center for Advanced Telecommunications Technologies at NYU. He co-founded the Trust-Hub (http://trust-hub.org) and organizes the Embedded Systems Challenge (https://csaw.engineering.nyu.edu/esc), the annual red team blue team event. He has a Ph.D. in computer science and engineering, from the UC San Diego and a B.E in ece from Andhra University. His research and education activities in hardware cybersecurity include trustworthy ICs; processors and cyber-physical systems; security-aware computer-aided design, test, verification, validation, and reliability; nano meets security; hardware security competitions, benchmarks and metrics; biochip security; additive manufacturing security. He has published over 240 articles in leading journals and conference proceedings. Karri's work on hardware cybersecurity received best paper nominations (ICCD 2015 and DFTS 2015) and awards (ACM TODAES 2018, ITC 2014, CCS 2013, DFTS 2013 and VLSI Design 2012). He received the Humboldt Fellowship and the National Science Foundation CAREER Award. He serves on the editorial boards of several IEEE and ACM Transactions (TIFS, TCAD, TODAES, ESL, D\&T, JETC). He served as an IEEE Computer Society Distinguished Visitor (2013-2015). He served on the Executive Committee of the IEEE/ACM Design Automation Conference leading the Security\@DAC initiative (2014-2017). He delivers invited keynotes, talks, and tutorials on Hardware Security and Trust (ESRF, DAC, DATE, VTS, ITC, ICCD, NATW, LATW, CROSSING, etc.). He co-founded the IEEE/ACM NANOARCH Symposium and served as program/general chair of conferences ( IEEE ICCD, IEEE HOST, IEEE DFTS, NANOARCH, RFIDSEC and WISEC). He serves on several program committees (DAC, ICCAD, HOST, ITC, VTS, ETS, ICCD, DTIS, WIFS).
\end{IEEEbiography}


\begin{IEEEbiography}[{\includegraphics[width=1in,height=1.25in,clip,keepaspectratio]{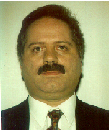}}]{Farshad Khorrami} received his Bachelors degrees in mathematics and electrical engineering in 1982 and 1984 respectively from The Ohio State University. He also received his Master's degree in mathematics and Ph.D. in electrical engineering in 1984 and 1988 from The Ohio State University. He is currently a professor of Electrical \& Computer Engineering Department at NYU, Brooklyn, NY where he joined as an assistant professor in Sept. 1988. His research interests include adaptive and nonlinear controls, robotics and automation, unmanned vehicles, cyber security for cyber-physical systems, embedded systems security, machine learning, and large-scale systems and decentralized control. He has published more than 280 refereed journal and conference papers in these areas. His book ``Modeling and Adaptive Nonlinear Control of Electric Motors'' was published by Springer Verlag in 2003. He also has thirteen U.S. patents on novel smart micro-positioners and actuators, control systems, security, and wireless sensors and actuators. He has developed and directed the Control/Robotics Research Laboratory at Polytechnic University (Now NYU).  He has also commercialized UAVs as well as development of auto-pilots for various unmanned vehicles. His research has been supported by the ARO, NSF, ONR, DARPA, ARL, AFRL, NASA, and several corporations. He has served as general chair and conference organizing committee member of several international conferences.  
\end{IEEEbiography}

\end{document}